\documentclass{article}

\usepackage{amssymb,amsfonts,amsmath}
\usepackage{enumerate,float}
\usepackage{color}
\usepackage{tikz}
\usepackage{subfigure}
\usetikzlibrary{arrows,snakes,backgrounds}

\def\be{\begin{eqnarray}}
\def\ee{\end{eqnarray}}
\def\nn{\nonumber}

\date{}

\def\tr{{\rm tr}\,}
\def\Tr{{\rm Tr}\,}

\def\horr{{{\smallsmile}\atop{\smallfrown}}}

\definecolor{red}{rgb}{1,0,0}
\definecolor{orange}{rgb}{1,0.5,0}
\definecolor{violet}{rgb}{0.7,0,1}

\usepackage{setspace}
\usepackage[utf8]{inputenc}
\usepackage[english]{babel}
\usepackage{amsfonts,amsmath,amssymb,mathtools}
\usepackage[left=2cm,right=2cm,top=2cm,bottom=2cm,bindingoffset=0cm]{geometry}
\usepackage{graphicx}
\usepackage{xcolor}
\usepackage{hyperref}
\usepackage{dsfont}
\usepackage{authblk}
\usepackage{slashed}
\usepackage{braket}
\usepackage{comment}
\usepackage{pb-diagram}
\usepackage{wasysym}
\usepackage{amsthm}
\usepackage{cancel}
\usepackage{tikz}\usetikzlibrary{tikzmark}
\usetikzlibrary{fadings}
\usetikzlibrary{positioning}
\usetikzlibrary{backgrounds} 
\usetikzlibrary{decorations.pathreplacing}
\usetikzlibrary{arrows}
\definecolor{airforceblue}{rgb}{0.36, 0.54, 0.66}	
\definecolor{beige}{rgb}{0.96, 0.96, 0.86}
\definecolor{bittersweet}{rgb}{1.0, 0.44, 0.37}
\definecolor{melon}{rgb}{0.99, 0.74, 0.71}
\definecolor{mustard}{rgb}{1.0, 0.86, 0.35}
\definecolor{lava}{rgb}{0.81, 0.06, 0.13}
\definecolor{magnolia}{rgb}{0.97, 0.96, 1.0}
\definecolor{lavendermist}{rgb}{0.9, 0.9, 0.98}
\definecolor{lavendergray}{rgb}{0.77, 0.76, 0.82}
\definecolor{palepink}{rgb}{0.98, 0.85, 0.87}
\definecolor{palesilver}{rgb}{0.79, 0.75, 0.73}
\definecolor{cadetgrey}{rgb}{0.57, 0.64, 0.69}
\definecolor{anti-flashwhite}{rgb}{0.95, 0.95, 0.96}
\colorlet{Light0anti-flashwhite}{anti-flashwhite!70!white}
\colorlet{Lightanti-flashwhite}{anti-flashwhite!50!white}
\colorlet{Light2anti-flashwhite}{anti-flashwhite!30!white}
\definecolor{linkcolor}{rgb}{0,0,1}
\definecolor{urlcolor}{rgb}{0,0,1}

\usepackage{tikz-cd}
\usetikzlibrary{decorations.markings}
\usetikzlibrary{positioning}
\usepackage{braids}
\usepackage{array}
\usepackage{ytableau}
\usepackage{longtable}
\usepackage{empheq}
\usepackage{multicol}
\usepackage{mathrsfs}
\usepackage{diagbox}
\usepackage[vcentermath]{youngtab}
\usepackage[natbib=true, backend = bibtex, style=numeric-comp, sorting=none,maxbibnames=99]{biblatex}
\hypersetup{pdfstartview=FitH,  linkcolor=linkcolor,urlcolor=urlcolor,citecolor=urlcolor, colorlinks=true}

\def\be{\begin{eqnarray}}
\def\ee{\end{eqnarray}}
\def\nn{\nonumber}

\def\bphi{\bar\phi}

\definecolor{red}{rgb}{1,0,0}
\definecolor{orange}{rgb}{1,0.5,0}
\definecolor{violet}{rgb}{0.7,0,1}

\begin{document}

\title{\bf Bipartite expansion beyond biparticity
}

\author[2,3]{{\bf A. Anokhina}\thanks{\href{mailto:anokhina@itep.ru}{anokhina@itep.ru}}}
\author[1,2,3,4]{{\bf E. Lanina}\thanks{\href{mailto:lanina.en@phystech.edu}{lanina.en@phystech.edu}}}
\author[1,2,3]{{\bf A. Morozov}\thanks{\href{mailto:morozov@itep.ru}{ morozov@itep.ru}}}

\vspace{5cm}

\affil[1]{Moscow Institute of Physics and Technology, 141700, Dolgoprudny, Russia}
\affil[2]{Institute for Information Transmission Problems, 127051, Moscow, Russia}
\affil[3]{NRC "Kurchatov Institute", 123182, Moscow, Russia}
\affil[4]{Saint Petersburg University, 199034, St. Petersburg, Russia}
\affil[5]{Institute for Theoretical and Experimental Physics, 117218, Moscow, Russia}
\renewcommand\Affilfont{\itshape\small}

\maketitle

\vspace{-7.0cm}

\begin{center}
	\hfill MIPT/TH-01/25\\
	\hfill ITEP/TH-01/25\\
	\hfill IITP/TH-01/25
\end{center}

\vspace{4.0cm}

\begin{abstract}

{
The recently suggested bipartite analysis extends the Kauffman planar decomposition to arbitrary $N$,
i.e. extends it from the Jones polynomial to the HOMFLY polynomial.
This provides a generic and straightforward non-perturbative calculus in an arbitrary Chern--Simons theory.
Technically, this approach is restricted to knots and links which possess bipartite realizations,
i.e. can be entirely glued from antiparallel lock (two-vertex) tangles
rather than single-vertex $\cal R$-matrices. 
However, we demonstrate that the resulting {\it positive decomposition} (PD),
i.e. the representation of the fundamental HOMFLY polynomials as
{\it positive integer} polynomials of the three parameters $\phi$, $\bar\phi$ and $D$,
exists for {\it arbitrary} knots, not only bipartite ones.
This poses new questions about the true significance
of bipartite expansion, which
appears to make sense far beyond its original scope,
and its generalizations to higher representations.
We have provided two explanations for the existence of the PD for non-bipartite knots. An interesting option is to resolve a particular bipartite vertex in a not-fully-bipartite diagram
and reduce the HOMFLY polynomial to a linear combination of those for smaller diagrams. If the resulting diagrams correspond to bipartite links, this option provides a PD even to an initially non-bipartite knot. Another possibility for a non-bipartite knot is to have a bipartite clone with the same HOMFLY polynomial providing this PD. 
We also suggest a promising criterium for the existence of a bipartite realization
behind a given PD, which is based on the study of the precursor Jones polynomials.
}
\end{abstract}

















\tableofcontents

\section{Introduction}

Today the knot  theory
is the main point of development for non-perturbative quantum field theory.
Relevant for knots is the $3d$ Chern--Simons QFT \cite{CS,Wit,Schwarz,MSmi,Anorev,Anorev2}.
It is topological, and therefore free of the ``admixture'' of particles
and complications related to Feynman diagrams \cite{FD,FD2,FD3,FD4,FD5} --
and it allows to concentrate on non-perturbative phenomena {\it per se}.
At the same time, it is much richer than just the theory of free fields
-- like the 2d CFT  intensively studied at the previous stage of cognition,
from \cite{BPZ} to \cite{GMMOS}.
In the Chern--Simons theory, we deal with a much wider variety of answers for the gauge-invariant observables
(averages of Wilson lines), and we are still far from putting them
in a clear and non-ambiguous order,
what is a typical problem for non-perturbative vacua.
From mathematical side, we deal with a  badly tamed world of topology
and combinatorics -- thus it does not help much for the needs of physics.

Non-perturbative observables in the three-dimensional Chern--Simons theory
are known as {\it knot polynomials} because they depend {\it polynomially} on essentially
non-perturbative variables $q:=e^{\hbar} := \exp\left(\frac{2\pi i}{g+N}\right)$
and $A:=q^N$.
Most probably, polynomiality is a lucky artefact of the Chern--Simons theory violated in other Yang-Mills theories,
but it is a useful artefact at the present state of affairs because it helps to avoid irrelevant speculations
often appearing in discussion of more transcendental answers.
The standard approach to study knot polynomials
is through the Reshetikhin--Turaev approach \cite{RT,RT2,MMM,MMM2,MMM3,MMM4,MMM5},
which implicitly uses the temporal gauge $A_0=0$ to make the action quadratic \cite{MSmi}
and maps 2d non-planar {\it knot diagrams}
into products and graded traces of quantum ${\cal R}$-matrices.
However, in the case of $N=2$ (for the gauge group $\mathfrak{su}(2)$, i.e. the Jones polynomial rather
than the HOMFLY polynomial and more general polynomials)
there is another possibility -- to use {\it Kauffman bracket}
and {\it planar decomposition} \cite{Kb} of knot diagrams.
This can be considered as arising from a peculiar choice of ${\cal R}$-matrix --
but in fact it is conceptually different and rather associated with
Khovanov {\it categorification} \cite{Kho,Kho2,Kho3} of these diagrams.
We refer to \cite{DM13,DM13-2,DM13-3,DM13-4,DM13-5,DM13-6,DM13-7} for physical aspects of this construction
and switch instead to a still different and fresh alternative.

In \cite{bipfund}, we started a systematic consideration of the HOMFLY calculus
for the set of knot diagrams often referred to as ``bipartite'' \cite{bip}
which are made entirely by gluing the {\it antiparallel lock} tangles.
This set is huge, though not exhaustive, i.e. it does not enumerate {\it all}
observables in the Chern-Simons theory. Instead, it allows {\it planar decomposition} for any $N$, not just for $N=2$
like the original Kauffman calculus.
Given the importance of this phenomenon, it looks suspicious that
it is restricted to the particular set of diagrams.
It is natural to ask whether and how this restriction can be lifted.

What we do in this paper, we try to extend {\it planar decomposition}
to arbitrary HOMFLY polynomials, i.e. to represent them as polynomials of
the peculiar variables $\phi,\bar\phi$ and $D$, defined in (\ref{bippars}) below,
with integer positive coefficients.
This is how the HOMFLY polynomials emerge from bipartite diagrams, but it appears that an {\it arbitrary} polynomial
with the {\it symmetricity}  property ${\rm Pol}(q^{-2})={\rm Pol}(q^2)$, satisfied by the fundamental HOMFLY polynomial,
has such a decomposition -- with no reference to biparticity.
This confirms our original optimism, but instead raises
a complementary, still interesting inverse question --
if we can judge if there is any bipartite diagram behind the given polynomial.

To better formulate this dilemma, we introduce two notions:
{\it positive decomposition} (PD) applicable to arbitrary symmetric polynomial,
without a reference to knot diagrams,
and {\it bipartite expansion} (BE) which is the particular PD following from planar decomposition
of the particular bipartite diagram.
Our {\bf main question is if one can decide that PD$\stackrel{?}{=}$BE}, i.e. when there is some bipartite diagram
behind the given polynomial.
The question is not simple, and in fact, it also touches the old but still open problem,
if any symmetric polynomial can be the HOMFLY polynomial of some knot.

To simplify our main question, we select a particular class of {\it chiral} PD, when there is no dependence on $\bar\phi$,
only on $\phi$ and $D$ (\!{\it anti-chiral} PD are instead independent of $\phi$).
On the other hand, chiral BE naturally arise from {\it chiral} bipartite diagrams where all locks are of the same orientation.
We can then ask \ {\bf if  $ \ {\rm PD}\stackrel{?}{=}{\rm BE}\Big|_{\rm chiral_{\phantom\int}}$} on the smaller set of chiral polynomials.
This question looks much simpler, because chiral PD is unique -- while it is ambiguous in non-chiral case,
see Section \ref{ambig} for details.
Still, it appears, that there are non-bipartite HOMFLY with chiral PD, the simplest examples are the knots $9_{35}$ and $9_{49}$.
In this case, we can say that {\bf PD = fake BE}.
We consider two possibilities (which do not exclude each other) to explain this phenomenon.
One, see Section \ref{sec:sum},
is that the knot is associated with a {\it sum} of bipartite diagrams, with $\phi$-dependent coefficients.
This is indeed the case both for $9_{35}$ and $9_{49}$.
Another option, discussed along with the other subjects in Section \ref{obsta},
 is to try to attribute {\it fake BE} to the existence of {\it clones} --
knots with coinciding (fundamental) HOMFLY, which are quite abundant.
The hope can be that a non-bipartite knot has a bipartite clone.
Indeed, it might be true for $9_{49}$, but we did not achieve full understanding of the situation, even in chiral case.
The problem is two-fold -- to find a clone and to judge if it is bipartite. Still, there is an example of the non-chiral non-bipartite knot $10_{103}$ having the bipartite clone $10_{40}$. 

In Section \ref{sec:prec}, we develop a powerful criterium, based on the idea of {\it precursor diagrams} \cite{bipfund},
when lock tangles are substituted by singe vertices and the coefficient $\phi$ by $(-q)$ from Kauffman-Jones expansion.
This is a well-defined procedure for any PD,
and it is always possible, if bipartite diagram exists --
thus the downgrade of every BE must be the Jones polynomial for some link.
This requirement appears quite restrictive, but its true power remains a question.

However, the main problem is that {\it chiral} PD is not available for all symmetric polynomials --
thus we need to consider the non-chiral case.
Then we face the following problems:
\begin{itemize}
\item{} PD is not unique.
\item{} PD is not fully algorithmic.
\begin{itemize} \item{} The origin of both difficulties is the same -- the three parameters are algebraically dependent,
$G:=\phi + \bphi+ \phi\bphi D=0$, and every decomposition is defined modulo $G$.
However, additions/subtractions of multiples of $G$ can change positivity.
An algorithmically derived decomposition often needs such a correction to become positive,
but positivity can be achieved in many ways.
Some of them are {\it minimal} -- loose positivity by any subtraction of $G$.
One could try to decrease ambiguity by asking for minimality.
But even minimal PD is not unique -- there are a lot of ``local minima''.

\end{itemize}

\item{} On the other hand, for a given knot, a bipartite diagram is not unique, and different non-chiral diagrams
can give different PD (equivalent modulo $G$).
\item{} It is unclear if these PD, all or some, should or should not be minimal.
\item{} Still, quite a number of knots is not bipartite, so their PD cannot be BE.
\item{} Given a PD, one can ask if it is BE.

\begin{itemize}
\item{} For chiral PD the precursor criterium is rather efficient,
but it works much worse in the non-chiral case.
Still, together with other ideas surveyed in Section \ref{obsta}, it allows
to identify many ``parasitic'' PD, which are not BE.
\end{itemize}
\end{itemize}

We discuss all these problems, consider examples, and try to explain the difficulties --
but without definite conclusions in most cases.
They remain for future work.

\bigskip

Three other subjects also deserve mentioning, of which only the first one is briefly considered
in this paper in Section \ref{reps}.

A lift to non-trivial representations, even symmetric, remains
somewhat difficult \cite{bipsym}, still possible --
and it will certainly be simplified during the further studies.
Since this planar calculus is new,
and the number of examples studied in \cite{bipfund,bipsym} was quite narrow\footnote{
It, however, included a very interesting Kanenobu family, which is quite difficult
to handle by usual methods.},
it is necessary to extend it to motivate more people to look at the subject --
thus, we use the chance to address this story from the perspective of PD and BE.
In fact, derivation of {\it colored} BE through PD, once developed,
can appear simpler than a direct calculation from
bipartite diagrams which involves somewhat tedious projector calculus.
At the moment, however, the very definition of PD for symmetric representations is obscure --
the coefficients are now made from more variables, and their specification is not fully fixed yet.

Given the close relation of BE to Kauffman bracket -- of which it is a direct generalization --
it is also natural to consider the associated Khovanov calculus,
which would substitute powers of $D$ to nilpotent maps between $D$-dimensional vector spaces.
A natural question is if this can also make sense for PD, not only for BE.
We just raise this question leaving the answer for future.
The most interesting part of the question is if this approach can give the same or different
answers from Khovanov-Rozansky matrix factorization \cite{Kho,Kho2,Kho3,Kras}.

The third potentially interesting observation is that the parameter $\phi$ is expressed through $z=q-q^{-1}$ rather than $q$ itself.
Thus, PD provides a new kind of expansion of the fundamental HOMFLY polynomial, in $z$ and $D$,
instead of the usual one in $q$ and $A$.
We call such expansions {\it semi-perturbative} because they are in non-perturbative parameters
$q=e^\hbar$ rather than $\hbar$, still, they are {\it expansions} --
which, moreover, has a clear meaning in the cases of  Kauffman expansion and our BE.
It deserves noting that there are alternative attempts to introduce meaningful $z$ expansions \cite{Ito} --
though in that case the second variable is $A$, rather than $D$.
This can point to the future role of semi-perturbative expansions as certain substitutes
of the pure perturbative Vassiliev calculus. 

\bigskip

The structure of the paper is as follows.

First, in Section \ref{sec:plad}, we remind the idea \cite{bipfund} of planar decomposition for bipartite diagrams.

Second, in Section \ref{sec:regsearch}, we introduce the notion of {\it positive decomposition} (PD)
of an {\it arbitrary} symmetric polynomial.
The central point of the present paper is the  formula (\ref{chirexpanswer}).
In the chiral case (no $\bphi$ dependence in the polynomial) it is automatically positive and unique.
In the non-chiral case, it can be {\it made} positive by additional operations.

However, the procedure in the non-chiral case is not unique, and we discuss ambiguities in Section \ref{ambig}.
Here, we face a question which remains open -- if the resulting polynomial has anything to do with any
bipartite diagram, and if yes, what is its relation to a given polynomial,
i.e. if it is the HOMFLY polynomial for this diagram.
In other words, we pose the question if the given PD is BE or fake BE. We provide some results of our investigations in Section~\ref{sec:expec}.

We begin from restrictions on positive polynomials, which {\it are} the bipartite HOMFLY polynomials.
Two such criteria are considered in Sections \ref{sec:D1} and \ref{sec:prec}.
But in fact, there is a very different option, surveyed in Section \ref{sec:sum}:
the resulting positive polynomial can be a {\it sum} of the HOMFLY polynomials for different diagrams.
This happens when in an original diagram, {\it not fully bipartite},
we decompose {\it some} lock tangles,
and each of the resulting smaller diagrams corresponds to a bipartite link 
(while an original one does not).

Finally, in Section \ref{reps}, we attempt to generalize the concept of positive decomposition
to symmetric representations, BE of which were studied in \cite{bipsym}.
This raises a new set of open problems, which help us to get a broader view on the subject.

In Sections \ref{summary} and \ref{conc}, we summarize what we learned about positive decompositions and
provide one more list of open questions.

\setcounter{equation}{0}
\section{Basics of planar decomposition 
}\label{sec:plad}

In this section, we introduce the notions of positive decomposition (in Section~\ref{sec:PD-def}), planar decomposition and bipartite expansion (in Section~\ref{sec:BE-def}) extensively used in the present paper. We consider the simplest examples of bipartite expansions in Section~\ref{sec:BE-simplest-ex}. In the HOMFLY case, the planar decomposition is applicable to the peculiar set of bipartite links and is compatible with the Kauffman bracket at $A=q^2$, see Section~\ref{sec:Kauffman}. The bipartite family seems quite abundant. Currently, only 12 knots of crossing numbers up to 10 are known to be non-bipartite. Their HOMFLY polynomials do not look very different from the bipartite ones, see Section~\ref{sec:bip-comm}, what gives hope that positive decomposition makes sense beyond the bipartite case.

\subsection{Definition of PD}\label{sec:PD-def}

Positive decomposition (PD) of the reduced (normalized) fundamental HOMFLY polynomial is
\be
\boxed{
H_\Box^{\cal L}(A,q) = {\rm Fr}\cdot \sum_{a,b,c}  D^a\phi^b\bar\phi^c
= A^{-2(n_\bullet - n_\circ)}\sum_k \sum_{i=0}^{n_\bullet} \sum_{j=0}^{n_\circ} {\cal N}_{ijk} D^k \phi^i\bphi^j
}
\label{bipexp}
\ee
where\footnote{In the text, we use the notations $\{x\}\equiv x-1/x$, $z=\{q\}$, $v=A^{-1}$.}
\be
\phi = A\{q\} = \frac{z}{v}\,, \ \ \ \ \bar\phi = -\frac{\{q\}}{A} = -zv\,, \ \ \ \  D =\frac{\{A\}}{\{q\}} = -\frac{v-v^{-1}}{z}\,,
\label{bippars}
\ee
and in the first equality all items come with unit coefficients.
However, some terms can coincide, and  if we  sum over different non-negative triples $i$, $j$, $k$,
then the coefficients are non-negative integers ${\cal N}_{ijk}$.
Non-negativity of powers $i,j,k$ and of the coefficients ${\cal N}_{ijk}$ are the two positivity properties of PD.

Since the three parameters $\phi$, $\bar\phi$, $D$ are not independent, the expansion (\ref{bipexp}) need not be unique for a given knot/link.
The expression~\eqref{bipexp} is defined only modulo
\be
G:=\phi+\bar\phi + \phi\bar\phi D
\ee
which vanishes on the locus (\ref{bippars}).
However, factorization modulo $G$ should preserve {\it positivity}.
The non-trivial framing factor ${\rm Fr}=A^{-2(n_\bullet - n_\circ)}$ is just a power of $A^2$,
it can also be written as a power of $1+\phi D=A^2$ or $1+\bar\phi D=A^{-2}$,
depending on whether the power is positive or negative --
and does not weaken the polynomiality and positivity constraints\footnote{By positive/negative polynomial we mean a polynomial, which has positive/negative coefficients in front of all terms.}.

When the expansion~\eqref{bipexp} does not contain $\bphi$, it is unique (see Section~\ref{sec:chiral}), and we call knots, which possess {\it such} a PD, {\it chiral}.
For $\phi$-independent PD, we use the term {\it anti-chiral}.
We will see that there are quite many chiral knots, but also quite many ones are non-chiral (see examples in Tables~\ref{tab:ch-PD-1}--\ref{tab:non-ch-PD-3}).
Naturally, the mirror image of a chiral knot is anti-chiral -- though exact realization of this property is somewhat delicate,
see Section \ref{partners}.
We will also see an interesting dilemma: for chiral knots the framing factor is  a {\it negative} power of $A^2$,
thus, the framing factor breaks either polynomiality or chirality --
therefore, we usually consider the framing factor separately.

\subsection{PD from bipartite diagrams}\label{sec:BE-def}

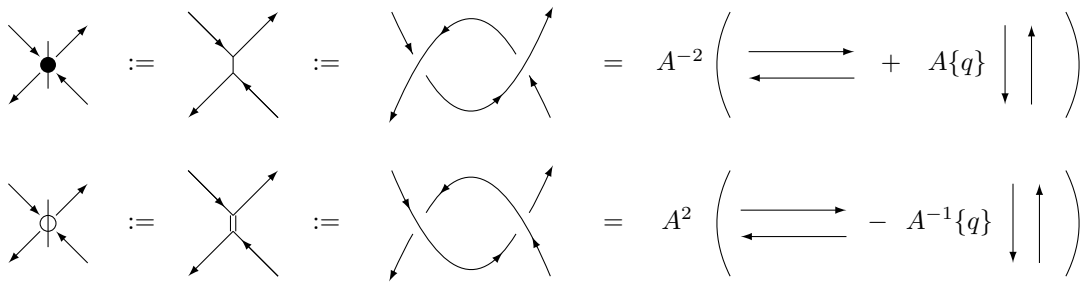
\begin{figure}[h!]
\begin{picture}(100,110)(-130,-85)

\put(20,0){
\put(-105,15){\vector(1,-1){12}} \put(-87,3){\vector(1,1){12}}
\put(-93,-3){\vector(-1,-1){12}} \put(-75,-15){\vector(-1,1){12}}
\put(-90,0){\circle*{6}}  \put(-90,-9){\line(0,1){18}}

\put(-60,-2){\mbox{$:=$}}
}

\put(0,0){\put(-17,20){\line(1,-1){17}}\put(-17,20){\vector(1,-1){14}}   \put(0,3){\vector(1,1){17}}
 \put(0,-3){\vector(-1,-1){17}}   \put(17,-20){\line(-1,1){17}} \put(17,-20){\vector(-1,1){14}}
 \put(0,-3){\line(0,1){6}}}

\put(10,0){

\put(20,-2){\mbox{$:=$}}

\qbezier(50,20)(55,9)(58,4) \qbezier(63,-4)(85,-40)(110,20)
\put(56,8){\vector(1,-2){2}} \put(90,-13){\vector(1,1){2}} \put(109,18){\vector(1,2){2}}
\qbezier(50,-20)(75,40)(97,4)  \qbezier(102,-4)(105,-9)(110,-20)
\put(104,-8){\vector(-1,2){2}} \put(70,13){\vector(-1,-1){2}} \put(51,-18){\vector(-1,-2){2}}

}

\put(20,0){
\put(120,-2){\mbox{$=$}}
\put(140,-2){\mbox{$A^{-2}$}}

\put(5,0){
\qbezier(163,-20)(153,0)(163,20)
\qbezier(290,-20)(300,0)(290,20)

\put(100,65){
\put(70,-60){\vector(1,0){40}}
\put(110,-70){\vector(-1,0){40}}
\put(120,-67){\mbox{$+\ \ \ A \{q\}$}}
\put(167,-50){\vector(0,-1){30}}
\put(177,-80){\vector(0,1){30}}
}
}
}

\put(0,-60){

\put(20,0){
\put(-105,15){\vector(1,-1){12}} \put(-87,3){\vector(1,1){12}}
\put(-93,-3){\vector(-1,-1){12}} \put(-75,-15){\vector(-1,1){12}}
\put(-90,0){\circle{6}}  \put(-90,-9){\line(0,1){18}}

\put(-60,-2){\mbox{$:=$}}
}

\put(0,0){
\put(0,0){\put(-17,20){\line(1,-1){17}}\put(-17,20){\vector(1,-1){14}}   \put(0,3){\vector(1,1){17}}
 \put(0,-3){\vector(-1,-1){17}}   \put(17,-20){\line(-1,1){17}} \put(17,-20){\vector(-1,1){14}}
\put(-1,-3){\line(0,1){6}} \put(1,-3){\line(0,1){6}}
}

\put(30,-2){\mbox{$:=$}}

\put(10,0){
\qbezier(50,20)(75,-40)(97,-4)  \qbezier(102,4)(105,9)(110,20)
\qbezier(50,-20)(55,-9)(58,-4) \qbezier(63,4)(85,40)(110,-20)
\put(55,9){\vector(1,-2){2}} \put(90,-13){\vector(1,1){2}} \put(109,18){\vector(1,2){2}}
\put(105,-9){\vector(-1,2){2}} \put(70,13){\vector(-1,-1){2}} \put(51,-18){\vector(-1,-2){2}}
}

\put(140,-2){\mbox{$=$}}

\put(25,0){
\put(137,-2){\mbox{$A^2$}}

\put(100,65){
\put(67,-60){\vector(1,0){40}}
\put(107,-70){\vector(-1,0){40}}
\put(115,-67){\mbox{$-\ \ A^{-1}\{q\}$}}
\put(-25,0){
\put(195,-50){\vector(0,-1){30}}
\put(205,-80){\vector(0,1){30}}
}
}}}
\put(25,0){\qbezier(163,-20)(153,0)(163,20)
\qbezier(290,-20)(300,0)(290,20)}

}

\end{picture}
\caption{\footnotesize
The planar decomposition of the positive (in the first line) and negative (in the second line) lock vertices in the \textit{topological} framing.
In \cite{bipfund}, we used the {\it vertical} framing, but the {\it topological} one is more convenient for our considerations,
thus, we use it throughout the present paper.
}\label{fig:pladeco}
\end{figure}

Originally, we deduced (\ref{bipexp}) in \cite{bipfund} from the study of bipartite diagrams, consisting entirely of the antiparallel
lock tangles from Fig.\,\ref{fig:pladeco}. The observation of \cite{bipfund} was that the HOMFLY polynomials, built from planar decomposition of bipartite diagrams, possess PD, and we call this specific version of PD a {\it bipartite expansion} (BE). Then, the sum in~\eqref{bipexp} goes over planar $2^{n_\bullet+ n_\circ}$ resolutions of $n_\bullet$ positive and $n_\circ$ negative antiparallel locks. 

BE is of course a direct generalization of the Kauffman bracket \cite{Kb}
and related Kauffman calculus for the Jones polynomials, reviewed in detail in \cite{DM13,DM13-2,DM13-3,DM13-4,DM13-5,DM13-6,DM13-7}.
The advantage is that it is now {\bf applicable to an arbitrary $N$}, i.e. realizes the dream of \cite{DM13,DM13-2,DM13-3,DM13-4,DM13-5,DM13-6,DM13-7}.
The disadvantage is that it refers to bipartite diagrams, and not all the knots are bipartite \cite{bipfund}.

A purpose of this paper is to {\bf lift this restriction and to introduce a positive decomposition for all $N$
and for all knots}.
We proceed to  this problem in Section \ref{sec:regsearch} below, where the extension will be actually to
arbitrary symmetric polynomials (it is still unknown if any of them can be the HOMFLY polynomials of some knots).
The problem for the follow up Sections \ref{ambig}--\ref{sec:sum} is to find if a chosen PD is a BE.
It is not clear to us how significant this question is, still it appears interesting by itself
and deserves a paper, or at least a significant part of it.

Another problem
is {\bf if BE (\ref{bipexp}) 
actually depends on a choice of a bipartite diagram}.
It does since, say, chirality implies that all the lock tangles have the same orientation,
and additional pairs lock--anti-lock can be easily added, violating chirality of  PD.
The question is therefore more delicate -- {\it how much} (whatever this can mean) BE
depends on a diagram, if any positive equivalent of the given answer can be obtained from some
Reidemeister-equivalent bipartite diagram?
In other words, do the bipartite Reidemeister orbits coincide with those obtained by factoring of
positive polynomials by $G$?

In the present section, we comment a little more on the bipartite case.

\subsection{Relation to Kauffman calculus at $N=2$}\label{sec:Kauffman}

\begin{figure}[h!]
\begin{picture}(100,80)(-200,-40)

\put(-60,-20){\line(1,1){40}}
\put(-20,-20){\line(-1,1){18}}
\put(-60,20){\line(1,-1){18}}

\put(10,-2){\mbox{$=$}}

\qbezier(30,20)(50,0)(70,20)
\qbezier(30,-20)(50,0)(70,-20)

\put(85,-2){\mbox{$- \ \ \ q $}}

\qbezier(125,20)(145,0)(125,-20)
\qbezier(150,20)(130,0)(150,-20)

\put(-68,22){\mbox{$i$}}  \put(-15,22){\mbox{$j$}}  \put(-68,-28){\mbox{$k$}}  \put(-15,-28){\mbox{$l$}}

\end{picture}
\caption{\footnotesize The  celebrated Kauffman bracket -- the planar decomposition of the ${\cal R}$-matrix vertex for the fundamental representation of $\mathfrak{sl}_q(2)$.
In this case ($N=2$), the conjugate of the fundamental representation is isomorphic to it,
thus, tangles in the picture has no orientation.
}
\label{fig:Kauff}
\end{figure}
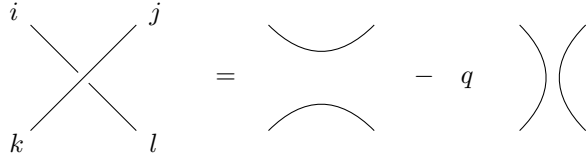

If we have a bipartite diagram, then the planar decomposition in Fig.\,\ref{fig:pladeco} implies that every lock tangle is substituted
by $\ \horr \ + \ \phi \ \cdot \ )(\ $, and the open ends at all locks are connected
according to the structure of the diagram.
At the same time, in Kauffman calculus (due to Fig.\,\ref{fig:Kauff}) the substitution of the same lock is rather
$\ \horr \ -2q \ \cdot \ )( \ + \ q^2D \ \cdot \ )( \ $.
However, at $N=2$, i.e. $A=q^2$, and $D=q+q^{-1}$ we get the exact equality
\be
\phi := Az = A(q-q^{-1})  \ \stackrel{N=2}{ =}\   q^2 D - 2q\,.
\ee
This means that for $N=2$, the Kauffman decomposition when applied to bipartite diagrams,
{\it literally}, tangle-by-tangle, reproduces the BE\footnote{However, there is a subtlety as the Kauffman bracket is related to \textit{unoriented} tangles, and the bipartite expansion is related to oriented \textit{tangles}. The difference reveals itself in a rather involved framing factor in the Kauffman case.}.

An obvious question is if this implies anything for Khovanov calculus.
It is left beyond the scope of \cite{bipfund} and of this paper as well.

\subsection{The simplest examples}\label{sec:BE-simplest-ex}

To illustrate the notion of BE we provide a few simple examples already at this stage.
There are many more examples in Section \ref{sec:regsearch} below. 

\medskip

\noindent $\bullet$ One bipartite crossing, $n=1$ with two different closures, providing unknot and Hopf link,
each with two different orientations.

{\bf The unknot $=\bigcirc\,$:}
{\small \be
\!\!\!\!\!\!\!\!\!\!\!\!\!\!\!\!\!\!\!\!\!\!\
\begin{aligned}
n_\bullet&=1 \quad
\boxed{H^{\bigcirc}_\Box = A^{-2}\cdot \frac{D+\phi D^2}{D} = A^{-2}\left(1 + \phi D\right) }
= A^{-2}\left( 1 + Az\cdot \frac{A-A^{-1}}{z}\right) =1  =  A^{-1}\cdot \frac{q\{Aq\}-q^{-1}\{A/q\}}{\{q^2\}}\,,   \\
n_\circ&=1 \quad
\boxed{H^{\overline{\bigcirc}}_\Box = A^{2}\cdot \frac{D+\bar\phi D^2}{D}
= A^{2}\left(1 + \bar\phi D\right)} = A^{2}\left( 1 - \frac{z}{A}\cdot \frac{A-A^{-1}}{z}\right) =1
 =  A^{+1}\cdot \frac{q^{-1}\{Aq\}-q\{A/q\}}{\{q^2\}}\,.
\end{aligned}
\ee}

{\bf The Hopf link:}
\be \label{Hopf}
\!\!\!\!\!\!\!\!\!\!\!\!\!\!\!\!\!\!\!\!\!\!\
\begin{aligned}
n_\bullet&=1 \quad
\boxed{H^{\rm Hopf}_\Box = A^{-2}\cdot \frac{\phi D + D^2}{D} =  A^{-2}\left(\phi + D\right)}
= A^{-2}\left( Az+\frac{A-A^{-1}}{z}\right) =  A^{-2}\cdot \frac{q^2\{Aq\}+q^{-2}\{A/q\}}{\{q^2\}}\,,   \\
n_\circ&=1 \quad
\boxed{H^{\overline{\rm Hopf}}_\Box = A^{2}\cdot \frac{\bar\phi D + D^2}{D}
=  A^{2}\left(\bar\phi + D\right)} = A^{2}\left( - \frac{z}{A} + \frac{A-A^{-1}}{z}\right)
= A^{+2}\cdot \frac{q^{-2}\{Aq\}+q^{2}\{A/q\}}{\{q^2\}}\,.
\end{aligned}
\ee
The last expressions in the rows are the standard ones for torus knots/links.
Note that they also have non-trivial framing factors, and for the unknot they are different from bipartite,
because the knot diagrams are different: bipartite one represents unknot as  a twist-knot diagram with two crossings
(one bipartite), while the torus one has just one crossing.
For Hopf the diagrams can look the same, but in fact the orientations are different:
the two components have opposite orientations in the bipartite realization and coincident in the torus one.
For Hopf the answers are the same (namely, reversing the mutual orientation of the components is equivalent to reversing the orientation of the crossings), but it is no longer true for $n>1$ \cite{katlas}, as an example see~\eqref{T[2,4]}.

\medskip

\noindent $\bullet$ Two bipartite crossings, $n=2$, in different positions and of different orientations.

{\bf The trefoil knot as a twist knot and its mirror image:}
{\footnotesize
\be\label{trefoil}
\begin{aligned}
\!\!\!\!\!\!\!\!\!\!\!\!\!\!\!
n_\bullet &= 2 \quad \boxed{H^{3_1}_\Box =   A^{-4}\cdot \frac{(1+\phi^2)D + 2\phi D^2}{D}} = A^{-4}\cdot (1+A^2z^2 +2Az D) =
-A^{-4}+A^{-2}(q^2+q^{-2}) =
 A^{-3}\cdot \frac{q^3\{Aq\}-q^{-3}\{A/q\}}{\{q^2\}}\,,  \\
 \!\!\!\!\!\!\!\!\!\!\!\!\!\!\!
 n_\circ &= 2 \quad \boxed{H^{\overline{3}_1}_\Box =   A^{4}\cdot \frac{(1+\bar\phi^2)D + 2\bar\phi D^2}{D}} = A^{4}\cdot (1+A^{-2}z^2 -2A^{-1}z D) =
-A^{4}+A^{2}(q^2+q^{-2}) =
 A^{+3}\cdot \frac{q^{-3}\{Aq\}-q^{3}\{A/q\}}{\{q^2\}}\,.
\end{aligned}
\ee
}

{\bf The figure-eight knot (which is equivalent to its mirror image):}
{\small \be\label{PD-4_1}
\!\!\!\!\!\!\!\!\!\!\!\!\!\!\!
n_\bullet=n_\circ=1 \quad \boxed{H^{4_1}_\Box =   \frac{(1+\phi\bar\phi)D + (\phi+\bar\phi) D^2}{D}} = 1- z^2 +(A-A^{-1})z D
= A^2+A^{-2} +1-q^2-q^{-2}
= 1+\{Aq\}\{A/q\}\,.
\ee}

{\bf The antiparallel torus link $T[2,4]$ and its mirror image}

{\small \begin{equation}\label{T[2,4]}
\begin{aligned}
n_\bullet&=2 \quad \boxed{H^{T[2,4]}_\Box=A^{-4}\frac{(1+\phi^2)D^2+2\phi D}{D}}=A^{-4}\cdot\left((1+A^2z^2)\textstyle{\frac{A-A^{-1}}{z}}+Az\right)=A^{-4}\cdot \frac{q^{4}\{Aq\}+q^{-4}\{A/q\}}{\{q^2\}}\,, \\
n_\circ&=2 \quad \boxed{H^{\overline{T[2,4]}}_\Box =A^4\frac{(1+\bphi^2)D^2+2\bphi D}{D}}=A^{-4}\cdot\left((1+A^{-2}z^2)\textstyle{\frac{A-A^{-1}}{z}}+A^{-1}z\right)=A^{4}\cdot \frac{q^{-4}\{Aq\}+q^{4}\{A/q\}}{\{q^2\}}\,.
\end{aligned}
\end{equation}}


\subsection{On bipartite abundancy}\label{sec:bip-comm}

Here we make some general comments about bipartite knots.

When exists, a bipartite realization is not necessarily the minimal one (i.e. has minimal number of crossings),
for example, the minimal bipartite diagram for the trefoil $3_1$ has four crossings
(it is its realization as a twist knot) while the crossing number of $3_1$ is equal to three.
This is always true when minimal realization contains odd number of crossings,
since bipartite diagrams have even.
The simplest illustration is the half of twist knots, $5_2$, $7_2$, $9_2$, $\ldots$ --
as bipartite ones they have $6$, $8$, $10$, $\ldots$ crossings.
But for another half of twist knots, $4_1$, $6_1$, $8_1$, $\ldots$, bipartite realization {\it is}
the minimal one.
The difference of intersection numbers between  minimal bipartite and minimal realizations
can be quite big, for example, it is $2(2n-2)-(2n-1)=2n-3$ for a torus knot $T[2,2n-1]$, whose minimal bipartite realization has $2n-2$ bipartite vertices.

In Rolfsen table, there currently known just 12 knots with no more than 10 intersections\footnote{For 11 crossings, we have $11a_{123}$, $11a_{135}$, $11a_{155}$, $11a_{173}$, $11a_{181}$, $11a_{196}$, $11a_{249}$, $11a_{277}$,
$11a_{291}$, $11a_{293}$,  $11a_{314}$, $11a_{317}$, $11a_{321}$, $11a_{366}$,
$11n_{126}$, $11n_{133}$, $11n_{167}$ which are non-bipartite.},
which \textit{definitely} do not possess bipartite realizations at all:
$9_{35}$, $9_{37}$, $9_{41}$, $9_{46}$, $9_{47}$, $9_{48}$, $9_{49}$, $10_{74}$, $10_{75}$, $10_{103}$, $10_{155}$, $10_{157}$.
As mentioned in \cite{bipfund}, a possible obstacle for biparticity is provided by
Alexander ideals \cite{bip}.
One can look in Appendix to \cite{bipfund} for application of this criterium to $9_{35}$.

However, Alexander ideals are not expressed through the HOMFLY polynomials
(at least an expression, if any, is not yet available).
The fundamental HOMFLY polynomials for the above 12 non-bipartite knots do not seem qualitatively different
from the bipartite ones, e.g., the cyclotomic coefficient of the differential expansion~\cite{DE,DEgen,Kononov_2016,Morozov_2018,Morozov_2019,morozov2020kntz,BM1,morozov2022differential} for the non-bipartite $9_{37}$:
\be
\frac{H_{\Box}^{9_{37}} - 1}{(v^2 + vz - 1)(v^2 - vz - 1)} =
-\frac{v^4z^2 + v^2z^2 - 2v^2 - 1}{v^4}\,,
\ee
does not look very different from that for the bipartite $7_3$:
\be
\frac{H_{\Box}^{7_3} - 1}{(v^2 + vz - 1)(v^2 - vz - 1)} = -v^4z^2-2v^4-v^2z^2 - 2v^2-1
\ee
(here recall that $v=A^{-1}$ and $z=q-q^{-1}$).

\setcounter{equation}{0}
\section{A search for positive decomposition: eq.(\ref{chirexpanswer}) and its applications}\label{sec:regsearch}

In this section, we change the starting point:
instead of a diagram we begin from the fundamental HOMFLY polynomial and ask if it possesses PD (\ref{bipexp}).
The answer is positive, but in fact much simpler:
{\it every} symmetric polynomial does.
We come to this conclusion in steps, through formula (\ref{chirexpanswer}) for chiral polynomials.
This can look like an unnecessary complication, but it keeps us closer to the historic track
and to the logic of knot calculus -- which we prefer in this particular paper.

\subsection{Generalities}

Given a HOMFLY polynomial, one can ask if it has a  {\it positive decomposition},
i.e., if it can be rewritten as a polynomial
\be
H_\Box^{\cal K} \ \stackrel{?}{=} \  A^{-2\,\mathrm{fr}}\sum_{k \geq 0} \sum_{i=0}^{n_\bullet} \sum_{j=0}^{n_\circ} {\cal N}_{ijk} D^k \phi^i\bphi^{\,j}
\label{PD}
\ee
with non-negative powers $i$, $j$, $k$ and non-negative integer coefficients ${\cal N}_{ijk}$.
It is a separate task to define the power  of the framing factor $\mathrm{fr}$. A priori, $\mathrm{fr}$ -- the algebraic number of bipartite vertices -- is unknown when we look just at the answer for the HOMFLY polynomial with an unknown bipartite realization.

Note that {\it positive} decomposition gives rise to certain framing, e.g., 
\be
H_{\Box}^{3_1} = v^4(2\phi D + 1+\phi^2).
\ee
If the non-negativity condition for the coefficients ${\cal N}_{ijk}$ is ignored, then, one can provide a decomposition (\ref{PD}) with any $\mathrm{fr}$, using that $\phi/(1+D\phi)=-\bphi$, e.g.,
\be
H_\Box^{9_{35}} =  v^2\Big(-\bar\phi^4D^4 -5\bar\phi^3D^3 + (\bar\phi^4-6\bar\phi^2)D^2 + (5\bar\phi^3-\bar\phi)D + 6\bar\phi^2
- \phi\bar\phi +1\Big).
\ee

\subsection{Positive decompositions for {\it chiral} bipartite knots}\label{sec:chiral}

Positive decompositions are easier to study in terms of the variable $z=q-q^{-1}$.
The HOMFLY polynomials of knots in the fundamental representation are functions of $A^2$, $q^2$ and invariant under the change $q\longrightarrow q^{-1}$
(we call such polynomials {\it symmetric}) and hence depend on $z^2$.
Symmetric function of $q^2$ is converted into that of $z$ by the following rule:
\begin{equation}
\begin{aligned}
F(q) &= f_0 + \sum_{k=1}^{\infty} f_k \left(q^{2k}+q^{-2k}\right)\quad \longrightarrow \quad \tilde F(z) = f_0 + \sum_{k=1}^\infty f_k\cdot \sum_{j=0}^k  \frac{2k(2k-j-1)!}{j!(2k-2j)!}\, z^{2(k-j)}\,. 
\end{aligned}
\end{equation}
The $q\longrightarrow q^{-1}$ invariance of reduced $H_\Box$ for {\it knots} implies that bipartite variables
enter in combinations $\phi^{2}, \phi\bar\phi, \bar\phi^2, \phi D, \bar\phi D$.
For links $H_\Box$ acquires a factor $(-)^{\# {\rm components}}$, thus for two component links
the dependence is on $D$, $\phi^{2k+1}$ etc\footnote{Generally, invariance is under  $q\longrightarrow -q^{-1}$, but it does not impose any constraints of the above type.}.

Below we consider the case when PD does not depend on $\bphi$. We call knots with such PD \textit{chiral}.
Hence, all negative signs in the HOMFLY polynomial, when it is expressed via $v=A^{-1}$ and $z$, come from $D$ entering the PD. It is easy to see that relatively high powers of $D$ are needed to ``adsorb'' all negative signs. 
Actually, the braid width does {\it not} directly restrict the power of $D$, because $\phi D+1 = A^{-2}$
is actually a monomial of zero width.

The conditions for a chiral positive decomposition for a knot $\cal K$
\be\label{chiral-PD}
H^{\cal K}_\Box = {\rm Fr} \cdot \sum_{a,b} M_{ab}\phi^{2a}(\phi D)^b = {\rm Fr} \cdot \sum_{a,b} L_{ab} \phi^{2a}(\phi D+1)^b =  {\rm Fr} \cdot \sum_{a,b} \frac{z^{2a}}{v^{2(a+b)}}
\ee
are:

\begin{itemize}

\item{} $M_{ab}$ are non-negative integers  (not $L_{ab}$ -- the counterexample is~\eqref{H-8_15}).

\item{} In the second sum with $L_{ab}$ (not in the first one with $M_{ab}$),
$c\leq a+b\leq c +{\rm braid \ index}-1$ with $c$ being some non-negative integer number -- a version of the Morton--Franks--Williams inequality~\cite{morton1986seifert,morton1988polynomials,franks1987braids}, which easily follows from the character expansion of the HOMFLY polynomial~\cite{MMM,MMM2}.

\item{}
$\sum\limits_{a,b} L_{ab} z^{2a}q^{\pm 2(a+b)} = q^{\pm 2{\rm fr}}$ as at $A=q$, $H^{\cal K}_\Box = H^{\bar{\cal K}}_\Box = 1\,$.

\end{itemize}

On the other hand, assuming {\bf chiral PD}, we would get the explicit formula for this chiral PD of the HOMFLY polynomial:
\be\label{chirexpanswer}
\boxed{{\rm  PD_{chiral}}(H^{\cal K}_\Box):=H^{\cal K}_\Box(v = X^{-1/2},\,z = X^{-1/2}\phi) \quad \text{with} \quad X=D\phi+1}
\ee
being a rational function of $X$, $\phi$. The denominator is a power of $X=v^{-2}=A^2$ and defines a framing factor Fr in~\eqref{chiral-PD}. 
The last substitution $X=D\phi+1$ is important,
before it the polynomial in $X$, $\phi$ is not positive
already for the trefoil.

For a generic knot, the polynomial (\ref{chirexpanswer}) in $D$, $\phi$ is not necessarily positive.
Also, there is a question whether positivity of this expansion of $H_\Box=H_{[1]}$
guarantees positivity for all higher $H_{[r]}$. The chirality requirement implies that positive decomposition is possible
(and unique modulo $\phi+\bar\phi + \phi\bar\phi D=0$).

On the other hand, not all chiral knots are bipartite. A possible reason is that a non-bipartite chiral knot can have a bipartite clone, i.e. a knot with the same HOMFLY polynomial. We have searched clones of all known chiral non-bipartite knots (see their list in Section~\ref{sec:bip-comm}) through the HOMFLY polynomials of all knots with the crossing number up to 16. The clones were not found for all these non-bipartite knots except for the knot $9_{49}$ whose clone is $16n_{350659}$. However, it is currently unknown whether the knot $16n_{350659}$ is bipartite. At least, the Alexander ideals do not forbid it to be bipartite. 
Another possible reason for the non-bipartite chirality is discussed in Section \ref{sec:sum}.

To illustrate how (\ref{chirexpanswer}) works, we provide an example of $H_\Box^{8_{15}}$, which turns to be chiral.
Then
\be\label{H-8_15}
H^{8_{15}}_\Box(X,\phi) =\frac{X\phi^4+2X^2\phi^2+2\phi^4+X^3+5X\phi^3+3X^2-3\phi^2-4X+1}{X^5}\,.
\ee
The denominator gives the framing factor, while the numerator becomes positive after the substitution of $X=D\phi + 1$.

For non-chiral $H_\Box^{4_1}$, invariant under the change $A\longrightarrow A^{-1}$,
\be\label{H-4_1}
H^{4_{1}}_\Box(X,\phi)=\frac{X^2-\phi^2-X+1}{X}\,.
\ee
The numerator of~\eqref{H-4_1} is {\it not} made positive by the same substitution.

\subsection{All chiral PD for knots with up to 10 crossings\label{sec:allchir}}
Now we can apply (\ref{chirexpanswer}) -- and list the cases (all knots of crossing numbers up to 10) when a chiral positive decomposition exists, i.e. all the coefficients are non-negative.
In the semilast column we write ``$+$'' if the bipartite knot diagram is known, ``$-$'' if the knot is definitely non-bipartite, and ``?'' otherwise.

The last column lists braid index -- the minimal number of strands in the knot diagram, which bounds from above the maximal
 difference of powers in $A$ or $v$.

We remind that one should appropriately choose between $A$ and $A^{-1}$ to fit into chiral rather than anti-chiral realization of the PD.
At the same time, existing tables of knot polynomials are not very accurate with this choice\footnote{Actually, the Knotinfo table \cite{katlas} presents the chiral partners for all knots in the table below but knots $10_{132}$ and $10_{145}$ when it presents the antichiral partners.}.
Therefore for each knot ${\cal K}$ we could list two implications of (\ref{chirexpanswer}) --
for $H^{\cal K}_\Box(A,q)$ and  $H^{\cal K}_\Box(A^{-1},q)$.
To save space we actually do it for the first three chiral knots (negative terms are put in boxes)
but they are easily extracted from (\ref{chirexpanswer}) in all other cases.
At most one of them can be positive -- then we call ${\cal K}$  a chiral knot, and only such are kept in {\it this} table in what follows.
Then the second HOMFLY polynomial, for the knot ${\bar{\cal K}}$, with the inverse $A$, is anti-chiral, i.e., given by the substitution $\phi\longrightarrow \bar\phi$
from the positive chiral expansion.
But reproducing it from the non-positive {\it partner} expression is a separate task
to be discussed in Section \ref{partners} below.
For some yet unclear reason all the ``wrong'' partners of chiral (positive) expressions have the unit framing --
despite this the framings of the chiral expressions themselves of both non-positive pairs in the non-chiral case
are usually non-trivial.

For each knot we also add a line, which names its {\it precursor diagram} (in the first column) and the corresponding Jones polynomial (in the second column) to be discussed much later in Section \ref{sec:prec}.
There are just two chiral examples, $9_{35}$ (definitely non-bipartite) and $10_{120}$ (unknown whether is bipartite) when a precursor diagram does not exist because the resulting reductions from the PDs to the hypothetical Jones polynomials do not correspond any link. Thus, there do not exist chiral bipartite diagrams of knots $9_{35}$ and $10_{120}$. This fact, however, does not prohibit these knots to have non-chiral bipartite diagrams.

If both expressions for PD are non-positive, we call ${\cal K}$ non-chiral, and PD for some of them are listed in the next Section \ref{nce}.

\begin{table}[h!]
\begin{tabular}{c|c|c|c|c}
{\rm knot} & {\rm fr} & (\ref{chirexpanswer}) & {\rm existence\ of}  & {\rm braid}  \\
&&&{\rm BP\ diagram} & {\rm index}  \\ \hline\hline
\raisebox{-2pt}{$3_1$} & \raisebox{-2pt}{2} &  \raisebox{-2pt}{$2D\phi+\phi^2+1$}   &  \raisebox{-2pt}{+} & \raisebox{-2pt}{2} \\ [1.1ex] \hline
 $\overline{3}_1$   & 0 &  $\boxed{-D^2\phi^2}+\phi^2+1$        &  + & 2 \\ \hline
\raisebox{-2pt}{$\bigcirc^2$} & & \raisebox{-2pt}{$q+q^{-1}$} & &
 \\ [1.1ex] \hline\hline
\raisebox{-2pt}{$5_1$} & \raisebox{-2pt}{4} & \raisebox{-2pt}{$3D^2\phi^2+4D\phi^3 +4D\phi + \phi^4 + 3\phi^2+1$}  & \raisebox{-2pt}{+} & \raisebox{-2pt}{2}  \\ [1.1ex]  \hline
  $\overline{5}_1$  & 0 &   $\boxed{-2D^3\phi^3-D^2\phi^4-3D^2\phi^2}+2D\phi^3 + \phi^4 + 3\phi^2+1$        &  + & 2 \\ \hline \raisebox{-2pt}{$\bigcirc$} &  & \raisebox{-2pt}{1} & & \\ [1.1ex] \hline\hline
\raisebox{-2pt}{$5_2$} & \raisebox{-2pt}{3} & \raisebox{-2pt}{$D^2\phi^2 + D\phi^3+3D\phi +2\phi^2+1$}  & \raisebox{-2pt}{+} & \raisebox{-2pt}{3} \\ [1.1ex] \hline
  $\overline{5}_2$  & 0 &    $\boxed{-D^3\phi^3-2D^2\phi^2}+D\phi^3+2\phi^2+1$        &  &   \\ \hline \raisebox{-2pt}{$\bigcirc$} &  & \raisebox{-2pt}{1} & & \\ [1.1ex] \hline\hline
\raisebox{-2pt}{$7_1$} & \raisebox{-2pt}{6} & \raisebox{-2pt}{$4 D^3 \phi^3+10 D^2 \phi^4+9 D^2 \phi^2+6 D \phi^5+16 D \phi^3+6 D \phi+\phi^6+5 \phi^4+6 \phi^2+1$}  & \raisebox{-2pt}{+} & \raisebox{-2pt}{2} \\ [1.1ex] \hline \raisebox{-2pt}{$\bigcirc$} &  & \raisebox{-2pt}{1} & & \\ [1.1ex] \hline\hline
\raisebox{-2pt}{$7_2$} & \raisebox{-2pt}{4}  & \raisebox{-2pt}{$D^3 \phi^3+D^2 \phi^4+3 D^2 \phi^2+3 D \phi^3+4 D \phi+3 \phi^2+1$}  & \raisebox{-2pt}{+} & \raisebox{-2pt}{4} \\ [1.1ex] \hline \raisebox{-2pt}{Hopf} &  & \raisebox{-2pt}{$q+q^5$} & & \\ [1.1ex] \hline\hline
\raisebox{-2pt}{$7_3$} &  \raisebox{-2pt}{5} & \raisebox{-2pt}{$D^3 \phi^3+3 D^2 \phi^4+5 D^2 \phi^2+D \phi^5+9 D \phi^3+5 D \phi+2 \phi^4+5 \phi^2+1$} & \raisebox{-2pt}{+} & \raisebox{-2pt}{3} \\ [1.1ex] \hline \raisebox{-2pt}{Hopf} &  & \raisebox{-2pt}{$q+q^5$} & & \\ [1.1ex] \hline\hline
\raisebox{-2pt}{$7_4$} &  \raisebox{-2pt}{4} & \raisebox{-2pt}{$D^2 \phi^4+2 D^2 \phi^2+4 D \phi^3+4 D \phi+4 \phi^2+1$}  & \raisebox{-2pt}{+} & \raisebox{-2pt}{4} \\ [1.1ex] \hline \raisebox{-2pt}{$3_1$} &  & \raisebox{-2pt}{$-q^8+q^6+q^2$} & & \\ [1.1ex] \hline\hline
\raisebox{-2pt}{$7_5$} & \raisebox{-2pt}{5}  &  \raisebox{-2pt}{$2 D^3 \phi^3+3 D^2 \phi^4+6 D^2 \phi^2+D \phi^5+8 D \phi^3+5 D \phi+2 \phi^4+4 \phi^2+1$} & \raisebox{-2pt}{+} & \raisebox{-2pt}{3} \\ [1.1ex] \hline \raisebox{-2pt}{$\bigcirc$} &  & \raisebox{-2pt}{1} & & \\ [1.1ex] \hline\hline
\raisebox{-2pt}{$8_{15}$} & \raisebox{-2pt}{5}  & \raisebox{-2pt}{$D^3 \phi^3+2 D^2 \phi^4+6 D^2 \phi^2+ D \phi^5+9 D \phi^3+5 D \phi+3 \phi^4+4 \phi^2+1$}  & \raisebox{-2pt}{+} & \raisebox{-2pt}{4} \\ [1.1ex] \hline
\raisebox{-2pt}{$\bigcirc^2$} &  & \raisebox{-2pt}{$q+q^{-1}$} & &
 \\ [1.1ex] \hline\hline
\raisebox{-2pt}{$8_{19}$} & \raisebox{-2pt}{6} & \raisebox{-2pt}{$5 D^3 \phi^3+10 D^2 \phi^4+10 D^2 \phi^2+6 D \phi^5+15 D \phi^3+6 D \phi+\phi^6+5 \phi^4+5 \phi^2+1$} & \raisebox{-2pt}{+} & \raisebox{-2pt}{3} \\ [1.1ex] \hline
\raisebox{-2pt}{$\bigcirc^2$} &  & \raisebox{-2pt}{$q+q^{-1}$} & &
 \\ [1.1ex] \hline\hline
\raisebox{-2pt}{$9_1$} &  \raisebox{-2pt}{8}  & \raisebox{-2pt}{$5 D^4 \phi ^4+20 D^3 \phi ^5+16 D^3 \phi ^3+21 D^2 \phi ^6+50 D^2 \phi ^4+18 D^2 \phi ^2+8 D \phi ^7+$} & \raisebox{-2pt}{+} & \raisebox{-2pt}{2} \\ [1.1ex]
& & $+36 D \phi ^5+40 D \phi ^3+8 D \phi +\phi ^8+7 \phi ^6+15 \phi ^4+10 \phi ^2+1$ & & \\ \hline
\raisebox{-2pt}{$\bigcirc^2$} &  & \raisebox{-2pt}{$q+q^{-1}$} & &
 \\ [1.1ex] \hline\hline
\raisebox{-2pt}{$9_2$} & \raisebox{-2pt}{5} & \raisebox{-2pt}{$D^4 \phi ^4+D^3 \phi ^5+4 D^3 \phi ^3+4 D^2 \phi ^4+6 D^2 \phi ^2+6 D \phi ^3+5 D \phi +4 \phi ^2+1$} & \raisebox{-2pt}{+} & \raisebox{-2pt}{5} \\ [1.1ex] \hline \raisebox{-2pt}{$3_1$} &  & \raisebox{-2pt}{$-q^8+q^6+q^2$} & & \\ [1.1ex] \hline\hline

\end{tabular}
\caption{\footnotesize The chiral PDs = BEs for the knots $3_1$, $5_1$, $5_2$, $7_1$, $7_2$, $7_3$, $7_4$, $7_5$, $8_{15}$, $8_{19}$, $9_1$, $9_2$. All of these knots are bipartite. Their precursor diagrams and the corresponding Jones polynomials are also indicated, see Section~\ref{sec:prec} for details.}
\label{tab:ch-PD-1}
\end{table}

\begin{table}[h!]
\begin{tabular}{c|c|c|c|c}
{\rm knot} & {\rm fr} & (\ref{chirexpanswer}) & {\rm existence\ of}  & {\rm braid}  \\
&&&{\rm BP\ diagram} & {\rm index}  \\ \hline\hline

\raisebox{-2pt}{$9_3$} & \raisebox{-2pt}{7} & \raisebox{-2pt}{$D^4 \phi ^4+6 D^3 \phi ^5+7 D^3 \phi ^3+5 D^2 \phi ^6+25 D^2 \phi ^4+12 D^2 \phi ^2+D \phi ^7+$} & \raisebox{-2pt}{+} & \raisebox{-2pt}{3} \\ [1.1ex]
& & $+15 D \phi ^5+28 D \phi ^3+7 D \phi +2 \phi ^6+9 \phi ^4+9 \phi ^2+1$ & & \\ \hline \raisebox{-2pt}{$\bigcirc$} &  & \raisebox{-2pt}{1} & & \\ [1.1ex] \hline\hline

\raisebox{-2pt}{$9_4$} & \raisebox{-2pt}{6} & \raisebox{-2pt}{$D^4 \phi ^4+3 D^3 \phi ^5+4 D^3 \phi ^3+D^2 \phi ^6+$} & \raisebox{-2pt}{+} & \raisebox{-2pt}{4} \\ [1.1ex]
& & $+11 D^2 \phi ^4+8 D^2 \phi ^2+3 D \phi ^5+16 D \phi ^3+6 D \phi +3 \phi ^4+7 \phi ^2+1$ & & \\ \hline \raisebox{-2pt}{$3_1$} &  & \raisebox{-2pt}{$-q^8+q^6+q^2$} & & \\ [1.1ex] \hline\hline

\raisebox{-2pt}{$9_5$} & \raisebox{-2pt}{5} & \raisebox{-2pt}{$D^3 \phi ^5+D^3 \phi ^3+5 D^2 \phi ^4+4 D^2 \phi ^2+9 D \phi ^3+5 D \phi +6 \phi ^2+1$} & \raisebox{-2pt}{+} & \raisebox{-2pt}{5} \\ [1.1ex] \hline \raisebox{-2pt}{$4_1$} &  & \raisebox{-2pt}{$q^4+q^{-4}-q^2-q^{-2}+1$} & & \\ [1.1ex] \hline\hline

\raisebox{-2pt}{$9_6$} & \raisebox{-2pt}{7} & \raisebox{-2pt}{$3 D^4 \phi ^4+7 D^3 \phi ^5+11 D^3 \phi ^3+5 D^2 \phi ^6+24 D^2 \phi ^4+14 D^2 \phi ^2+D \phi ^7+$} & \raisebox{-2pt}{+} & \raisebox{-2pt}{3} \\ [1.1ex]
& & $+14 D \phi ^5+24 D \phi ^3+7 D \phi +2 \phi ^6+8 \phi ^4+7 \phi ^2+1$ & & \\ \hline
\raisebox{-2pt}{$\bigcirc^2$} &  & \raisebox{-2pt}{$q+q^{-1}$} & &
 \\ [1.1ex] \hline\hline

\raisebox{-2pt}{$9_7$} &  \raisebox{-2pt}{6}  & \raisebox{-2pt}{$2 D^4 \phi ^4+3 D^3 \phi ^5+7 D^3 \phi ^3+D^2 \phi ^6+10 D^2 \phi ^4+$} & \raisebox{-2pt}{+} & \raisebox{-2pt}{4} \\ [1.1ex]
& & $+10 D^2 \phi ^2+3 D \phi ^5+13 D \phi ^3+6 D \phi +3 \phi ^4+5 \phi ^2+1$ & & \\ \hline \raisebox{-2pt}{$\bigcirc$} &  & \raisebox{-2pt}{1} & & \\ [1.1ex] \hline\hline

\raisebox{-2pt}{$9_9$} &  \raisebox{-2pt}{7}  & \raisebox{-2pt}{$2 D^4 \phi ^4+7 D^3 \phi ^5+9 D^3 \phi ^3+5 D^2 \phi ^6+25 D^2 \phi ^4+13 D^2 \phi ^2+$} & \raisebox{-2pt}{+} & \raisebox{-2pt}{3} \\ [1.1ex]
& & $+D \phi ^7+14 D \phi ^5+26 D \phi ^3+7 D \phi +2 \phi ^6+8 \phi ^4+8 \phi ^2+1$ & & \\ 
\hline \raisebox{-2pt}{$\bigcirc$} &  & \raisebox{-2pt}{1} & & \\ [1.1ex] \hline\hline

\raisebox{-2pt}{$9_{10}$} &  \raisebox{-2pt}{6}  & \raisebox{-2pt}{$2 D^3 \phi ^5+2 D^3 \phi ^3+D^2 \phi ^6+11 D^2 \phi ^4+7 D^2 \phi ^2+4 D \phi ^5+$} & \raisebox{-2pt}{+} & \raisebox{-2pt}{4} \\ [1.1ex] & & $+18 D \phi ^3+6 D \phi +4 \phi ^4+8 \phi ^2+1$ & & \\ \hline \raisebox{-2pt}{$3_1$} &  & \raisebox{-2pt}{$-q^8+q^6+q^2$} & & \\ [1.1ex] \hline\hline

\raisebox{-2pt}{$9_{13}$} &  \raisebox{-2pt}{6}  & \raisebox{-2pt}{$2 D^3 \phi ^5+3 D^3 \phi ^3+D^2 \phi ^6+11 D^2 \phi ^4+8 D^2 \phi ^2+4 D \phi ^5+$} & \raisebox{-2pt}{+} & \raisebox{-2pt}{4} \\ [1.1ex] & & $+17 D \phi ^3+6 D \phi +4 \phi ^4+7 \phi ^2+1$ & & \\ \hline \raisebox{-2pt}{Hopf} &  & \raisebox{-2pt}{$q+q^5$} & & \\ [1.1ex] \hline\hline

\raisebox{-2pt}{$9_{16}$} &  \raisebox{-2pt}{7}  & \raisebox{-2pt}{$4 D^4 \phi ^4+8 D^3 \phi ^5+13 D^3 \phi ^3+5 D^2 \phi ^6+24 D^2 \phi ^4+15 D^2 \phi ^2+$} & \raisebox{-2pt}{+} & \raisebox{-2pt}{3} \\ [1.1ex]
& & $+D \phi ^7+13 D \phi ^5+22 D \phi ^3+7 D \phi +2 \phi ^6+7 \phi ^4+6 \phi ^2+1$ & & \\ \hline
\raisebox{-2pt}{$\bigcirc^2$} &  & \raisebox{-2pt}{$q+q^{-1}$} & &
 \\ [1.1ex] \hline\hline
 
\raisebox{-2pt}{$9_{18}$} &  \raisebox{-2pt}{6}  & \raisebox{-2pt}{$D^4 \phi ^4+2 D^3 \phi ^5+5 D^3 \phi ^3+D^2 \phi ^6+10 D^2 \phi ^4+9 D^2 \phi ^2$} & \raisebox{-2pt}{+} & \raisebox{-2pt}{4} \\ [1.1ex]
& & $+4 D \phi ^5+15 D \phi ^3+6 D \phi +4 \phi ^4+6 \phi ^2+1$ & & \\ \hline \raisebox{-2pt}{$\bigcirc$} &  & \raisebox{-2pt}{1} & & \\ [1.1ex] \hline\hline

\raisebox{-2pt}{$9_{23}$} &  \raisebox{-2pt}{6}  & \raisebox{-2pt}{$D^4 \phi ^4+2 D^3 \phi ^5+6 D^3 \phi ^3+D^2 \phi ^6+10 D^2 \phi ^4+$} & \raisebox{-2pt}{+} & \raisebox{-2pt}{4}  \\ [1.1ex]
& & $+10 D^2 \phi ^2+4 D \phi ^5+14 D \phi ^3+6 D \phi +4 \phi ^4+5 \phi ^2+1$ & & \\ \hline
\raisebox{-2pt}{$\bigcirc^2$} &  & \raisebox{-2pt}{$q+q^{-1}$} & &
 \\ [1.1ex] \hline\hline

\raisebox{-2pt}{$9_{35}$} &  \raisebox{-2pt}{5}  & \raisebox{-2pt}{$D^3 \phi ^5+5 D^2 \phi ^4+3 D^2 \phi ^2+10 D \phi ^3+5 D \phi +7 \phi ^2+1$} & \raisebox{-2pt}{\boxed{-}} & \raisebox{-2pt}{5} \\ [1.1ex] \hline
\raisebox{-2pt}{\boxed{-}} &  & \raisebox{-2pt}{$q^{6k-1}(-q^8+2 q^6+2 q^2-1)$, $k\in \mathbb{Z}$} & &
 \\ [1.1ex] \hline\hline

\end{tabular}
\caption{\footnotesize The chiral PDs for the knots $9_3$, $9_4$, $9_5$, $9_6$, $9_7$, $9_9$, $9_{10}$, $9_{13}$, $9_{16}$, $9_{18}$, $9_{23}$, $9_{35}$. The knot $9_{35}$ is non-bipartite due to its Alexander ideals~\cite{bip}. The hypothetical precursor Jones polynomial for the knot $9_{35}$ could correspond to some 2-component link. However, there is no such link among all known links of crossing numbers up to and including 11.}
\label{tab:ch-PD-2}
\end{table}

\begin{table}[h!]
\begin{tabular}{c|c|c|c|c}
{\rm knot} & {\rm fr} & (\ref{chirexpanswer})  & {\rm existence\ of}  & {\rm braid}  \\
&&&{\rm BP\ diagram} & {\rm index}  \\ \hline\hline

\raisebox{-2pt}{$9_{38}$} &  \raisebox{-2pt}{6} & \raisebox{-2pt}{$D^3 \phi ^5+4 D^3 \phi ^3+D^2 \phi ^6+10 D^2 \phi ^4+9 D^2 \phi ^2+$} & \raisebox{-2pt}{?} & \raisebox{-2pt}{4} \\ [1.1ex] & & $+5 D \phi ^5+16 D \phi ^3+6 D \phi +5 \phi ^4+6 \phi ^2+1$ & & \\ \hline
\raisebox{-2pt}{$\bigcirc^2$} &  & \raisebox{-2pt}{$q+q^{-1}$} & &
 \\ [1.1ex] \hline\hline

\raisebox{-2pt}{$9_{49}$} &  \raisebox{-2pt}{5}  & \raisebox{-2pt}{$2 D^2 \phi ^4+4 D^2 \phi ^2+D \phi ^5+10 D \phi ^3+5 D \phi +3 \phi ^4+6 \phi ^2+1$} & \raisebox{-2pt}{\boxed{-}} & \raisebox{-2pt}{4} \\ [1.1ex] \hline \raisebox{-2pt}{Hopf} &  & \raisebox{-2pt}{$q+q^5$} & & \\ [1.1ex] \hline\hline

\raisebox{-2pt}{$10_{49}$} & \raisebox{-2pt}{7}   & \raisebox{-2pt}{$D^4 \phi ^4+4 D^3 \phi ^5+9 D^3 \phi ^3+4 D^2 \phi ^6+24 D^2 \phi ^4+14 D^2 \phi ^2+$} & \raisebox{-2pt}{+} & \raisebox{-2pt}{4} \\ [1.1ex]
& & $+D \phi ^7+17 D \phi ^5+26 D \phi ^3+7 D \phi +3 \phi ^6+10 \phi ^4+7 \phi ^2+1$ & & \\ \hline \raisebox{-2pt}{$\bigcirc$} &  & \raisebox{-2pt}{1} & & \\ [1.1ex] \hline\hline

\raisebox{-2pt}{$10_{53}$} & \raisebox{-2pt}{6} & \raisebox{-2pt}{$D^3 \phi ^5+3 D^3 \phi ^3+D^2 \phi ^6+9 D^2 \phi ^4+9 D^2 \phi ^2+$} & \raisebox{-2pt}{+} & \raisebox{-2pt}{5} \\ [1.1ex] & & $+5 D \phi ^5+17 D \phi ^3+6 D \phi +6 \phi ^4+6 \phi ^2+1$ & & \\ \hline \raisebox{-2pt}{$\bigcirc$} &  & \raisebox{-2pt}{1} & & \\ [1.1ex] \hline\hline

\raisebox{-2pt}{$10_{55}$} & \raisebox{-2pt}{6} & \raisebox{-2pt}{$D^4 \phi ^4+2 D^3 \phi ^5+5 D^3 \phi ^3+D^2 \phi ^6+9 D^2 \phi ^4+10 D^2 \phi ^2+4 D \phi ^5+$} & \raisebox{-2pt}{+} & \raisebox{-2pt}{5} \\ [1.1ex]
& & $+15 D \phi ^3+6 D \phi +5 \phi ^4+5 \phi ^2+1$ & & \\ \hline \raisebox{-2pt}{$\bigcirc$} &  & \raisebox{-2pt}{1} & & \\ [1.1ex] \hline\hline

\raisebox{-2pt}{$10_{63}$} &   \raisebox{-2pt}{6} & \raisebox{-2pt}{$D^4 \phi ^4+2 D^3 \phi ^5+4 D^3 \phi ^3+D^2 \phi ^6+9 D^2 \phi ^4+9 D^2 \phi ^2+4 D \phi ^5+$} & \raisebox{-2pt}{+} & \raisebox{-2pt}{5} \\
& & $+16 D \phi ^3+6 D \phi +5 \phi ^4+6 \phi ^2+1$ & & \\ \hline \raisebox{-2pt}{$\bigcirc^2 \;\cup$ Hopf} &  & \raisebox{-2pt}{$(q+q^{-1})(q+q^5)$} & & \\ [1.1ex] \hline\hline

\raisebox{-2pt}{$10_{66}$} &  \raisebox{-2pt}{7}  & \raisebox{-2pt}{$2 D^4 \phi ^4+5 D^3 \phi ^5+10 D^3 \phi ^3+4 D^2 \phi ^6+24 D^2 \phi ^4+14 D^2 \phi ^2+$} & \raisebox{-2pt}{+} & \raisebox{-2pt}{4} \\
& & $+D \phi ^7+16 D \phi ^5+25 D \phi ^3+7 D \phi +3 \phi ^6+9 \phi ^4+7 \phi ^2+1$ & & \\ \hline
\raisebox{-2pt}{$\bigcirc^2$} &  & \raisebox{-2pt}{$q+q^{-1}$} & &
 \\ [1.1ex] \hline\hline

\raisebox{-2pt}{$10_{80}$} & \raisebox{-2pt}{7} & \raisebox{-2pt}{$2 D^4 \phi ^4+5 D^3 \phi ^5+11 D^3 \phi ^3+4 D^2 \phi ^6+24 D^2 \phi ^4+15 D^2 \phi ^2+$} & \raisebox{-2pt}{?} & \raisebox{-2pt}{4} \\ [1.1ex]
& & $+D \phi ^7+16 D \phi ^5+24 D \phi ^3+7 D \phi +3 \phi ^6+9 \phi ^4+6 \phi ^2+1$ & & \\ \hline \raisebox{-2pt}{$\bigcirc$} &  & \raisebox{-2pt}{1} & & \\ [1.1ex] \hline\hline

\raisebox{-2pt}{$10_{101}$} &  \raisebox{-2pt}{6}  & \raisebox{-2pt}{$D^3 \phi ^5+2 D^3 \phi ^3+D^2 \phi ^6+8 D^2 \phi ^4+8 D^2 \phi ^2+5 D \phi ^5+$} & \raisebox{-2pt}{?} & \raisebox{-2pt}{5} \\ [1.1ex] & & $+18 D \phi ^3+6 D \phi +7 \phi ^4+7 \phi ^2+1$ & & \\ \hline \raisebox{-2pt}{$\bigcirc$} &  & \raisebox{-2pt}{1} & & \\ [1.1ex] \hline\hline

\raisebox{-2pt}{$10_{120}$} &  \raisebox{-2pt}{6}  & \raisebox{-2pt}{{\small $3 D^3 \phi ^3+D^2 \phi ^6+7 D^2 \phi ^4+9 D^2 \phi ^2+6 D \phi ^5+17 D \phi ^3+6 D \phi +8 \phi ^4+6 \phi ^2+1$}} & \raisebox{-2pt}{?} & \raisebox{-2pt}{5} \\ [1.1ex] \hline \raisebox{-2pt}{{\color{red}\boxed{-}}} & & \raisebox{-2pt}{$q^{-6}(q^8-q^2+1)$} & & \\ [1.1ex] \hline\hline

\raisebox{-2pt}{$10_{124}$} &  \raisebox{-2pt}{8}  & \raisebox{-2pt}{$7 D^4 \phi ^4+21 D^3 \phi ^5+20 D^3 \phi ^3+21 D^2 \phi ^6+49 D^2 \phi ^4+20 D^2 \phi ^2+$} & \raisebox{-2pt}{+} & \raisebox{-2pt}{3} \\ [1.1ex]
& & $+8 D \phi ^7+35 D \phi ^5+36 D \phi ^3+8 D \phi +\phi ^8+7 \phi ^6+14 \phi ^4+8 \phi ^2+1$ & & \\ \hline \raisebox{-2pt}{$\bigcirc$} &  & \raisebox{-2pt}{1} & & \\ [1.1ex] \hline\hline

\raisebox{-2pt}{$10_{128}$} & \raisebox{-2pt}{7} & \raisebox{-2pt}{$2 D^4 \phi ^4+6 D^3 \phi ^5+10 D^3 \phi ^3+5 D^2 \phi ^6+24 D^2 \phi ^4+14 D^2 \phi ^2+$} & \raisebox{-2pt}{+} & \raisebox{-2pt}{4} \\ [1.1ex]
& & $+D \phi ^7+15 D \phi ^5+25 D \phi ^3+7 D \phi +2 \phi ^6+9 \phi ^4+7 \phi ^2+1$ & & \\ \hline \raisebox{-2pt}{$\bigcirc$} &  & \raisebox{-2pt}{1} & & \\ [1.1ex] \hline\hline

\end{tabular}
\caption{\footnotesize The chiral PDs for the knots $9_{38}$, $9_{49}$, $10_{49}$, $10_{53}$, $10_{55}$, $10_{63}$, $10_{66}$, $10_{80}$, $10_{101}$, $10_{120}$, $10_{124}$, $10_{128}$. The knot $9_{49}$ is non-bipartite. For the knots $9_{38}$, $10_{80}$, $10_{101}$, $10_{120}$, the biparticity is unknown: neither the Alexander ideals nor the precursor check do not detect non-biparticity of these knots. The hypothetical precursor Jones polynomial for the knot $10_{120}$ could correspond to some knot. However, there is no such knot among all known links of crossing numbers up to and including 16. Thus, $10_{120}$ is a candidate to non-bipartite knots, which was currently unknown (not restricted by the Alexander ideals).
}
\label{tab:ch-PD-3}
\end{table}

\newpage \pagebreak \cleardoublepage

\begin{table}[h!]
\begin{tabular}{c|c|c|c|c}
{\rm knot} & {\rm fr} & (\ref{chirexpanswer})  & {\rm existence\ of}  & {\rm braid}  \\
&&&{\rm BP\ diagram} & {\rm index}  \\ \hline\hline

\raisebox{-2pt}{$10_{132}$} &  \raisebox{-2pt}{4} & \raisebox{-2pt}{$3 D^2 \phi ^2+4 D \phi ^3+4 D \phi +\phi ^4+3 \phi ^2+1$} & \raisebox{-2pt}{+} & \raisebox{-2pt}{4} \\ [1.1ex] \hline \raisebox{-2pt}{$\bigcirc$} &  & \raisebox{-2pt}{1} & & \\ [1.1ex] \hline\hline

\raisebox{-2pt}{$10_{134}$} & \raisebox{-2pt}{7} & \raisebox{-2pt}{$3 D^4 \phi ^4+7 D^3 \phi ^5+12 D^3 \phi ^3+5 D^2 \phi ^6+24 D^2 \phi ^4+15 D^2 \phi ^2+$} & \raisebox{-2pt}{+} & \raisebox{-2pt}{4} \\ [1.1ex]
& & $+D \phi ^7+14 D \phi ^5+23 D \phi ^3+7 D \phi +2 \phi ^6+8 \phi ^4+6 \phi ^2+1$ & & \\ \hline \raisebox{-2pt}{$\bigcirc$} &  & \raisebox{-2pt}{1} & & \\ [1.1ex] \hline\hline

\raisebox{-2pt}{$10_{139}$} & \raisebox{-2pt}{8} & \raisebox{-2pt}{$6 D^4 \phi ^4+21 D^3 \phi ^5+18 D^3 \phi ^3+21 D^2 \phi ^6+50 D^2 \phi ^4+19 D^2 \phi ^2+$} & \raisebox{-2pt}{+} & \raisebox{-2pt}{$3$} \\ [1.1ex]
& & $+8 D \phi ^7+35 D \phi ^5+38 D \phi ^3+8 D \phi +\phi ^8+7 \phi ^6+14 \phi ^4+9 \phi ^2+1$ & & \\ \hline
\raisebox{-2pt}{$\bigcirc^2$} &  & \raisebox{-2pt}{$q+q^{-1}$} & &
 \\ [1.1ex] \hline\hline

\raisebox{-2pt}{$10_{142}$} & \raisebox{-2pt}{7} & \raisebox{-2pt}{$D^4 \phi ^4+6 D^3 \phi ^5+8 D^3 \phi ^3+5 D^2 \phi ^6+25 D^2 \phi ^4+13 D^2 \phi ^2+D \phi ^7+$} & \raisebox{-2pt}{+} & \raisebox{-2pt}{4} \\ [1.1ex]
& & $+15 D \phi ^5+27 D \phi ^3+7 D \phi +2 \phi ^6+9 \phi ^4+8 \phi ^2+1$ & & \\ \hline \raisebox{-2pt}{$\bigcirc^2 \;\cup$ Hopf} &  & \raisebox{-2pt}{$(q+q^{-1})(q+q^5)$} & & \\ [1.1ex] \hline\hline

\raisebox{-2pt}{$10_{145}$} & \raisebox{-2pt}{5} & \raisebox{-2pt}{$2 D^3 \phi ^3+4 D^2 \phi ^4+5 D^2 \phi ^2+D \phi ^5+8 D \phi ^3+5 D \phi +\phi ^4+5 \phi ^2+1$} & \raisebox{-2pt}{?} & \raisebox{-2pt}{4} \\ [1.1ex] \hline \raisebox{-2pt}{$3_1$} &  & \raisebox{-2pt}{$-q^8+q^6+q^2$} & & \\ [1.1ex] \hline\hline

\raisebox{-2pt}{$10_{152}$} & \raisebox{-2pt}{8} & \raisebox{-2pt}{$8 D^4 \phi ^4+22 D^3 \phi ^5+22 D^3 \phi ^3+21 D^2 \phi ^6+49 D^2 \phi ^4+21 D^2 \phi ^2+$} & \raisebox{-2pt}{?} & \raisebox{-2pt}{3} \\ [1.1ex]
& & $+8 D \phi ^7+34 D \phi ^5+34 D \phi ^3+8 D \phi +\phi ^8+7 \phi ^6+13 \phi ^4+7 \phi ^2+1$ & & \\ \hline \raisebox{-2pt}{$\bigcirc$} &  & \raisebox{-2pt}{1} & & \\ [1.1ex] \hline\hline

\raisebox{-2pt}{$10_{154}$} & \raisebox{-2pt}{6} & \raisebox{-2pt}{{\small $4 D^3 \phi ^3+9 D^2 \phi ^4+10 D^2 \phi ^2+6 D \phi ^5+16 D \phi ^3+6 D \phi +\phi ^6+6 \phi ^4+5 \phi ^2+1$}} & \raisebox{-2pt}{?} & \raisebox{-2pt}{4} \\ [1.1ex] \hline \raisebox{-2pt}{$\bigcirc$} &  & \raisebox{-2pt}{1} & & \\ [1.1ex] \hline\hline

\raisebox{-2pt}{$10_{161}$} & \raisebox{-2pt}{6} & \raisebox{-2pt}{{\small $3 D^3 \phi ^3+9 D^2 \phi ^4+8 D^2 \phi ^2+6 D \phi ^5+17 D \phi ^3+6 D \phi +\phi ^6+6 \phi ^4+7 \phi ^2+1$}} & \raisebox{-2pt}{?} & \raisebox{-2pt}{3} \\ [1.1ex] \hline \raisebox{-2pt}{$\bigcirc$} &  & \raisebox{-2pt}{1} & & \\ [1.1ex] \hline\hline

\ldots &&&& 
\end{tabular}
\caption{\footnotesize The chiral PDs for the knots $10_{132}$, $10_{134}$, $10_{139}$, $10_{142}$, $10_{145}$, $10_{152}$, $10_{154}$, $10_{161}$. For the knots $10_{145}$, $10_{152}$, $10_{154}$, $10_{161}$, the biparticity is unknown.
}
\label{tab:ch-PD-4}
\end{table}

\subsection{On non-chiral partners of chiral HOMFLY
\label{partners}}

An amusing exercise is to take the second, ``wrong'' expressions for the bipartite-chiral knots
and demonstrate that they provide anti-chiral expressions.
In other words, to see that the non-positive second chiral formula differs from the
anti-chiral version of the positive first one by a multiple of  $(\phi\bar\phi D+\phi+\bar\phi)=0$.
We should substitute the framing factor, which is always of the opposite chirality:
a power of $\bar\phi D+1$ for chiral case,
and a power of $\phi D+1$ for the anti-chiral case.
For example, for $3_1$
\be
\boxed{-\phi^2 D^2} + \phi^2 + 1 \  \longrightarrow  \
\boxed{-\phi^2 D^2} + \phi^2 + 1 + \phi^2D\Big(\phi\bar\phi D\ \boxed{+\phi}+\bar\phi\Big)
= \phi^2\bar\phi D^3 + \phi\bar\phi D^2 + \phi^2+1
\ee
is already positive, but further adding of
$(\phi D^2+\phi\bar\phi D + 2D +\bar\phi-\phi)\Big(\phi\bar\phi D+\phi+\bar\phi\Big)$
not only preserves positivity, but also converts it into the anti-chiral version of the
chiral formula:
\be
H_\Box^{\overline{3}_1}=\underbrace{(\phi D+1)^2}_{\rm framing}(2\bar\phi D + \bar\phi^2+1)
\ee
As already mentioned, an alternative formulation is that
\begin{equation}
\begin{aligned}
&{\rm wrong \ PD}(H^{3_1}_\Box) - {\rm antichiral\ dual \ PD}(H^{3_1}_\Box)
:= \\ 
&=\left(\boxed{-\phi^2 D^2} + \phi^2 + 1\right) - (\phi D+1)^2 (2\bar\phi D + \bar\phi^2+1)
\sim  \phi\bar\phi D+\phi+\bar\phi\,.
\end{aligned}
\end{equation}
One can easily check the last statement for any other bipartite-chiral knots.

\subsection{Non-chiral examples}\label{nce}

In this section, we list the results of
formal application of (\ref{chirexpanswer}) to all non-chiral knots of crossing numbers up to 8.
By definition, this gives $\bphi$-independent expressions.
With the two (illustrative) exceptions of $6_1$ and $6_2$, we present only one of the two PDs (for $H_\Box(A,q)$ and $H_\Box(A^{-1},q)$)
-- but both ones are non-positive.
Negative terms are put in boxes. For each knot, we also list its reduction from PD to the hypothetical Jones polynomial and indicate the corresponding precursor diagram, see details in Section~\ref{sec:prec}. The column ``existence of BP diagram'' indicates whether a knot has a bipartite realization\footnote{In fact, all knots with crossing numbers up to 8 have bipartite diagrams~\cite{bip,8_18}.}.

Note that, non-chiral non-bipartite knots possess PDs too. It again could mean that non-bipartite knots have bipartite clones definitely having PDs. This is actually the case for the knot $10_{103}$ having the bipartite clone $10_{40}$. All non-chiral non-bipartite knots are $9_{41}$ (clones $14n_{8554}$, $14n_{11429}$), $10_{103}$ (clones $10_{40}$ -- bipartite, $12n_{412}$, $14n_{7777}$, $15n_{54585}$, $16n_{224374}$, $16n_{362384}$), $10_{155}$ (clones $11n_{37}$, $16n_{17578}$, $16n_{233419}$), $10_{157}$ (clones $15n_{100300}$, $16n_{774625}$), $11a_{196}$ (clones $11a_{216}$, $13n_{1852}$, $15n_{51862}$, $15n_{51899}$, $15n_{54587}$, $15n_{84036}$, $15n_{85377}$, $15n_{136105}$, $16n_{224342}$, $16n_{233788}$, $16n_{244566}$, $16n_{380841}$, $16n_{501489}$), $11a_{317}$ (clones $16n_{6664}$, $16n_{7281}$, $16n_{450453}$), $11a_{321}$ (clones $16n_{739381}$, $16n_{739863}$), $11n_{133}$ (clone not found among knots of crossing numbers up to and including 16), $9_{37}$ (clone not found), $9_{46}$ (clone not found), $9_{47}$ (clone not found), $9_{48}$ (clone not found), $10_{74}$ (clone not found), $10_{75}$ (clones $15n_{44448}$, $15n_{62861}$), $11a_{135}$ (clones $13n_{145}$, $15n_{68216}$, $16n_{2760}$, $16n_{2814}$), $11a_{155}$ (clone $16n_{543312}$), $11a_{173}$ (clone $15n_{44038}$), $11a_{181}$ (clone not found), $11a_{249}$ (clones $13n_{2892}$, $16n_{18209}$, $16n_{278906}$), $11a_{277}$ (clones $16n_{234826}$, $16n_{245331}$, $16n_{368047}$), $11a_{293}$ (clone $14n_{24141}$), $11a_{314}$ ($15n_{138596}$), $11n_{167}$ (clone not found). 

To make the expressions from this table positive, we use the same trick and compensate negative terms by positive multiples of $\phi\bar\phi D+\phi+\bar\phi=0$
and subtract its other multiples, if this is allowed by positivity. For example, in the case of knot $4_1$ we add the underlined term
\be
(\bar\phi D+1)\Big( D^2\phi^2+D\phi\  \boxed{-\phi^2} + 1\Big) =
(\bar\phi D+1)\Big(D^2\phi^2+D\phi\  \boxed{-\phi^2} + 1\Big)
+ \underline{ \phi(\boxed{\phi\bar\phi D +\phi}+\bar\phi)} = \nn \\
= 1+\phi\bar\phi +  (\phi+\bar\phi)D + \underline{\underline{(\phi\bar\phi D+\phi+\bar\phi)\phi D^2}}\,.
\ee
Note that we took into account the framing factor, which can be always substituted by a {\it positive} power of either
$\phi D+1 = v^{-2}$ or $\bar\phi D+1 = v^2$.
One can now eliminate the double-underlined combination proportional to $(\phi\bar\phi D+\phi+\bar\phi)=0$
without breaking the positivity.
This leaves the answer for the PD that is obtained from the standard (bipartite) diagram\footnote{Note that one half of twisted knots $(2k+1)_2$ are chiral, while the other half $(2k)_1$ are not.
This is simply seen from their knot diagrams -- either all locks are the same or one is mirror to the others.
Clearly in the latter case, there is just one $\bar\phi$ in the PD.} of the knot $4_1$:
\be
H_\Box^{4_1} = 1+\phi\bar\phi +  (\phi+\bar\phi)D\,.
\ee
In this case, we were additionally lucky, because just one underlined term was sufficient to restore positivity --
though more algorithmically we could add $\underline{ (1+\bar\phi D)\phi(\phi\bar\phi D\ \boxed{ +\phi}+\bar\phi)}$
and subtract $\underline{\underline{(\phi\bar\phi D+\phi+\bar\phi)(\phi D^2 + \phi\bar\phi D  )}}$.

However, the above procedure gives rise to many questions we address to in the next section.

\begin{table}[h!]
\begin{tabular}{c|c|c|c|c}
{\rm knot} & {\rm Fr} & (\ref{chirexpanswer})  & {\rm existence\ of}  & {\rm braid}  \\
&&&{\rm BP\ diagram} & {\rm width}  \\ \hline \hline

$4_1$ & 1 &  $D^2\phi^2+D\phi\  \boxed{-\phi^2} + 1$      & $+$ & 3 \\ \hline \raisebox{-2pt}{Hopf} &  & \raisebox{-2pt}{$q+q^5$} & & \\ [1.1ex] \hline\hline

$6_1$ & 1 & $D^3\phi^3+2D^2\phi^2 \ \boxed{-D\phi^3} + D\phi \ \boxed{-2\phi^2}+1$ & $+$ & 4 \\ \hline
   $\overline{6}_1$ & 2 & $D^3\phi^3 + 3D^2\phi^2 \ \boxed{-D\phi^3} + 2D\phi \ \boxed{-2\phi^2} +1$ & + & 4 \\ \hline \raisebox{-2pt}{$3_1$} &  & \raisebox{-2pt}{$-q^8+q^6+q^2$} & & \\ [1.1ex] \hline\hline

$6_2$ & 1 & $D^3\phi^3+D^2\phi^4+D^2\phi^2 \ \boxed{-D\phi^3 - \phi^4}+D\phi \ \boxed{-\phi^2}+1$ & $+$ & 3 \\ \hline
   $\overline{6}_2$ & 3 & $2D^3\phi^3+D^2\phi^4+4D^2\phi^2 \ \boxed{-D\phi^2-\phi^4} + 3D\phi \ \boxed{-\phi^2} + 1$ & + & 3 \\ \hline \raisebox{-2pt}{$\bigcirc$} &  & \raisebox{-2pt}{1} & & \\ [1.1ex] \hline\hline

\end{tabular}
\caption{\footnotesize Formal application of~\eqref{chirexpanswer} to the bipartite non-chiral knots $4_1$, $6_1$, $6_2$.}
\label{tab:non-ch-PD-1}
\end{table}


\begin{table}[h!]
\begin{tabular}{c|c|c|c|c}
{\rm knot} & {\rm Fr} & (\ref{chirexpanswer}) & {\rm existence\ of}  & {\rm braid}  \\
&&&{\rm BP\ diagram} & {\rm width}  \\ \hline \hline

$6_3$ & 2  & $\boxed{-D^3 \phi ^3-D^2 \phi ^4} +D \phi ^3+2 D \phi +\phi ^4+\phi ^2+1$  & $+$ & 3 \\ \hline \raisebox{-2pt}{$\bigcirc$} &  & \raisebox{-2pt}{1} & & \\ [1.1ex] \hline\hline

$7_6$ &  3 & $D^3 \phi ^3+D^2 \phi ^4+2 D^2 \phi ^2+3 D \phi \, \boxed{-\phi ^4} +\phi ^2+1$  & $+$ & 4 \\ \hline \raisebox{-2pt}{$\bigcirc$} &  & \raisebox{-2pt}{1} & & \\ [1.1ex] \hline\hline

$7_7$ & 2  &  $D^4 \phi ^4+2 D^3 \phi ^3 \boxed{-2 D^2 \phi ^4} +2 D^2 \phi ^2 \boxed{-2 D \phi ^3} +2 D \phi +\phi ^4 \,\boxed{-\phi ^2} +1$ & $+$ & 4 \\ \hline \raisebox{-2pt}{$3_1$} &  & \raisebox{-2pt}{$-q^8+q^6+q^2$} & & \\ [1.1ex] \hline\hline

$8_{1}$ & 3  & $D^4 \phi ^4+4 D^3 \phi ^3\boxed{-D^2 \phi ^4}+6 D^2 \phi ^2\boxed{-3 D \phi ^3}+3 D \phi \boxed{-3 \phi ^2}+1$ & + & 5 \\ \hline \raisebox{-2pt}{$L4a_{1}$} &  & \raisebox{-2pt}{$q^9+q^5-q^3+q$} & & \\ [1.1ex] \hline\hline

$8_{2}$ & 1 & {\small $D^4 \phi ^4+3 D^3 \phi ^5+D^3 \phi ^3+D^2 \phi ^6+2 D^2 \phi ^4\boxed{-3 D \phi ^5-D \phi ^3}+D \phi\, \boxed{-\phi ^6-3 \phi ^4} +1$} & + & 3 \\ \hline \raisebox{-2pt}{$\bigcirc$} &  & \raisebox{-2pt}{1} & & \\ [1.1ex] \hline\hline

$8_{3}$ & 2 & $D^4 \phi ^4+4 D^3 \phi ^3\boxed{-D^2 \phi ^4}+5 D^2 \phi ^2\boxed{-4 D \phi ^3}+2 D \phi\, \boxed{-4 \phi ^2}+1$ & + & 5 \\ \hline \raisebox{-2pt}{$4_1$} &  & \raisebox{-2pt}{$q^4+q^{-4}-q^2-q^{-2}+1$} & & \\ [1.1ex] \hline\hline

$8_{4}$ & 2 & {\small $D^4 \phi ^4+D^3 \phi ^5+4 D^3 \phi ^3+D^2 \phi ^4+4 D^2 \phi ^2\boxed{-D \phi ^5-4 D \phi ^3}+2 D \phi\, \boxed{-2 \phi ^4-3 \phi ^2}+1$} & + & 4 \\ \hline \raisebox{-2pt}{Hopf} &  & \raisebox{-2pt}{$q+q^5$} & & \\ [1.1ex] \hline\hline

\raisebox{-2pt}{$8_{5}$} & \raisebox{-2pt}{5} & \raisebox{-2pt}{$4 D^4 \phi ^4+4 D^3 \phi ^5+11 D^3 \phi ^3+D^2 \phi ^6+4 D^2 \phi ^4+11 D^2 \phi ^2-$} & \raisebox{-2pt}{+} & \raisebox{-2pt}{3} \\ [1.1ex] & & $\boxed{-3 D \phi ^5-D \phi ^3}+5 D \phi\, \boxed{-\phi ^6-3 \phi ^4-\phi ^2}+1$ & & \\ \hline
\raisebox{-2pt}{$\bigcirc^2$} &  & \raisebox{-2pt}{$q+q^{-1}$} & &
 \\ [1.1ex] \hline\hline

$8_{6}$ & 1 & {\small $D^4 \phi ^4+D^3 \phi ^5+3 D^3 \phi ^3+D^2 \phi ^4+2 D^2 \phi ^2\boxed{-D \phi ^5-3 D \phi ^3}+D \phi\, \boxed{-2 \phi ^4-2 \phi ^2}+1$} & + & 4 \\ \hline \raisebox{-2pt}{$\bigcirc$} &  & \raisebox{-2pt}{1} & & \\ [1.1ex] \hline\hline

$8_{7}$ & 4 & $\boxed{-D^4 \phi ^4-3 D^3 \phi ^5-D^2 \phi ^6-D^2 \phi ^4}+4 D^2 \phi ^2+3 D \phi ^5+$ & + & 3 \\ & & \raisebox{-2pt}{$+4 D \phi ^3+4 D \phi +\phi ^6+3 \phi ^4+2 \phi ^2+1$} & & \\ [1.1ex] \hline \raisebox{-2pt}{$\bigcirc$} &  & \raisebox{-2pt}{1} & & \\ [1.1ex] \hline\hline

$8_{8}$ & 3 & {\small $\boxed{-D^4 \phi ^4-D^3 \phi ^5-2 D^3 \phi ^3-D^2 \phi ^4}+D^2 \phi ^2+D \phi ^5+3 D \phi ^3+3 D \phi +2 \phi ^4+2 \phi ^2+1$} & + & 4 \\ \hline \raisebox{-2pt}{$\bigcirc$} &  & \raisebox{-2pt}{1} & & \\ [1.1ex] \hline\hline

\raisebox{-2pt}{$8_{9}$} & \raisebox{-2pt}{3} & \raisebox{-2pt}{$2 D^4 \phi ^4+3 D^3 \phi ^5+5 D^3 \phi ^3+D^2 \phi ^6+D^2 \phi ^4+5 D^2 \phi ^2-$} & \raisebox{-2pt}{+} & \raisebox{-2pt}{3} \\ [1.1ex] & & $\boxed{-3 D \phi ^5-4 D \phi ^3}+3 D \phi\, \boxed{-\phi ^6-3 \phi ^4-2 \phi ^2}+1$ & & \\ \hline \raisebox{-2pt}{$\bigcirc$} &  & \raisebox{-2pt}{1} & & \\ [1.1ex] \hline\hline

$8_{10}$ & 4 & $\boxed{-2 D^4 \phi ^4-3 D^3 \phi ^5-2 D^3 \phi ^3-D^2 \phi^6}+3 D^2 \phi ^2+3 D \phi ^5+$ & + & 3 \\ & & \raisebox{-2pt}{$+6 D \phi ^3+4 D \phi +\phi ^6+3 \phi ^4+3 \phi ^2+1$} & & \\ [1.1ex] \hline
\raisebox{-2pt}{$\bigcirc^2$} &  & \raisebox{-2pt}{$q+q^{-1}$} & &
 \\ [1.1ex] \hline\hline

$8_{11}$ & 1 & $D^4 \phi ^4+D^3 \phi ^5+2 D^3 \phi ^3+D^2 \phi ^4+D^2 \phi ^2\boxed{-D \phi ^5-2 D \phi ^3}+D \phi\, \boxed{-2 \phi ^4-\phi ^2}+1$ & + & 4 \\ \hline
\raisebox{-2pt}{$\bigcirc^2$} &  & \raisebox{-2pt}{$q+q^{-1}$} & &
 \\ [1.1ex] \hline\hline

\end{tabular}
\caption{\footnotesize Formal application of~\eqref{chirexpanswer} to the bipartite non-chiral knots $6_3$, $7_6$, $7_7$, $8_1$, $8_2$, $8_3$, $8_4$, $8_5$, $8_6$, $8_7$, $8_8$, $8_9$, $8_{10}$, $8_{11}$.}
\label{tab:non-ch-PD-2}
\end{table}

\begin{table}[h!]
\begin{tabular}{c|c|c|c|c}
{\rm knot} & {\rm Fr} & (\ref{chirexpanswer}) & {\rm existence\ of}  & {\rm braid}  \\
&&&{\rm BP\ diagram} & {\rm width}  \\ \hline \hline

$8_{12}$ & 2 & $D^4 \phi ^4+3 D^3 \phi ^3\boxed{-2 D^2 \phi ^4}+4 D^2 \phi ^2\boxed{-3 D \phi ^3}+2 D \phi +\phi ^4\,\boxed{-3 \phi ^2}+1$ & + & 5 \\ \hline \raisebox{-2pt}{$4_1$} &  & \raisebox{-2pt}{$q^4+q^{-4}-q^2-q^{-2}+1$} & & \\ [1.1ex] \hline\hline

$8_{13}$ & 3 & $\boxed{-D^3 \phi ^5-2 D^2 \phi ^4}+2 D^2 \phi ^2+D \phi ^5+D \phi ^3+3 D \phi +2 \phi ^4+\phi ^2+1$ & + & 4 \\ \hline \raisebox{-2pt}{Hopf} &  & \raisebox{-2pt}{$q+q^5$} & & \\ [1.1ex] \hline\hline

$8_{14}$ & 1 & $D^3 \phi ^5+2 D^2 \phi ^4\boxed{-D \phi ^5}+D \phi\, \boxed{-2 \phi ^4}+1$ & + & 4 \\ \hline \raisebox{-2pt}{$\bigcirc$} &  & \raisebox{-2pt}{1} & & \\ [1.1ex] \hline\hline

$8_{16}$ & 2 & {\footnotesize $\boxed{-D^4 \phi ^4-2 D^3 \phi ^5-2 D^3 \phi ^3-D^2 \phi ^6-D^2 \phi ^4}+2 D \phi ^5+2 D \phi ^3+2 D \phi +\phi ^6+2 \phi ^4+\phi ^2+1$} & + & 3 \\ \hline \raisebox{-2pt}{$\bigcirc$} &  & \raisebox{-2pt}{1} & & \\ [1.1ex] \hline\hline

\raisebox{-2pt}{$8_{17}$} & \raisebox{-2pt}{3} & \raisebox{-2pt}{$D^4 \phi ^4+2 D^3 \phi ^5+3 D^3 \phi ^3+D^2 \phi ^6+D^2 \phi ^4+4 D^2 \phi ^2-$} & \raisebox{-2pt}{+} & \raisebox{-2pt}{3} \\ [1.1ex] & & $\boxed{-2 D \phi ^5-2 D \phi ^3}+3 D \phi\, \boxed{-\phi ^6-2 \phi ^4-\phi ^2}+1$ & & \\ \hline \raisebox{-2pt}{$\bigcirc$} &  & \raisebox{-2pt}{1} & & \\ [1.1ex] \hline\hline

$8_{18}$ & 3 & $\boxed{-D^4 \phi ^4}+D^3 \phi ^5\boxed{-D^3 \phi ^3}+D^2 \phi ^6+2 D^2 \phi ^4+2 D^2 \phi ^2\boxed{-D \phi ^5}+$ & + & 3 \\ & & $+2 D \phi ^3+3 D \phi\, \boxed{-\phi ^6-\phi ^4}+\phi ^2+1$ & & \\ \hline
\raisebox{-2pt}{$\bigcirc^3$} &  & \raisebox{-2pt}{$(q+q^{-1})^2$} & & 
 \\ [1.1ex] \hline\hline

$8_{20}$ & 1 & $\boxed{-2 D^3 \phi ^3-D^2 \phi ^4-2 D^2 \phi ^2}+2 D \phi ^3+D \phi +\phi ^4+2 \phi ^2+1$ & + & 3 \\ \hline
\raisebox{-2pt}{$\bigcirc^2$} &  & \raisebox{-2pt}{$q+q^{-1}$} & & 
 \\ [1.1ex] \hline\hline

$8_{21}$ & 0 & $D^3 \phi ^3+D^2 \phi ^4\,\boxed{-D \phi ^3-\phi ^4}+1$ & + & 3 \\ \hline
\raisebox{-2pt}{$\bigcirc^2$} &  & \raisebox{-2pt}{$q+q^{-1}$} & & 
 \\ [1.1ex] \hline\hline

\ldots &&&&\\

\end{tabular}
\caption{\footnotesize Formal application of~\eqref{chirexpanswer} to the bipartite non-chiral knots $8_{12}$, $8_{13}$, $8_{14}$, $8_{16}$, $8_{17}$, $8_{18}$, $8_{20}$, $8_{21}$.}
\label{tab:non-ch-PD-3}
\end{table}

\setcounter{equation}{0}
\section{Non-chiral ambiguity
\label{ambig}}

This section attempts to describe the situation with non-chiral PD.
We begin with the set of questions in Section \ref{oque} and end with our current expectations
about the answers to them in Section \ref{sec:expec}.
This is commented by the discussion of the factorization problem, the notion of local minima
and their seeming {\it irrelevance} to the ambiguity puzzle.

\subsection{Open questions \label{oque}}

Here, we formulate questions on the non-chiral case. Some answers are discussed in other subsections, and the outcome is formulated in Section~\ref{sec:expec}.

\medskip

\noindent $\bullet$ If there are different non-chiral bipartite diagrams describing a given knot,
do they provide different polynomials in $\phi,\bar\phi,D$ --
of course, equivalent modulo $G=\phi\bar\phi D + \phi + \bar\phi$? (See Sections~\ref{sec:minDiagPD}, \ref{sec:ambiguity}, \ref{sec:expec}.)

\medskip

\noindent $\bullet$ Are these answers ``minimal'', i.e. can one further subtract a positive polynomial that is a multiple of $G$ without losing the positivity? (See Sections~\ref{sec:minDiagPD}, \ref{sec:ambiguity}, \ref{sec:expec}.)

\medskip

\noindent $\bullet$ Depending on the outcome of the previous question,
are there diagrams reproducing {\it all} ``local minima''
or -- if not only minimal answers can emerge --
are there diagrams reproducing {\it all} allowed polynomials of $\phi,\bar\phi,D$? (See Sections~\ref{sec:minDiagPD}, \ref{sec:ambiguity}, \ref{sec:expec}.)

\medskip

\noindent $\bullet$ Can a chiral bipartite knot be also represented by a non-chiral Reidemeister-equivalent bipartite diagram,
which provide $\bar\phi$-dependent answer? (See Sections~\ref{sec:minDiagPD}, \ref{sec:ambiguity}, \ref{sec:expec}.)

\medskip

\noindent $\bullet$ How is the anti-chiral ``dual'' of a chiral answer reproduced from a ``wrong''\ non-positive partner
of the chiral expansion, when it exists? (See Section \ref{partners}.)

\medskip

\noindent $\bullet$ What happens for $H_{[2]}$ and higher $H_{[r]}$?
The colored HOMFLY polynomial do not possess the $q\leftrightarrow q^{-1}$ symmetry, thus, do not depend on $z$ only --
but does it mean that all the $\psi$-parameters (see~\cite{bipsym}) are needed for their decomposition?
Can they be unambiguously defined for chiral knots?
What is the ambiguity in the non-chiral case, does it exceed that for the fundamental representation?
In what sense? (See Section \ref{reps}.)

\medskip

\noindent $\bullet$ What is the meaning of PD in the case when there are {\it no} bipartite realizations
for a knot? (See Section \ref{sec:sum}.)


\subsection{Factorization problem for polynomials\label{sec:facprob}}

The origin of the problems in the non-chiral case is very simple --
the lack of a constructive solution to the factorization problem in the generic algebraic geometry.

Taking an integer number modulo another one is a well-defined operation,
$ \ a\, {\bf mod}\, p\ $ belongs to the segment $[0,p-1]$ and is unambiguous.
However, the factorization w.r.t. a linear combination of several variables is different and can have different ``local minima''.
For example, $u^2+uv+w^2$ modulo $u+v$ would have $u^2$ and $v^2$ as two different positive minima.
There is no way to prefer one against another.
Likewise $(u+v)^2+(u+v)w+w^2$ modulo $u+v+w$ would have $(u+v)^2$ and $w^2$ as different local minima.
In the case of non-chiral knot diagrams, we are exactly in this situation
since we need to factorize the HOMFLY polynomial w.r.t. $\phi\bar\phi D + \phi + \bar\phi$,
which pays the role of $u+v+w$ in the above example.
Thus, a result of the ``minimization'' can be ambiguous, and we do not always get a
unique answer for a ``minimal PD''.
This raises a bunch of questions.

\subsection{Are polynomials, related to the diagrams, minimal?}\label{sec:minDiagPD}

One may ask whether PDs related to bipartite diagrams (which we call BEs) are local minimums in the sense of Section \ref{sec:facprob}.
The answer is yes in some cases, but generally no, and it is too hard to answer exactly in most cases. Below, we illustrate various possibilities with the simplest examples. By $PD^{\mathcal{K}}$ we mean the PD for a knot $\mathcal{K}$ that is BE for its standard rational-knot diagram with even entries \cite{bipHOM}.

As soon as a PD depends on all three variables $D$, $\phi$, $\bphi$, it is defined up to a multiple of $G=\phi+\bphi+D\phi\bphi$. In particular cases, a BE can be a ``local minimum'', i.e. only \textit{addition} of a multiple of $G$ that is a positive polynomial leaves the answer a positive polynomial. E.g., this is the case of
\begin{equation}
PD_{4_1}=1+D(\phi+\bphi)+\phi\bphi\,.
\end{equation}
Generally, a BE is non-minimal, i.e. the \textit{subtraction} of a multiple of $G$ that is a positive polynomial may leave the answer a positive polynomial. 

\paragraph{Chiral PD with framing.}
Formally speaking, even a chiral BE can be non-minimal if one includes the framing factor of $(1+D\bar{\phi})^\mathrm{fr}$ in the expression. E.g., 
\begin{equation}\label{PD-3_1}
PD_{3_1}=(1+D\bar{\phi})^2(1+2D\phi+\phi^2)=1+\phi^2+2D\bphi\phi^2+D^2\bphi\phi(1+\bphi\phi)+D^3\bphi^2\phi+D(2+D\bphi)G\,.
\end{equation}
However, the original answer factorizes as $(1+\phi)^k(1+\bphi)^k$ (here $k=2$) for $D=1$, as any PD that is a BE must
(the $D=1$ criterium from Section \ref{sec:D1}), while the results of subtraction of $D(2+D\bphi)G$ (as well as of $DG$, $2DG$, $D^2\bphi G$, and $D(1+D\bphi)G$) do not. 
There are also positive polynomials that are products of $G$ and a non-positive polynomials, such as $(\phi+\bphi)^3+(D\phi\bphi)^3$, $\phi^3+(1+D\phi)^3\bphi^3$, $\bphi^3+(1+D\bphi)^3\phi^3$, but subtractions of their multiples cannot leave the PD~\eqref{PD-3_1} positive due to its degree in $\phi,\bphi,D$.
In this sense, the above PD for the trefoil knot \textit{is} a ``local minimum'' provided that the $D=1$ criteria is satisfied. The case of the chiral knot $5_2$ can be considered in a similar way with the same conclusion. At the same time, the BE for the chiral knot $5_1$ contains $80$ monomials such that each one, when subtracted with the factor of $G$, leaves the $PD$ a positive polynomial. Potential candidates for ``local minima'' contain at least all their combinations, which cannot be examined for the $D=1$ criterium within a reasonable time. 

\begin{figure}[h!]
\includegraphics[width=15cm]{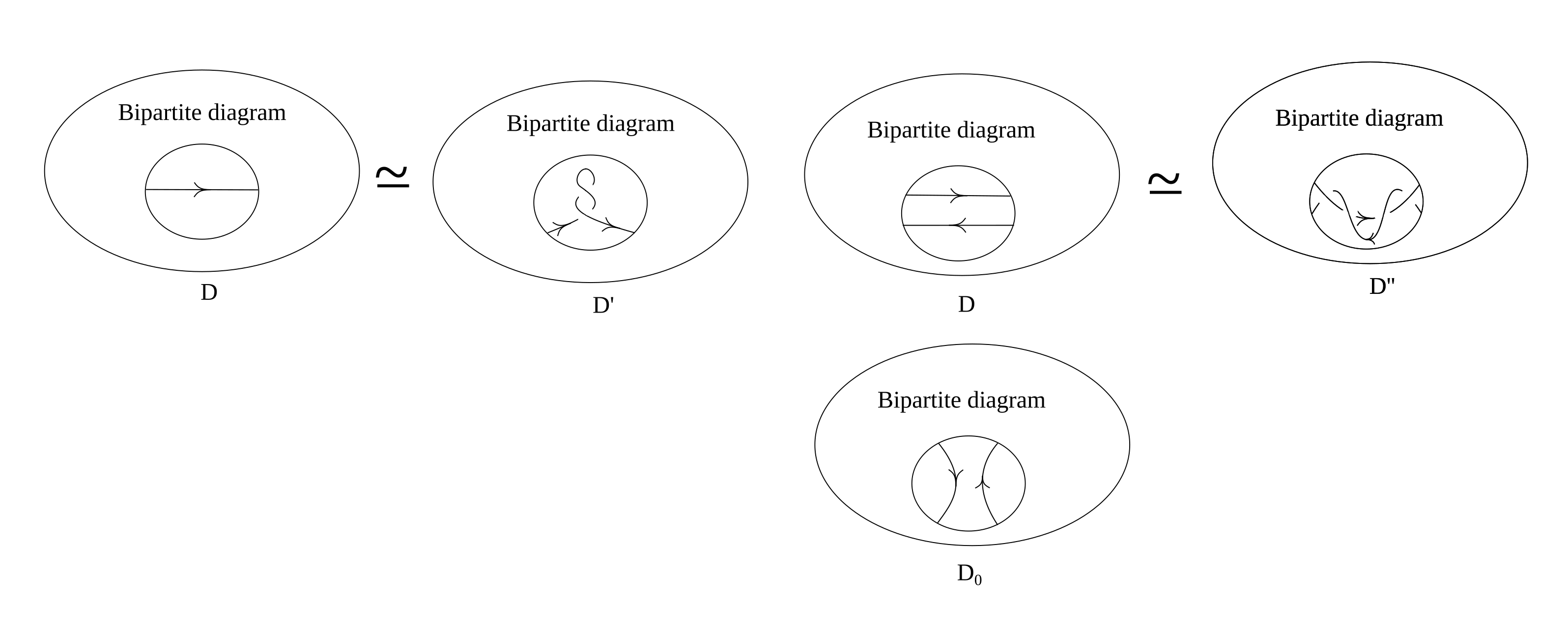}
\caption{\footnotesize Analogues of the first and the second Reidemeister moves for bipartite diagrams.}\label{fig:BipRei}
\end{figure}

\paragraph{Framing in the non-chiral case.}

In the generic non-chiral case, there is also an ambiguity in a framing factor (which may affect the minimality of an answer), even if we know a bipartite diagram. Namely, if we write the PD from a bipartite diagram with $n_\bullet$ positive and $n_\circ$ negative bipartite vertices, the framing factor $A^{-2(n_\bullet - n_\circ)}$ can be distributed in many ways between the powers of $(1+D\phi)$ and $(1+D\bphi)$.  A choice of the framing factor of the form $(1+D\phi)^{-n_\bullet}(1+D\bphi)^{-n_\circ}$ guarantees the ``bipartite first Reidemeister move'' (Fig.\,\ref{fig:BipRei}) invariance, i.e. contracting of a doubly twisted loop of any orientation gives the factor of $1$. With this choice of the framing factor, PD gives $1$ for $D=1$ (see Section \ref{sec:D1}).
On the other hand, this form breaks the polynomiality of an answer, and its polynomial counterpart $(1+D\bphi)^{n_\bullet}(1+D\phi)^{n_\circ}$ makes an answer rather complicated. Instead of that, one can take the framing factor in the form $(1+D\bphi)^{n_\bullet-n_\circ}$ (for $n_\bullet-n_\circ\ge0$) or $(1+D\phi)^{n_\bullet-n_\circ}$ (for $n_\bullet-n_\circ<0$). 
All three forms of the framing factor are equivalent modulo $G$ as $(1+D\phi)(1+D\bphi)=1+G$. Below, when examining the minimality of the PDs, we prefer to write the framing factor in the last, ``minimal'', form.

\paragraph{Non-chiral PD with the ``minimal'' framing.}
In the simplest case, such as
\begin{eqnarray}
PD_{6_1}&=&(D\bphi+1)\big(D(\bphi+D)\phi^2+(2D+2\bphi)\phi+D\bphi+1\big),\\
PD_{6_2}&=&(D\bphi+1)^2\big((\bphi+D)\phi^3+(2D^2+3D\bphi+1)\phi^2+(D^2\bphi+3D+2\bphi)\phi+D\bphi+1\big),\\
PD_{6_3}&=&(D\phi+1)^2\big(D^2(D+\phi)\bphi^3+3D(D+\phi)\bphi^2+(3D+3\phi)\bphi+D\phi+1\big),
\end{eqnarray}
$PD_{7_6}$, and $PD_{7_7}$, plain enumeration of all variants to subtract a multiple of $G$ leaving the PD positive (as above) is available and shows that the above expressions are ``local minima'' provided that the $D=1$ criteria is satisfied (see Section \ref{sec:D1}). However, the number of possible subtractions grows too fast to proceed straightforwardly.  

\begin{figure}[h!]
$\begin{array}{ccccccc}
&\multicolumn{5}{c}{\text{Signs of vertices}}\\
\text{Knot}&1&2&3&4&5&6\\
12n881&+&+&+&+&+&+\\
11n162&-&+&+&+&+&+\\
10_{140}&-&+&+&-&+&+\\
12n145,&-&+&+&+&-&+\\
12n838\\
12n442&-&-&+&+&+&+\\
4_1\#4_1,&-&+&+&-&-&+\\
12n462\\
4_1\#\overline{3}_1&+&-&+&-&-&+\\
12a1202&+&+&+&-&-&-
\end{array}$
\begin{tabular}{cc}
\includegraphics[width=4.5cm]{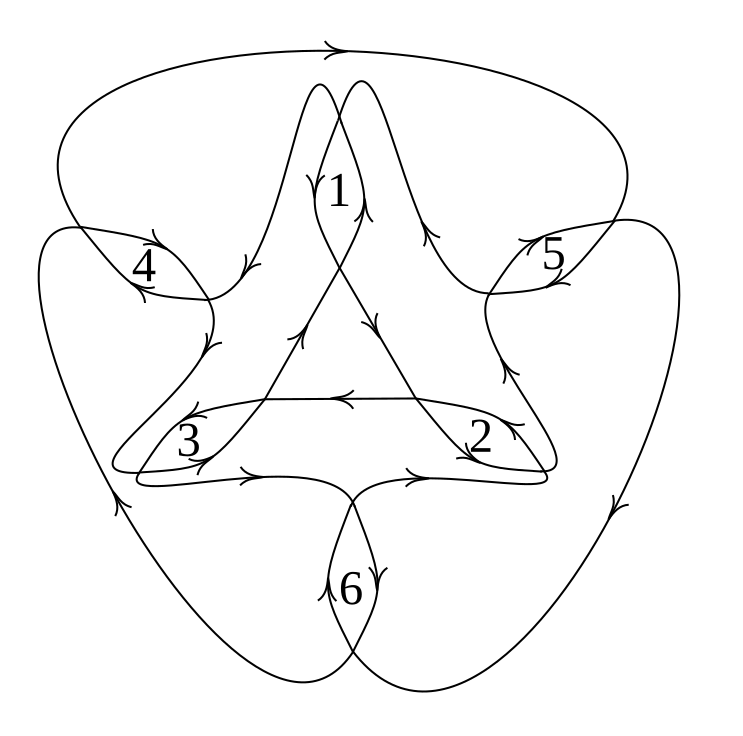}
&
\includegraphics[width=4.5cm]{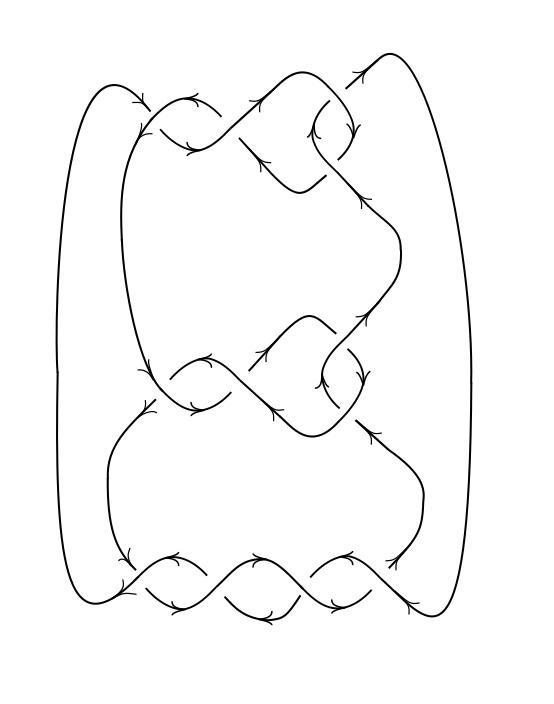}\\
I&II
\end{tabular}
\caption{\footnotesize I. The diagram that reproduces one of 8 bipartite knots depending on signs of the bipartite vertices. Table on the left give knots up to 12 crossings with the corresponding HOMFLY polynomial (a diagram with any other combination of signs is equivalent to one of these 8 ones). One of these knots is $10_{140}$. II. The knot $10_{140}$ as the Montensinos knot $K(\frac{2}{3},-\frac{2}{3},\frac{1}{4})$. The two diagrams may correspond to the two ``local minima'' of PD.}\label{fig:BP6}
\end{figure}

\paragraph{Two candidates for ``local minima'' for the knot $10_{140}$.}
This is the simplest case where we have two bipartite diagrams of the same knot, and they give rise to different PD.
For digram in Fig.\,\ref{fig:BP6}.I:
\begin{eqnarray}
PD_{10_{140}}^{\rm (I)}=\big((D\bphi+1)^2\phi^4+4(D+\bphi)(D\bphi+1)\phi^3+((2D^2+4)\bphi^2+(2D^3+10D)\bphi+3D^2+3)\phi^2+\nn\\
+(6D^2\bphi+4D\bphi^2+4D+2\bphi)\phi+2D\bphi+\bphi^2+1\big)(D\bphi+1)^{-2}(D\phi+1)^{-4}\,.
\end{eqnarray}
For the digram in Fig.\,\ref{fig:BP6}.II:
\begin{eqnarray}
PD_{10_{140}}^{\rm (II)}=\big(D(D+\bphi)(D\bphi+1)\phi^4+(D^4\bphi+(\bphi^2+1)D^3+5D^2\bphi+(3\bphi^2+3)D+2\bphi)\phi^3+(3D^3\bphi+\nn\\+(3\bphi^2+3)D^2+9D\bphi+3\bphi^2+3)\phi^2+(6D^2\bphi+4D\bphi^2+4D+2\bphi)\phi+2D\bphi+\bphi^2+1\big)(D\bphi+1)^{-2}(D\phi+1)^{-4}\,.
\end{eqnarray}
The difference is
\begin{eqnarray}
PD_{10_{140}}^{\rm (II)}-PD_{10_{140}}^{\rm (I)}=(D^2-1)(\phi+\bphi+D\phi\bphi)(\phi+\bphi+DD)\phi^2\,.
\end{eqnarray}
This means that neither of PDs is ``more minimal'' than the other. In particular, they both could be ``local minimuma'' provided that  the $D=1$ criterium is satisfied (see Section \ref{sec:D1}).

\begin{figure}[h!]
\includegraphics[width=17cm]{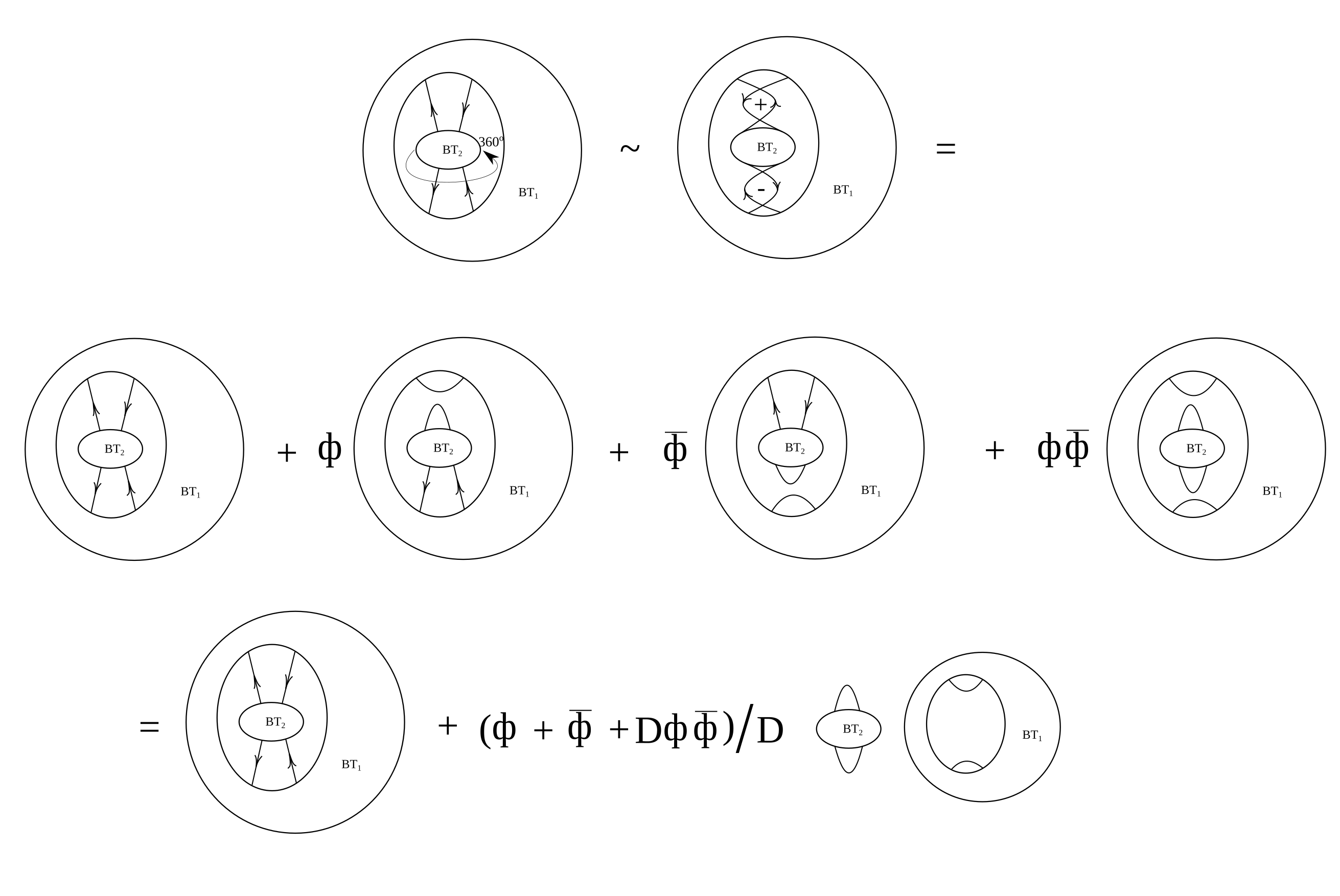}
\caption{\footnotesize Example of an equivalence transformation of a bipartite diagram that changes PD to a positive polynomial that is a multiple of $\phi+\bphi+D\phi\bphi$. An initial bipartite diagram (living inside a big circle) is arbitrarily split into two bipartite 4-tangles ${\rm BT}_1$ and ${\rm BT}_2$. The new answer for the HOMFLY polynomial (coming from the last line) is not a ``local minimum''.} 
\label{fig:BipEquiv}
\end{figure}

\paragraph{The general answer is non-minimal.}
Applying the transformation in Fig.\,\ref{fig:BipEquiv} to a bipartite diagram, we get a new bipartite diagram. The PD read from a new diagram differs from that read from an old diagram by adding $(\phi+\bphi+D\phi\bphi)*(\text{a positive polynomial})$. The second answer is then non-minimal.

\subsection{Ambiguity in non-chiral PD}\label{sec:ambiguity}

A straightforward application of (\ref{chirexpanswer}) in the general case gives (up to the framing factor) a mixed sign polynomial in $D$ and $\phi$. There are infinitely many ways to obtain a positive polynomial by adding multiples of $(\phi+\bphi+D\phi\bphi)$. The most straightforward way is to
\begin{itemize}
\item{Take the PD from (\ref{chirexpanswer}) with the framing factor in the polynomial form, $A^{-2\mathrm{fr}}\to(1+D\bphi)^{\mathrm{fr}}$.}
\item{Notice that all addends with the negative sign are proportional to $\phi$.}
\item{Substitute each addend with the negative sign with its product by $-\bphi(1+D\phi)/\phi$.}
\end{itemize}
I.e., one makes the change
\begin{equation}
PD=(1+D\bphi)^{\mathrm{fr}}(PD_+-\phi PD_-)\quad \to \quad \widetilde{PD}=(1+D\bar\phi)^{\mathrm{fr}}\big(PD_++\bphi(1+\phi D)PD_-\big)\,,\label{PosPol}
\end{equation}
where both $PD_+$ and $PD_-$ are positive polynomials in $D$, $\phi$.
Among other possibilities, one can take instead of (\ref{PosPol}), e.g.,
\begin{equation}
PD=(1+D\bar\phi)^{\mathrm{fr}-1}\big(PD_+(1+D\bar\phi)-\phi(1+D\bar\phi) PD_-\big)\quad \to \quad
\widetilde{\widetilde{PD}}=(1+D\bar\phi)^{\mathrm{fr}-1}\big(PD_+(1+D\bar\phi)+\bar\phi PD_-\big). \label{PosPol1}
\end{equation}
Such $\widetilde{PD}$ or $\widetilde{\widetilde{PD}}$ generates infinitely many PDs which are obtained by adding $(\phi+\bphi+\phi\bar\phi D)Q$, for a positive polynomial $Q(D,\phi,\bphi)$. Moreover, the answers (\ref{PosPol}), (\ref{PosPol1}) are generally not ``local minima'', i.e. there is a huge number of positive polynomials $Q$ such that the subtraction of $(\phi+\bphi\phi D)Q$ from $\widetilde{PD}$ or $\widetilde{\widetilde{PD}}$ gives a positive polynomial.

E.g., already $\widetilde{PD}$ from (\ref{PosPol}) for the knot $6_1$ contains $10$ different monomials that can be subtracted with the factor of $(\phi+\bphi+D\phi\bar\phi)$ so that the resulting polynomial remains positive. Some of these monomials can be subtracted with numeric coefficients greater than 1, so that one can construct overall $2^{17}$ different polynomials of these monomials. Enumeration of these polynomials selects $9792$ ones whose subtraction with the factor of $G=(\phi+\bphi+D\phi\bar\phi)$ still gives positive polynomials, but only two of them satisfy the $D=1$ criteria (see Section~\ref{sec:D1}), and one of them is the PD related to the standard (bipartite) diagram of $6_1$. But if one applies the same method already to the knot $6_2$, there are $2^{54}$ polynomials to be tested for $D=1$ criterium, and the described procedure cannot be done within a reasonable time.

One can perform the same method starting from $\widetilde{\widetilde{PD}}$ instead of $\widetilde{PD}$, and this notable reduces the number of possible subtractions. When applied to knots $6_1$, $8_1$, $10_1$, this version of the method reproduces in each case the $(D=1)$-correct polynomial that is the PD related to the standard (bipartite) diagram of the knot. However, for other non-chiral knots with $6$ and $7$ crossings, 
no $(D=1)$-correct polynomials have been found among differences of $\widetilde{\widetilde{PD}}$ and positive multiples of $G$, i.e. all such polynomials (including the polynomials related to the standard diagrams) are not ``more minimal'' than $\widetilde{\widetilde{PD}}$.

\section{Expectations about relations between PD and BE
}\label{sec:expec}

In this subsection, we give one more summary of what we expect to be true after a study of a big variety of examples.
Here, all examples themselves are not given (because of being a rather huge amount of data), only the outcome is present. However, one can easily consider examples by his/her own as all the methodology is provided and all the needed HOMFLY polynomials are listed in~\cite{katlas,knotinfo}. Examples of redrawing Montesinos knots and their BEs will be systematized and provided in a separate text~\cite{bipex}.

\begin{figure}[h!]
\begin{center}
\includegraphics[width=13cm]{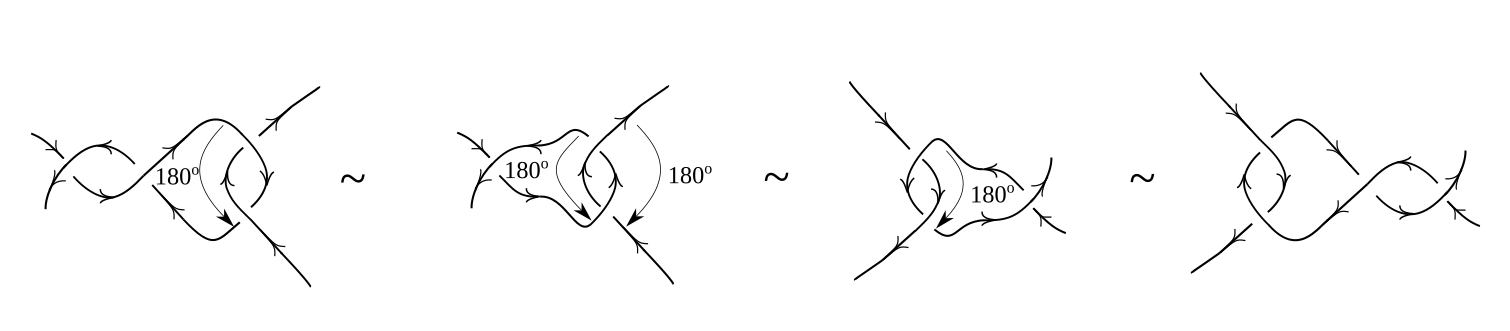}
\end{center}
\caption{\footnotesize Example of an equivalence transformation of a bipartite diagram that does not change PD \cite{BipEq}.}\label{fig:BipEq}
\end{figure}

\begin{itemize}
\item{} In the chiral case, the answer calculated from a chiral bipartite diagram is unique and coincides with~\eqref{chirexpanswer}.

This statement is not quite accurate because we can always add trivial loops (Fig.\,\ref{fig:BipRei}),
but they contribute as powers of $\frac{1+\phi D}{A^2}=1$ and do not affect the HOMFLY polynomial in the topological framing.

\item{} A chiral diagram is not unique, even the one with minimal number of bipartite vertices. At least, one can apply equivalence transformation in Fig.\,\ref{fig:BipEq}, which affects a bipartite diagram but not the PD.

\item{} A chiral knot can also have a chiral clone (see Section \ref{sec:chiral}) with the same BE, but with non-equivalent bipartite diagrams, e.g., as knots $5_1$ and $10_{132}$.

\item{} In the non-chiral case, we have a lot of Reidemeister-equivalent bipartite diagrams,
which provide the HOMFLY polynomials differing not just by the framing but by multiples of $G=\phi+\bphi+D\phi\bphi$,
i.e. possessing {\it a priori} different BEs, see e.g., Fig.\,\ref{fig:BipEquiv}.

In the non-chiral case, there is no algorithm like (\ref{chirexpanswer}) which unambiguously selects a particular formula
from this variety, i.e. there is no canonical way to attribute a BE to a given HOMFLY polynomial.

\item{} One can wonder, if {\it any} polynomial from the family of PDs equivalent modulo $G$ is associated with a given knot,
i.e. if one can find a Reidemeister-equivalent bipartite diagram with BE given by any addition of multiple of $G$.
In Section \ref{obsta}, we find serious restrictions, still, they leave a wide ambiguity.

\item{} The full set of equivalence transformations of bipartite diagrams is unknown, even for those ones with the minimal number of bipartite vertices.

\item{}  Despite factorization by $G$ does not provide a canonical representative in a given equivalence class,
one could hope that at least {\it ``local minima''} are somehow distinguished.
BE in the most cases are \textit{not} ``local minima'' w.r.t. subtraction of arbitrary $G$ multiples, but at least in simple cases the BE are ``local minima'' under additional restriction (see Section \ref{sec:D1}). 
On the other hand, our counterexamples to minimality in the latter sense (in Figs.\,\ref{fig:BipEquiv} and \ref{fig:BipRei}) 
require for ``extra windings'' and could be excluded
by a generalization of ``topological framing'' to bipartite knots --
like the standard topological framing, which trivialises the first Reidemeister move.
Hence, it is still an open question if ``local minima'' do not arise in less trivial situations.

\end{itemize}

All the mentioned questions remain for further investigation.

\setcounter{equation}{0}
\section{Possible obstacles to the existence of a bipartite diagram
\label{obsta}
}

The moral of the above considerations is that positive decomposition (\ref{bipexp})
of the fundamental HOMFLY polynomial exists for all knots, moreover, for {\it non-chiral} knots
it is ambiguous and requires further investigation.
At the same time, the original  motivation and the very discovery of PD came from the
study of {\it bipartite} diagrams made from antiparallel lock tangles in Fig.\,\ref{fig:pladeco}.
Usually, it is quite non-trivial to understand if the bipartite realization exists
for a given knot.
The best-known criterium involves the analysis of the Alexander {\it ideals} \cite{bip,bipfund},
which, in variance with the Alexander {\it polynomials}, are not immediately extracted
from the HOMFLY polynomials.
A natural question now is if the existence of such a diagram can be discovered from the study of the HOMFLY polynomials,
i.e. {\bf is there anything special about the fundamental HOMFLY polynomials coming from bipartite diagrams}?

In this section, we discuss two possible criteria to exclude particular PDs from the set of possible BEs.
One of them,  the $D=1$ reduction, efficiently restricts the choice within $G$-equivalence classes.
Another one, based on the {\it precursor Jones polynomials} \cite{bipfund}, 
allows to exclude the whole set of PDs of the fixed degree (not greater than 16) in $\phi$, $\bphi$ inside the same class.

\subsection{
$D=1$ reduction}\label{sec:D1}

If a knot ${\cal K}$ is realized as a bipartite diagram, then (\ref{PD}),
\be\label{sum}
H_\Box^{\cal K} \ \stackrel{?}{=} \  \mathrm{Fr}\cdot\sum_{k \geq 0} \sum_{i=0}^{n_\bullet} \sum_{j=0}^{n_\circ} {\cal N}_{ijk} D^k \phi^i\bphi^j,\qquad
\mathrm{Fr}=(1+\bphi D)^{\mathrm{fr}},\ \mathrm{fr} \geq 0, \quad \text{or} \quad \mathrm{Fr}=(1+\phi D)^{-\mathrm{fr}},\ \mathrm{fr}<0\,,
\ee
has additional properties.
This decomposition is obtained by resolving each lock tangle in two ways,
weighted with the coefficients $1$ and $\phi$ or $1$ and $\bphi$, depending on the orientation of a lock, see Fig.\,\ref{fig:pladeco}, 
and the power of $D$ is equal to the number of non-intersecting planar cycles in this resolution.
For $D=1$, i.e., $A=q^{\pm 1}$, the contributions of all the cycles become the same,  thus, the sum~\eqref{sum} reduces to just
\be
H_\Box^{\cal K}(A=q^{\pm 1}) = {\rm Fr}\cdot(1+\phi)^{n_\bullet}(1+\bphi)^{n_\circ}
\label{D=1PD}
\ee
where  $n_\bullet$ and $n_\circ$ are just the quantities of vertices with different orientations in a diagram.
Since for PD that is related to a bipartite diagram, $\mathrm{fr} = n_\bullet -n_\circ$,  this formula further reduces to
\be
H_\Box^{\cal K}(A=q^{\pm 1}) =   \Big((1+\phi) (1+\bphi)\Big)^{\max(n_\bullet,n_\circ)} = 1\ \mod\ G\,.
\ee
This imposes the additional constraint on PD coming from a bipartite diagram, which is quite severe --
and not satisfied by many expressions, obtained  in the previous sections.
At the same time, if we do not have a bipartite diagram, $\max(n_\bullet,n_\circ)$ is substituted with some integer that is not constrained, since different powers of $(1+\phi) (1+\bphi)$
differ by multiples of $(1+\phi) (1+\bphi) - 1 = \phi+\bphi+\phi\bphi \ \stackrel{D=1}{=} \ 0 \mod\ G$.

This criterium does not yet help {\it against} the chiral expressions for $9_{35}$:
\be \label{935PD}
H^{9_{35}}_\Box = A^{10}\Big(\phi^5D^3+5\phi^4D^2 +10\phi^3D + \phi^2(3D^2+7) +5\phi D  +1\Big)
\ \stackrel{D=1}{=} \ \Big((1+\bphi)(1+\phi)\Big)^5
\ee
-- despite there must be no bipartite diagram for this knot.

Actually, the PD for the knot $9_{35}$ is not forbidden even by the strengthened criterium discussed below.

Let us begin from the chiral case and do not consider framing.
We can look at powers of $\phi$.
Every vertex can contribute $1$ or $\phi$  to the PD.
The highest power defines the number of bipartite vertices, denote it by $M$.
Since we consider knots (not links) the zeroth power comes with coefficient $1$
(it contributes one cycle, i.e. $D$, but since we consider the reduced HOMFLY polynomial, we divide by $D$).
Each flip from $1$ to $\phi$ changes the power of $D$ by one -- either increase it or decrease.
Therefore the first  terms in PD, coming from a chiral bipartite diagram, are
\be
1 + M  \phi D + \phi^2 \left(m_{2,2}D^2 + m_{2,0}\right) + \phi^3 \left(m_{3,3}D^3 + m_{3,1}D \right)+ \ldots
= \sum_{k=0}^M  \phi^k \sum_{i=0}^{k/2} m_{k,k-2i} D^{k-2i}
\ee
where $m_{2,2}+m_{2,0} = \frac{M(M-1)}{2}$ and $\sum_{i=0}^{k/2}m_{k,k-2i} = \frac{M!}{k!(M-k)!}$ --
for $D=1$ these sum rules bring us back to the chiral version of (\ref{D=1PD}).
However, this does not exhaust all the restrictions.
In particular, $m_{3,1}$ gets two contributions -- from $m_{2,2}$ and $m_{2,0}$, and it cannot be smaller that $m_{2,0}$:
\be
m_{3,1} \geq m_{2,0}\,.
\ee
(But nothing prevents $m_{3,3}$ from vanishing.)
Expressions like
\be
1 + M  \phi D + \phi^2 \left\{\left(\frac{M(M-1)}{2}-m_{2,0}\right)D^2 + m_{2,0} \right\} + \frac{M(M-1)(M-2)}{6}\phi^3 D
+ \ldots
\ee
are allowed, and (\ref{935PD}) is exactly like this. 

Another example
\be
H^{9_{49}}_\Box = A^{10}\Big(\phi^5D + \phi^4(2D^2+3) +10\phi^3D + \phi^2(4D^2+6) +5\phi D  +1\Big)
\ \stackrel{D=1}{=} \ \Big((1+\bphi)(1+\phi)\Big)^5
\label{949PD}
\ee
is slightly more complicated, but also not forbidden by this enhanced criterium despite the fact that $9_{49}$ is non-bipartite.

\subsection{A viable alternative: precursor check}\label{sec:prec}

\begin{figure}[h!]
\centering
\includegraphics[width=11cm]{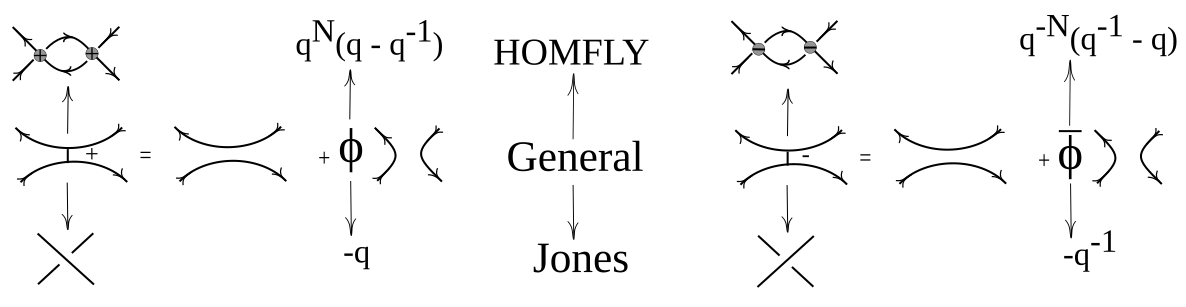}
\caption{\footnotesize
The Kauffman and lock decompositions can be generalized to describe both the bipartite HOMFLY polynomial and the precursor Jones polynomial, picture from~\cite{bipfund}.
}\label{fig:gen-pol}
\end{figure}

A precursor diagram is obtained from a bipartite diagram by shrinking lock vertices into ordinary single vertices. The planar technique reveals the duality between bipartite and precursor diagrams, as dictated by the similarity between the Kauffman bracket decomposition of a single crossing in Fig.\,\ref{fig:Kauff} and the planar decomposition of the lock element in Fig.\,\ref{fig:pladeco}. Thus, a polynomial written in a general form
\begin{equation}
    P^{\,\rm bipartite}_\Box = \mathrm{Fr}\cdot\sum^{2^v} D^a  \phi^b \bar \phi^c
\end{equation}
gives rise to both the bipartite HOMFLY polynomial and the precursor Jones polynomial (more details in~\cite{bipfund})
\begin{equation}\label{H-J-gen}
\begin{aligned}
    H^{\,\rm bipartite}_\Box &= A^{-2(n_\bullet-n_\circ)}\sum^{2^v} D^a \phi^b \bar \phi^c\,, \qquad \phi=A\{q\}\,,\; \bphi=-A^{-1}\{q\}\,, \; D=\frac{\{A\}}{\{q\}}\,, \\
    J^{\,\rm precursor}_\Box &= (-1)^{\frac{1}{2}(w+n_\bullet-n_\circ)}\cdot q^{-\frac{3}{2}w-\frac{1}{2}(n_\bullet-n_\circ)}\sum^{2^v} D^a (-q)^b (-q)^{-c}\,, \qquad D=q+q^{-1}\,,
\end{aligned}
\end{equation}
where the framing factors are present to restore topological invariance, $w$ is the writhe number of a precursor diagram, $v$ is both the number of bipartite vertices and the number of vertices in the corresponding precursor diagram.

This fact can be used to check the relevance of the posititve decomposition of the HOMFLY polynomial. The degree of the HOMFLY polynomial in $\phi$, $\bphi$ (to be denoted by $M$) is equal to the number of bipartite vertices in a hypothetical corresponding bipartite knot diagram, and in turn, equal to the number of crossings in the precursor diagram. Thus, if the HOMFLY PD corresponds to some bipartite diagram, then it must reduce to the precursor Jones of a link with the crossing number $\leq M$.

\paragraph{\bf Chiral case.} Let us demonstrate this trick on the trefoil knot. Its bipartite diagram corresponds to a disjoint sum of two unknots, which is really seen by the bipartite HOMFLY polynomial:
\begin{equation}
    A^{4} \cdot H^{3_1}_\Box = 1+\phi^2 + 2\phi D \quad \overset{\phi\,\rightarrow\, -q,\, D\,\rightarrow\, q+q^{-1}}{\longrightarrow} \quad {\rm Fr}_J^{-1} \cdot J^{\circ \circ}_\Box = -q(q+q^{-1})
\end{equation}
where ${\rm Fr}_J=(-1)^{\frac{1}{2}(w+n_\bullet-n_\circ)}\cdot q^{-\frac{3}{2}w-\frac{1}{2}(n_\bullet-n_\circ)}=-q^{-1}\,$ in this case, but if an initial bipartite diagram is unavailable, ${\rm Fr}_J$ is unknown. Note that we do not get the unity because we deal with the polynomials normalized only via one component (i.e., divided only by $D$ but not by $D$ to the power of the number of components in a link).

A case of the most interest are non-bipartite knots. Consider the knot $9_{35}$ and its PD HOMFLY polynomial~\eqref{935PD}. Its degree is $5$ = number of bipartite vertices = number of precursor vertices in a hypothetical diagram. Thus, the HOMFLY polynomial reduction
\be
A^{10}\cdot H^{9_{35}}_\Box = \Big(\phi^5D^3+5\phi^4D^2 +10\phi^3D + \phi^2(3D^2+7) +5\phi D  +1\Big)
\; \overset{\phi\,\rightarrow\, -q,\, D\,\rightarrow\, q+q^{-1}}{\longrightarrow} \;\nn \\ 
\stackrel{?}{\longrightarrow} \ {\rm Fr}_J^{-1} \cdot J_\Box^{?} =-q^8+2 q^6+2 q^2-1
\ee
must correspond to some link with the crossing number at most $5$ if the PD~\eqref{935PD} correspond to a bipartite diagram. Sorting through all such Jones polynomials, we get sure that there is no such value. Taking into account that the chiral PD is unique, we conclude that $9_{35}$ does not have a chiral bipartite diagram.

There is a subtlety. The chiral PD is unique. However, there exist infinitely many non-chiral PDs coming from the chiral one by addition of polynomials with positive coefficients proportional to $\phi + \bphi + \phi \bphi D$ equal zero on the locus $\phi=A\{q\}$, $\bar\phi=-A^{-1}\{q\}$, $D=\frac{\{A\}}{\{q\}}$. Still, this polynomial identically vanishes under the precursor substitutions $\phi\rightarrow -q,\, D\rightarrow q+q^{-1}$, and thus, does not change the answer for the Jones polynomial of a hypothetical precursor diagram. This means that all PDs coming from a chiral knot reduce to the same hypothetical Jones polynomial, and the precursor check should be applied to the chiral PD only. But the number of bipartite vertices, and thus, single vertices in the precursor diagram can grow, and we need other checks for hypothetical precursor Jones to conclude if a given knot is bipartite or non-bipartite (not necessary \textit{chiral}). In other words, we can prohibit only PDs of fixed degrees (up to 16 for knots and up to 11 for links) in $\phi$, $\bphi$ but not all of them.

\paragraph{\bf Non-chiral case.} The precursor argument does not spoil even in the non-chiral case due to the analogous reasoning. Instead of the existence of a chiral diagram, the precursor criterium now checks whether a PD of a given degree in $\phi$, $\bar\phi$ can correspond to a bipartite diagram, or, in other words, whether a PD corresponds to a bipartite diagram with no more than a given number of vertices. Let us briefly remind the algorithm to obtain \textit{a} positive non-chiral PD. One just makes substitutions from $A\,,q$ variables to $\phi,\, D$ variables~\eqref{chirexpanswer} but ends up with non-positive polynomials. Then, choosing a framing factor and adding an appropriate positive polynomial again divisible by $\phi + \bphi + \phi \bphi D$, one arrives to a positive non-chiral polynomial. These manipulations, however, again does not affect (up to a framing factor) the Jones polynomial of a hypothetical precursor diagram, but the degree in $\phi$, $\bphi$ again can grow. Thus, one can stop at the result of the change of $A\,,q$ variables to $\phi,\, D$ variables in the HOMFLY polynomial~\eqref{chirexpanswer}.  
Then, one makes the Jones reduction in the latter polynomial and checks if the resulting polynomial is the Jones polynomial of some link. Using the program from~\cite{katlas}, one can look through knots with the crossing numbers up to 16 and links with the crossing numbers up to 11. Thus, if the hypothetical precursor Jones polynomial is not found among links of up to some fixed crossing number, then the whole bunch of PDs of up to the same degree in $\phi$, $\bphi$ would be prohibited. 

Few examples are in order. For the figure-eight knot $4_1$, we have:
{\small \begin{equation}
   H_\Box^{4_1} = A^{-2}\left(D^2\phi^2+D\phi -\phi^2 + 1\right) = 1+\phi\bar\phi +  (\phi+\bar\phi)D \quad \overset{\phi\,\rightarrow\, -q,\, \bar\phi\,\rightarrow\, -q^{-1},\,D\rightarrow q+q^{-1}}{\longrightarrow} \quad {\rm Fr}_J^{-1} \cdot J^{\rm Hopf}_\Box = -q^{-2}\left(q^4+1\right)
\end{equation}}

\noindent what reflects the fact that the precursor of $4_1$ is the Hopf link. Now take a look at the HOMFLY PD of the knot $9_{37}\,$:
\begin{equation}
\begin{aligned}
    H_\Box^{9_{37}} = A^{-4}\big(-D^3 \phi ^5+2 D^3 \phi ^3-2 D^2 \phi ^4+4 D^2 \phi ^2+D \phi ^5-2 D \phi ^3&+2 D \phi +2 \phi ^4-3 \phi ^2+1 \big)
    \quad \overset{\phi\,\rightarrow\, -q,\,D\,\rightarrow\, q+q^{-1}}{\longrightarrow} \quad \\
&\stackrel{?}{\longrightarrow} \     {\rm Fr}_J^{-1} \cdot J^{?}_\Box = q^8-2 q^6-2 q^2+1\,.
\end{aligned}
\end{equation}
Among all links with crossing number up to $11$ 
there is no one having such a Jones polynomial. Thus, a PD for the knot $9_{37}$ of the degree up to $11$ in $\phi$, $\bphi$ does not correspond to a bipartite diagram, and the knot $9_{37}$ does not have a bipartite diagram with no more than $11$ bipartite vertices. 

\paragraph{\bf Candidates to fake precursor Jones polynomials.} However, one can notice even more. There is a very restricted number (namely, 5 plus 5 their mirror ones) of candidates to fake precursor Jones polynomials coming from (possibly) non-bipartite knots of the crossing number up to 11. We have thought how to fully forbid them to be the Jones polynomials in order to say that the found by us knots are definitely not bipartite. First, the fully reduced Jones polynomial at $q=1$ must be equal to one. This fact tells us how many components the hypothetical link must have. Second, the differential expansion must hold. This restricts the framing factor of the hypothetical Jones polynomials although it cannot be obtained from the bipartite HOMFLY polynomial. 

As we see, these restrictions cannot forbid our polynomials. We just state the result of imposing these conditions and the computer search:
\begin{itemize}
    \item $q^{-6}(-1 + 2 q^2 - q^4 + 2 q^6 - 2 q^8 + q^{10})$ could correspond to a knot but has not been found among the Jones polynomials for knots with crossing number up to and including 16;
    \item $-q^{6k-1}(1 - 2 q^2 - 2 q^6 + q^8)$, $k\in \mathbb{Z}$, could correspond to a 2-component link but has not been found among the Jones polynomials for 2-component links with crossing number up to and including 11;
    \item $-q^6(-2 + q^6)$ could correspond to a knot but has not been found among the Jones polynomials for knots with crossing number up to and including 16;
    \item $q^{6k+3}(1 - q^2 + q^4 - q^6 + 2 q^8)$, $k\in \mathbb{Z}$, could correspond to a 2-component link but has not been found among the Jones polynomials for 2-component links with crossing number up to and including 11;
    \item $q^{-2}(1 - q^6 + q^8)$ could correspond to a knot but has not been found among the Jones polynomials for knots with crossing number up to and including 16.
\end{itemize}


\noindent One may suppose that these particular polynomials cannot be the Jones polynomials of any links by some yet unknown reason. Then, the precursor argument in fact checks whether a given knot can have a bipartite realization at all.

\paragraph{Check for known non-bipartite knots.} State further results of the precursor check. The non-bipartite chiral knot $9_{49}$ withstands this test, its hypothetical precursor is the Hopf link. The corresponding PD~\eqref{949PD} could be a PD of some chiral bipartite clone\footnote{Recall that we call links clones if they have the same HOMFLY polynomial.} of $9_{49}$ (perhaps, for already found $16n_{350659}$ with the yet unknown biparticity), if such a knot exists. All other known chiral non-bipartite knots $9_{35}$, $11a_{123}$, $11a_{291}$, $11a_{366}$, $11n_{126}$ do not have precursors of 16 crossings in the knot cases and 11 crossings in the link cases. In other words, they seem to have fake precursor Jones polynomials.

Among all known non-chiral non-bipartite knots $9_{41}$, $10_{103}$ (clones $10_{40}$ -- bipartite), $10_{155}$, $10_{157}$, $11a_{196}$, $11a_{317}$, $11a_{321}$, $11n_{133}$ are non-chiral that are not forbidden via the precursor check. Moreover, some of them have clones which might be bipartite, see Section~\ref{nce}, and the knot $10_{103}$ has definitely bipartite clone $10_{40}$. Forbidden ones, up to 16 bipartite crossings in the knot precursor case and up to 11 bipartite crossings in the link precursor case, are $9_{37}$, $9_{46}$, $9_{47}$, $9_{48}$, $10_{74}$, $10_{75}$, $11a_{135}$, $11a_{155}$, $11a_{173}$, $11a_{181}$, $11a_{249}$, $11a_{277}$, $11a_{293}$, $11a_{314}$, $11n_{167}$.

\paragraph{\bf Candidates to non-bipartite knots.} We have searched through all knots of crossing numbers up to 11 and, by our precursor check, we have found the following chiral candidates to non-bipartite knots: $10_{120}$, $11a_{292}$, $11a_{362}$. Non-chiral candidates are 
$10_{96}$, $10_{97}$, $11a_{103}$, $11a_{128}$, $11a_{201}$, $11a_{209}$, $11a_{214}$, $11a_{219}$, $11a_{278}$, $11a_{280}$, $11a_{294}$, $11a_{296}$, $11a_{324}$, $11n_{100}$, $11n_{101}$, $11n_{102}$, $11n_{123}$, $11n_{139}$, $11n_{140}$, $11n_{141}$, $11n_{142}$, $11n_{150}$, $11n_{155}$, $11n_{161}$, $11n_{168}$, $11n_{170}$. All these knots have hypothetical precursor Jones polynomials of the above 5 types (or their mirrors). Thus, if one proves that these polynomials cannot be the Jones polynomials of some links, these candidates will become definitely non-bipartite knots.

\paragraph{\bf Precursors vs Alexander ideals.} The above examples show that the precursor check, if could be treated as a biparticy criterium, would not be equivalent to the argument with the Alexander ideals~\cite{bip}. 
And currently, it is unclear whether the precursor check could be stronger or weaker than the Alexander one. 
Both types of  examples exist -- 
of non-bipartite knots, not distinguished by the precursor Jones polynomial but distinguished by Alexander ideals,
and candidates for non-bipartite knots not distinguished by the Alexander ideals.

\setcounter{equation}{0}
\section{Bipartite calculus for non-bipartite diagrams
\label{sec:sum}}

A non-bipartite knot diagram can still have some antiparallel locks.
Then, they can be decomposed by the rule in Fig.\,\ref{fig:pladeco},
and the original diagram becomes substituted by a sum of simpler ones
with the coefficients made from $\phi,\bphi$ and $D$.
Moreover, these simpler diagrams can be already equivalent to bipartite ones,
and this provides a new way to obtain a PD of the original one.
Thus, {\bf bipartite calculus can be applied even to non-bipartite diagrams}!
But note that the diagram identities we write in this section in fact hold for the HOMFLY polynomials, not for the diagrams themselves.

\begin{figure}[h!]
\centering
\includegraphics[width=10.5cm]{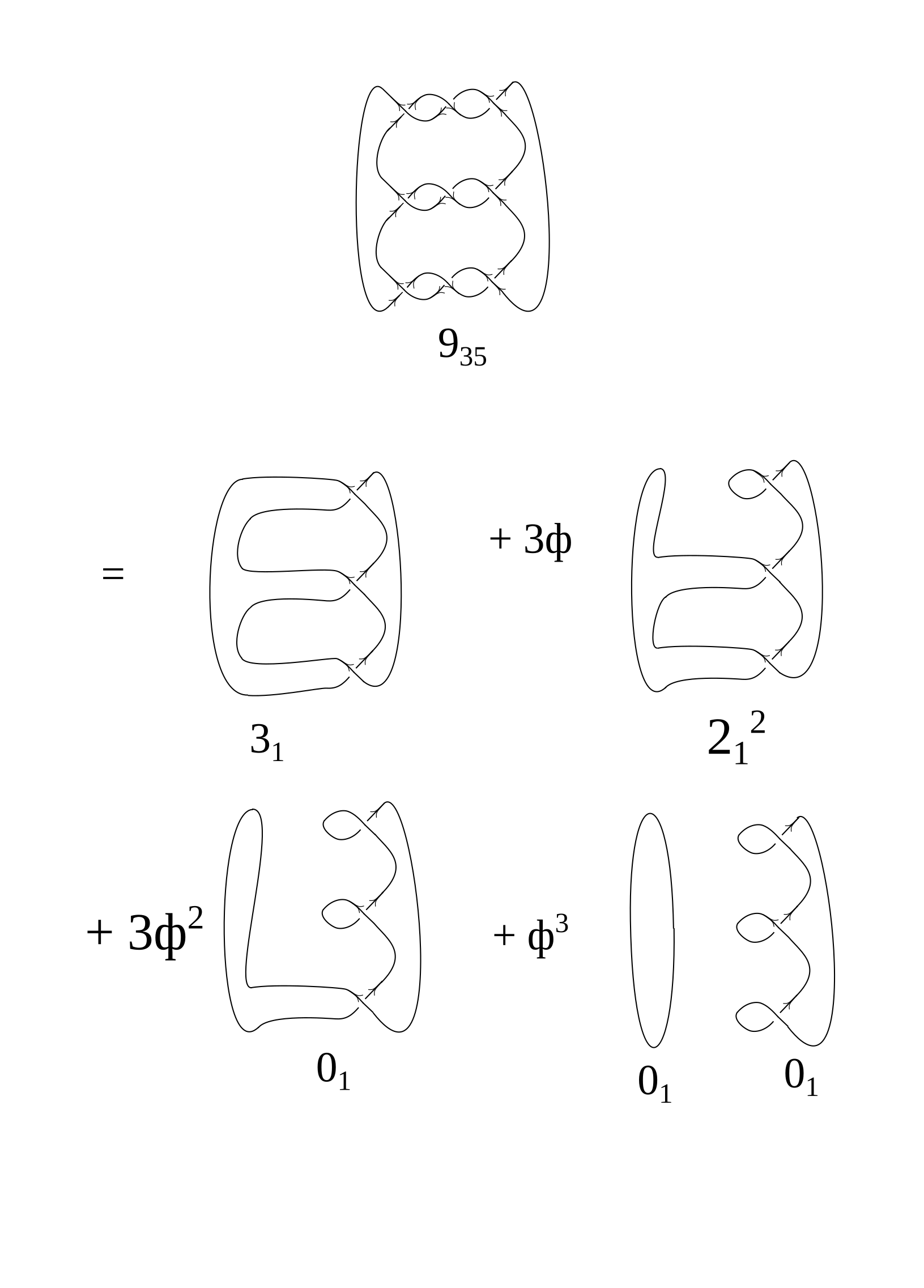}
\caption{\footnotesize
The three antiparallel locks in the non-bipartite diagram for $9_{35}$
can be decomposed with the help of Fig.\,\ref{fig:pladeco},
what substitutes a single diagram by a sum of eight ones.
Each of these smaller diagrams is Reidemeister-equivalent to a bipartite one,
and this provides a PD = fake BE for $H_\Box^{9_{35}}$.
Since all the diagrams are chiral, this explains why $H_\Box^{9_{35}}$ has
a chiral PD, despite being a non-bipartite knot -- what we already know from Section \ref{sec:allchir}.
The diagrams represent the polynomials in the \textbf{topological} framing. $2_1^2$ is the denotation from~\cite{linkinfo} for the Hopf link, the knot $0_1$ is the unknot.
}\label{fig:P333}
\end{figure}

A typical example is shown in Fig.\,\ref{fig:P333} -- and it explains why
the non-bipartite $9_{35}$ possesses a chiral PD.

The same reasoning can be applied to $9_{41}$, $9_{46}$, $9_{49}$ and, perhaps, to other non-bipartite knots.
The argument is not restricted to chiral PDs and can work in the non-chiral case as well.
Still, the generality and true power of it remain open questions.

\subsection{Framing in this section}

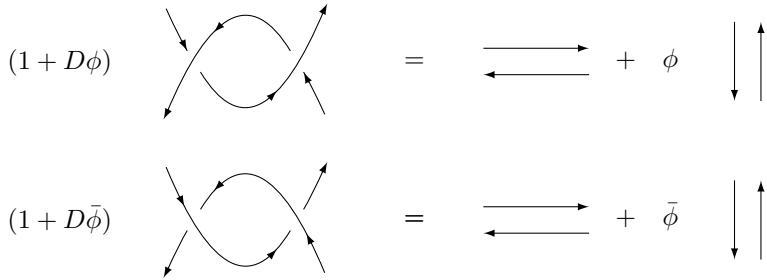
\begin{figure}[h!]
	\begin{picture}(60,120)(-80,-85)
			
		\put(-10,-3){$(1+D\phi)$}	
		
			\qbezier(50,20)(55,9)(58,4) \qbezier(63,-4)(85,-40)(110,20)
			\put(56,8){\vector(1,-2){2}} \put(90,-13){\vector(1,1){2}} \put(109,18){\vector(1,2){2}}
			\qbezier(50,-20)(75,40)(97,4)  \qbezier(102,-4)(105,-9)(110,-20)
			\put(104,-8){\vector(-1,2){2}} \put(70,13){\vector(-1,-1){2}} \put(51,-18){\vector(-1,-2){2}}

		\put(0,0){
			\put(140,-2){\mbox{$=$}}
			
			\put(100,65){
				\put(70,-60){\vector(1,0){40}}
				\put(110,-70){\vector(-1,0){40}}
				\put(120,-67){\mbox{$+\ \ \ \phi$}}
				\put(165,-50){\vector(0,-1){30}}
				\put(175,-80){\vector(0,1){30}}
			}
		}
		
	\put(0,-60){

\put(0,0){

\put(-10,-3){$(1+D\bar\phi)$}	

\put(0,0){
\qbezier(50,20)(75,-40)(97,-4)  \qbezier(102,4)(105,9)(110,20)
\qbezier(50,-20)(55,-9)(58,-4) \qbezier(63,4)(85,40)(110,-20)
\put(55,9){\vector(1,-2){2}} \put(90,-13){\vector(1,1){2}} \put(109,18){\vector(1,2){2}}
\put(105,-9){\vector(-1,2){2}} \put(70,13){\vector(-1,-1){2}} \put(51,-18){\vector(-1,-2){2}}
}

\put(140,-2){\mbox{$=$}}

\put(0,0){
			\put(140,-2){\mbox{$=$}}
			
			\put(100,65){
				\put(70,-60){\vector(1,0){40}}
				\put(110,-70){\vector(-1,0){40}}
				\put(120,-67){\mbox{$+\ \ \ \bar\phi$}}
				\put(165,-50){\vector(0,-1){30}}
				\put(175,-80){\vector(0,1){30}}
			}
		}
}

}	
		
	\end{picture}
	\caption{\footnotesize
		The planar decomposition of the lock vertices in the \textit{topological} framing.
No extra factor is needed to restore the topological invariance.
If we want a polynomial expression, we rather multiply the r.h.s. by $(1+D\bphi)$, using the fact that $(1+D\phi)(1+D\bphi)=1$
-- but then we loose chirality.
In this section, we prefer to preserve it for the beauty of the formulas.
	}\label{fig:toppladeco}
\end{figure}

In this section, we deal with many diagrams that are equivalent up to the first Reidemeister move (contracting a loop). 
This is one of the reasons why we choose to use the topological framing 
(instead of the vertical one, which 
was exploited in \cite{bipfund}).
We repeat here  Fig.\,\ref{fig:pladeco} in a  slightly different form of Fig.\,\ref{fig:toppladeco},
which is used in this section,
with $A^{-2}$ at the r.h.s. substituted by $1+\phi D$ at the l.h.s.
This allows to write everything in terms of the variables $D,\phi$ (and $\bphi$ for the opposite orientation), 
without the additional (and unnecessary) variable $A$ in the farming factors.

\subsection{Positive decompositions for chiral non-bipartite knots}\label{sec:PD-ch-non-bip}

If a non-bipartite diagram includes a bipartite vertex, the two resolutions of this vertex \textit{may} give diagrams whose bipartite forms are already known. Then we say that an original diagram is expanded over corresponding bipartite diagrams.

If all diagrams are chiral, then the PD of related knots are uniquely defined by their HOMFLY polynomial via (\ref{chirexpanswer}). The expansion of bipartite vertices (if any) in a diagram implies then a relation on the corresponding PD.

\paragraph{Pretzel knot $9_{35}\sim P(-3,-3,-3)$.}
The first non-bipartite knot \cite{katlas} is the knot $9_{35}$, which is also the pretzel knot $P(-3,-3,-3)$. According to \cite{bip}, a pretzel knot $P(p,q,r)$ with odd $p,q,r$ and their greatest common divisor greater than 1 cannot be bipartite. Yet, the odd $p,q,r$ mean that all ``handles'' of the pretzel knot are antiparallel and hence contain bipartite vertices.

Fig.\,\ref{fig:P333} demonstrates resolutions of three bipartite vertices in $P(-3,-3,-3)$ according to Fig.\,\ref{fig:toppladeco} and represents the following identity for the HOMFLY polynomials
\begin{eqnarray}\label{H935}
H_\Box^{9_{35}}(1+D\phi)^3=H_\Box^{3_1}+3\phi H_\Box^{\rm Hopf}+3\phi^2+D\phi^3\,,
\end{eqnarray}
where each of three factors of $(1+D\phi)$ in the l.h.s. stands for resolving one bipartite vertex.
Alternatively, one can resolve only one bipartite vertex and already get the expansion into the bipartite knot and link,
\begin{eqnarray}
H_\Box^{9_{35}}(1+D\phi)=H_\Box^{7_4}+\phi H_\Box^{6_3^2\{1\}},\qquad 6_3^2\{1\}\sim T[2,6]\text{ in antiparallel orientation}\,.
\end{eqnarray}


\noindent In this section, we use specific denotations for links -- ${\rm crossing \ number}^{\,\rm number \ of \ components}_{\,\rm link \ number}\{{\rm orientation \ of \ components}\}$, being in accordance with~\cite{linkinfo}.  

Identity (\ref{H935}) may look more transparent in the form
\begin{equation}\label{H-9-35}
    H_\Box^{9_{35}}=(1+D\bar\phi)^3 \left(H_\Box^{3_1}+3\phi H_\Box^{\rm Hopf}+3\phi^2+D\phi^3\right)
\end{equation}
Let us demonstrate that (\ref{H-9-35}) indeed holds. We know BEs for the Hopf link~\eqref{Hopf} and for the trefoil~\eqref{trefoil}:
\begin{equation}
\begin{aligned}
    H_\Box^{\rm Hopf}&=(1+D\bar \phi)(D+\phi)\,, \\
    H_\Box^{3_1}&=(1+D\bar \phi)^2(1+2\phi D+\phi^2)\,.
\end{aligned}
\end{equation}
Substitution of these HOMFLY polynomials into~\eqref{H-9-35} leads to
{\small \begin{equation}
\begin{aligned}
    H_\Box^{9_{35}}&=(1+D\bar\phi)^3\left((1+D\bar \phi)^2(1+2\phi D+\phi^2) + 3\phi\underbrace{(1+D\phi)(1+D\bar \phi)}_{=1}(1+D\bar \phi)(D+\phi) + \underbrace{(1+D\phi)^2(1+D\bar \phi)^2}_{=1}(3\phi^2+D\phi^3)\right)=\\
    &=(1+D\bar\phi)^5\Big(\phi^5D^3+5\phi^4D^2 +10\phi^3D + \phi^2(3D^2+7) +5\phi D  +1\Big)
\end{aligned}
\end{equation}}

\noindent where we also added unities represented by multiples of $(1+D\phi)(1+D\bar \phi)=1$ in order to recover the chiral BE~\eqref{935PD}.

\begin{figure}[h!]
\centering
\includegraphics[width=4cm]{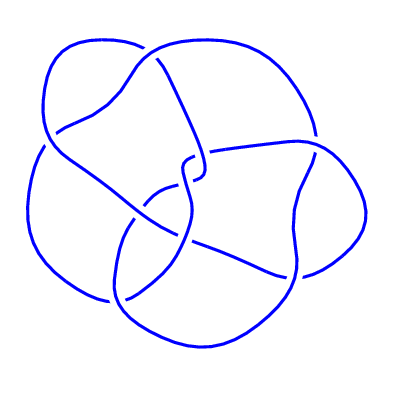} \ \ \ \ \ \ \ \ \ \ \ \ \ \ \
\includegraphics[width=4cm]{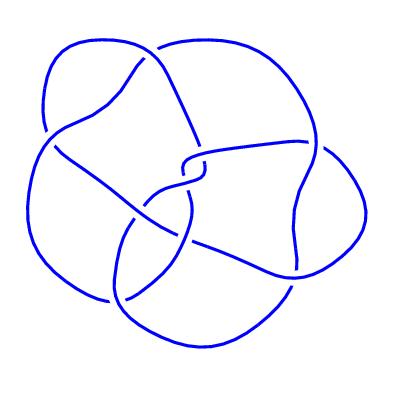}
\caption{\footnotesize
Diagrams of the knots $9_{49}$ (the left one) and $9_{41}$ (the right one) from \cite{knotinfo}.
To find the bipartite vertices in these diagrams, one selects an orientation.
\label{fig:P949}}
\end{figure}

\paragraph{Non-pretzel knot $9_{49}$.}
The other definitely non-bipartite \cite{bip} chiral knot up to 9 crossings is $9_{49}$. According to \cite{knotinfo}, it is not a pretzel knot, yet its standard digram
has three bipartite vertices (one can see it by selecting an orientation on the knot diagram in Fig.\,\ref{fig:P949}). Resolving \textit{any} one of these vertices in the standard diagram of the knot $9_{49}$, we get
\begin{eqnarray}
H_\Box^{9_{49}}(1+D\phi)&=&H_\Box^{5_1}+\phi H_\Box^{6_2^2\{1\}}\,,\nn\\
H_\Box^{9_{49}}(1+D\phi)^2&=&H_\Box^{5_1}(1+D\phi)+\phi H_\Box^{4_1^2\{1\}}+\phi^2 H_\Box^{3_1}\,,
\end{eqnarray}
where the last line we obtain by expansion of the bipartite vertex in $6_2^2\{1\}$. The substitution of the corresponding BEs to the l.h.s. of these equations leads to the PD for the knot $9_{49}$~\eqref{949PD}.

\subsection{Positive decompositions for non-chiral non-bipartite knots}

As we have seen in Section \ref{nce}, (\ref{chirexpanswer}) gives a PD only for some knots we call chiral. In the opposite case, we can either relate a knot with a non-positive decomposition in $(D,\phi)$ given by (\ref{chirexpanswer}), or compute the BE in $(D,\phi,\bphi)$ straightforwardly from a bipartite knot diagram (if one is known) as explained in \cite{bipfund}. Only the first option can be applied to a non-bipartite knot. Then, if a non-bipartite diagram is expanded over bipartite ones (in the sense of the above section), one can write the corresponding relation for the non-positive decomposition. Note that if we expand over a negative bipartite vertex, we should write $-\phi/(1+D\phi)$ instead of $\bphi$ and $1/(1+D\phi)$ instead of $(1+\bar{\phi} D)$ in such relations.

On the other hand, one can repeat the trick from the previous section and use an expansion of a non-chiral non-bipartite diagram over bipartite ones to \textit{define} a PD in $(D,\phi,\bphi)$ for a non-chiral non-bipartite diagram. Unlike general non-chiral PDs discussed in Section\,\ref{ambig}, such an expression has a sense as a combination of BEs. In particular, it satisfies the $D=1$ criterium from Section \ref{sec:D1}, which (as we saw) is very hard to satisfy by guessing the answer.

First, we illustrate the method by comparing similar expansions over the bipartite chiral diagrams for chiral and non-chiral bipartite knots. 

\paragraph{The simplest examples.}
E.g., the standard diagram of knot $4_1$ differs from the twist diagram of knot $3_1$ by inverting one bipartite vertex. The expansions over this vertex for these two diagrams are
\begin{equation}
\begin{aligned}
H_\Box^{3_1}(1+D\phi)&=&1+\phi H_\Box^{\rm Hopf},\\
H_\Box^{4_1}(1+D\bar\phi)&=&1+\bar\phi H_\Box^{\rm Hopf}.
\end{aligned}
\end{equation}
Hence, if one knows the first expansion and the described relation of the diagrams, one immediately gets the second expansion.
The explicit BE for $H_\Box^{\rm Hopf}$ can be easily calculated~\eqref{Hopf}, and the second expansion gives the explicit expression for $H_\Box^{4_1}$~\eqref{PD-4_1}.
The two following cases are similar,
\begin{equation}
\begin{aligned}
H_\Box^{5_1}(1+D\phi)&=&H_\Box^{3_1}+\phi H_\Box^{4_1^2\{1\}},\\
H_\Box^{6_2}(1+D\bar\phi)&=&H_\Box^{3_1}+\bphi H_\Box^{4_1^2\{1\}},
\end{aligned}
\end{equation}
and
\begin{equation}
\begin{aligned}
H_\Box^{5_2}(1+D\phi)&=&1+\phi H_\Box^{4_1^2\{0\}},\\
H_\Box^{6_1}(1+D\bar\phi)&=&1+\bphi H_\Box^{4_1^2\{0\}}.
\end{aligned}
\end{equation}

Now, we proceed with non-chiral non-bipartite knots. Below, we present two expressions in each case. The first expression is the identity for the non-positive decomposition defined by (\ref{chirexpanswer}), which in a sense explains a presence of a PD for the non-bipartite knot. The second expression uses the first one to relate the non-chiral non-bipartite knot with a PD.

\paragraph{Pretzel knot $P[-3,-3,3]\sim \bar{9}_{46}$.}
This case is similar to that one of $9_{35}\sim P(-3,-3,-3)$.
Resolving a bipartite vertex, we get
\begin{equation}
\begin{aligned}
H_\Box^{\bar{9}_{46}}/(1+D\phi)&=H_\Box^{3_1}-\phi/(1+D\phi)H_\Box^{6_3^2\{1\}}\,,\\
H_\Box^{\bar{9}_{46}}(1+D\bphi)&=H_\Box^{3_1}+\bphi H_\Box^{6_3^2\{1\}}.
\end{aligned}
\end{equation}
Instead, one can resolve all three bipartite vertices and get
\begin{equation}
\begin{aligned}
H_\Box^{\bar{9}_{46}}(1+D\phi)&=1+2D\phi-\phi/(1+D\phi)H_\Box^{\rm Hopf}+\phi^2-2\phi^2/(1+D\phi)-D\phi^3/(1+D\phi)\,,\\
H_\Box^{\bar{9}_{46}}(1+D\phi)&=1+2D\phi+\bphi H_\Box^{\rm Hopf}+\phi^2+2\phi\bphi+D\phi^2\bphi\,.
\end{aligned}
\end{equation}

\paragraph{Non-pretzel knot $9_{41}$.}
As an example of definitely non-bipartite, non-chiral and non-pretzel knot of up to 9 crossings we take the knot $9_{41}$.
\begin{equation}
\begin{aligned}
H_\Box^{9_{41}}/(1+D\phi)&=H_\Box^{7_7}-\phi/(1+D\phi)H_\Box^{7_6^2\{1\}}\,, \\
H_\Box^{9_{41}}(1+D\bphi)&=H_\Box^{7_7}+\bphi H_\Box^{7_6^2\{1\}}\,.
\end{aligned}
\end{equation}
In turn, we can determine $H_\Box^{7_7}$ and $H_\Box^{7_6^2\{1\}}$ with the help of further planar decomposition.
Instead that, we can notice that the pair of standard diagrams of the chiral knot $9_{49}$ and the non-chiral knot $9_{41}$ differs by the inversion of the three pairs of the bipartite vertices (Fig.\,\ref{fig:P949}). Using the expansion over these vertices for the chiral knot, we get the similar expansion for its non-chiral counterpart:
\begin{equation}
\begin{aligned}
H_\Box^{9_{49}}&=&1+3\phi H_\Box^{\rm Hopf}+3\phi^2H_\Box^{3_1}+\phi^3DH_\Box^{3_1},\\
H_\Box^{9_{41}}&=&1+3\bphi H_\Box^{\rm Hopf}+3\bphi^2H_\Box^{3_1}+\bphi^3DH_\Box^{3_1},
\end{aligned}
\end{equation}
which can be used to compute explicit form of $H_\Box^{9_{41}}$ as a function of $D$, $\phi$, $\bphi$.

\setcounter{equation}{0}
\section{Symmetric representation
\label{reps}}

In \cite{bipsym}, we extended bipartite decompositions from fundamental to arbitrary symmetric representations $[r]$.
It is natural to ask if this consideration can be extended to the {\it planar decompositions}, discussed in the present paper.
This is not so simple, because instead of a simple formula (\ref{PD})
we now have
\be
H_{[r]}^{\cal K} = {\rm Fr}_{[r]}\cdot \sum_{\vec c,\vec b} K_{\vec b,\vec c} \prod_{i,j=1}^r \psi_i^{b_i}\bar\psi_j^{c_j}
\ee
with many parameters $\psi_i$ and $\bar\psi_j$ instead of just a couple $\phi$, $\bphi$.
Coefficients no longer depend on $z$, they can be polynomials of independent variables $q$ and $q^{-1}$.
Most important, the coefficients $K_{\vec b,\vec c}(A,q)$ are now not just the powers of a single dimension, but include
many structures made from various combinations of projectors.
All this makes the hope for a formula like (\ref{chirexpanswer}), even in the chiral case,
somewhat elusive.
We do not go into detailed discussion in this paper, and provide just an example of the first symmetric representation $[2]$
for some representatives of a chiral ray of twist knots
$3_1,5_2,7_2,\ldots$ (the opposite ray $4_1,6_1,8_1,\ldots$ is non-chiral) and the torus knot $5_1$.
It illustrates the problem and can also provide insights into its possible resolution in the future.
At this moment {\bf the issue of PD for non-fundamental representations is obscure}.

\subsection{Basics of planar decomposition in representation $[2]$}\label{sec:planar-dec-[2]}

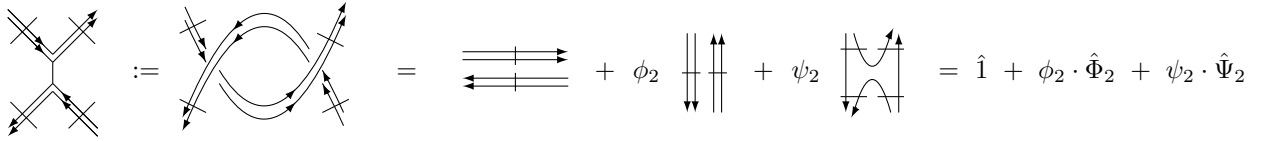
\begin{figure}[h!]
\begin{picture}(100,80)(70,-40)

\put(100,0){

\put(-17,21){\line(1,-1){17}}\put(-17,24){\line(1,-1){17}}   \put(0,4){\vector(1,1){17}}
\put(-17,21){\vector(1,-1){14}} \put(-17,24){\vector(1,-1){14}}
\put(0,7){\vector(1,1){17}}
\put(0,-4){\vector(-1,-1){17}}   \put(17,-21){\line(-1,1){17}} \put(17,-21){\vector(-1,1){14}}
\put(0,-7){\vector(-1,-1){17}}   \put(17,-24){\line(-1,1){17}} \put(17,-24){\vector(-1,1){14}}
\put(0,4){\line(0,-1){8}}

\put(-16,11){\line(1,1){10}}  \put(-16,-11){\line(1,-1){10}}
\put(16,11){\line(-1,1){10}}  \put(16,-11){\line(-1,-1){10}}

\put(30,-2){\mbox{$:=$}}

\qbezier(50,20)(55,9)(58,4) \qbezier(63,-4)(85,-40)(110,20)
\put(56,8){\vector(1,-2){2}} \put(90,-13){\vector(1,1){2}} \put(109,18){\vector(1,2){2}}
\qbezier(50,-20)(75,40)(97,4)  \qbezier(102,-4)(105,-9)(110,-20)
\put(104,-8){\vector(-1,2){2}} \put(70,13){\vector(-1,-1){2}} \put(51,-18){\vector(-1,-2){2}}

\qbezier(50,25)(55,14)(58,9) \qbezier(63,1)(85,-35)(110,25)
\put(56,13){\vector(1,-2){2}} \put(90,-8){\vector(1,1){2}} \put(109,23){\vector(1,2){2}}
\qbezier(50,-15)(75,45)(97,9)  \qbezier(102,1)(105,-4)(110,-15)
\put(104,-3){\vector(-1,2){2}} \put(70,18){\vector(-1,-1){2}} \put(51,-13){\vector(-1,-2){2}}

\put(48,15){\line(2,1){10}}  \put(48,-10){\line(2,-1){10}}
\put(110,10){\line(-2,1){10}}  \put(112,-10){\line(-2,-1){10}}

\put(130,-2){\mbox{$=$}}

\put(85,65){
\put(70,-60){\vector(1,0){40}} 
\put(70,-57){\vector(1,0){40}}
\put(110,-70){\vector(-1,0){40}}  \put(110,-67){\vector(-1,0){40}}
\put(90,-55){\line(0,-1){7}}   \put(90,-65){\line(0,-1){7}}
\put(120,-67){\mbox{$+ \ \ \phi_2$}}
\put(155,-50){\vector(0,-1){30}} \put(158,-50){\vector(0,-1){30}}
\put(165,-80){\vector(0,1){30}}  \put(168,-80){\vector(0,1){30}}
\put(153,-65){\line(1,0){7}}   
\put(163,-65){\line(1,0){7}}

\put(180,-67){\mbox{$+ \ \ \psi_2$}}

\put(-15,0){
\put(230,-50){\vector(0,-1){30}}
\put(250,-80){\vector(0,1){30}}
\qbezier(233,-50)(240,-75)(247,-50)  \put(246,-52){\vector(1,2){2}}
\qbezier(233,-80)(240,-55)(247,-80)  \put(234,-78){\vector(-1,-2){2}}
\put(228,-56){\line(1,0){10}}   \put(242,-56){\line(1,0){10}}
\put(228,-74){\line(1,0){10}}   \put(242,-74){\line(1,0){10}}
}

\put(250,-67){\mbox{$ = \ \hat{1} \ + \ \phi_2 \cdot \hat{\Phi}_2 \ + \ \psi_2 \cdot \hat{\Psi}_2 $}}
}}

\end{picture}
\caption{\footnotesize  The lock tangle projected to the representation $[2]$
and its planar decomposition. The coefficients are given by~\eqref{planar-coefs-[2]}.
} \label{fig:proje2lock}
\end{figure}

Let us briefly formulate the basics of planar decomposition for the representation $[2]$. Instead of the lock tangle in Fig.\,\ref{fig:pladeco}, one considers the 2-cabled~\cite{MMM3} lock tangle in Fig.\,\ref{fig:proje2lock} (and its mirror) and inserts projectors on the representation $[2]$ at each open end of the tangle. When gluing together a knot from 2-cabled locks and expanding them to obtain the bipartite expansion of the HOMFLY polynomial in the representation $[2]$, one gets intricate contractions of projectors coming from the $\hat{\Psi}_2$ diagram. 

\begin{figure}[h!]
\begin{picture}(100,205)(-50,-20)

\put(-30,83){\mbox{$\hat \Pi^n  \ :=$}}
\put(20,0){\vector(0,1){170}}
\put(50,170){\vector(0,-1){170}}

\put(25,10){\vector(0,1){10}}
\put(45,20){\vector(0,-1){10}}
\qbezier(25,20)(25,30)(35,30)
\qbezier(45,20)(45,30)(35,30)
\qbezier(25,10)(25,0)(35,0)
\qbezier(45,10)(45,0)(35,0)
\put(15,13){\line(1,0){15}} \put(40,18){\line(1,0){15}}

\put(0,40){
\put(25,10){\vector(0,1){10}}
\put(45,20){\vector(0,-1){10}}
\qbezier(25,20)(25,30)(35,30)
\qbezier(45,20)(45,30)(35,30)
\qbezier(25,10)(25,0)(35,0)
\qbezier(45,10)(45,0)(35,0)
\put(15,13){\line(1,0){15}} \put(40,18){\line(1,0){15}}
}

\put(0,80){
\put(25,10){\vector(0,1){10}}
\put(45,20){\vector(0,-1){10}}
\qbezier(25,20)(25,30)(35,30)
\qbezier(45,20)(45,30)(35,30)
\qbezier(25,10)(25,0)(35,0)
\qbezier(45,10)(45,0)(35,0)
\put(15,13){\line(1,0){15}} \put(40,18){\line(1,0){15}}
}

\put(30,120){\mbox{$\ldots$}}

\put(0,140){
\put(25,10){\vector(0,1){10}}
\put(45,20){\vector(0,-1){10}}
\qbezier(25,20)(25,30)(35,30)
\qbezier(45,20)(45,30)(35,30)
\qbezier(25,10)(25,0)(35,0)
\qbezier(45,10)(45,0)(35,0)
\put(15,13){\line(1,0){15}} \put(40,18){\line(1,0){15}}
}

\put(12,-2){\mbox{$i$}} \put(52,-2){\mbox{$\bar j$}}\put(12,172){\mbox{$i'$}}\put(52,172){\mbox{$\bar j'$}}

\put(15,13){\line(1,0){15}} \put(40,18){\line(1,0){15}}

\put(20,30){\vector(0,1){2}}  \put(50,140){\vector(0,-1){2}}


\put(180,0){
\put(-90,83){\mbox{$\Pi_n:=(\tr\otimes \tr)\, \hat \Pi^n   \ :=$}}
\put(20,15){\vector(0,1){140}}
\put(50,150){\vector(0,-1){140}}

\qbezier(20,150)(20,175)(35,175)
\qbezier(50,150)(50,175)(35,175)
\qbezier(20,15)(20,-5)(35,-5)
\qbezier(50,15)(50,-5)(35,-5)

\put(25,10){\vector(0,1){10}}
\put(45,20){\vector(0,-1){10}}
\qbezier(25,20)(25,30)(35,30)
\qbezier(45,20)(45,30)(35,30)
\qbezier(25,10)(25,0)(35,0)
\qbezier(45,10)(45,0)(35,0)
\put(15,13){\line(1,0){15}} \put(40,18){\line(1,0){15}}

\put(0,40){
\put(25,10){\vector(0,1){10}}
\put(45,20){\vector(0,-1){10}}
\qbezier(25,20)(25,30)(35,30)
\qbezier(45,20)(45,30)(35,30)
\qbezier(25,10)(25,0)(35,0)
\qbezier(45,10)(45,0)(35,0)
\put(15,13){\line(1,0){15}} \put(40,18){\line(1,0){15}}
}

\put(0,80){
\put(25,10){\vector(0,1){10}}
\put(45,20){\vector(0,-1){10}}
\qbezier(25,20)(25,30)(35,30)
\qbezier(45,20)(45,30)(35,30)
\qbezier(25,10)(25,0)(35,0)
\qbezier(45,10)(45,0)(35,0)
\put(15,13){\line(1,0){15}} \put(40,18){\line(1,0){15}}
}

\put(30,120){\mbox{$\ldots$}}

\put(0,140){
\put(25,10){\vector(0,1){10}}
\put(45,20){\vector(0,-1){10}}
\qbezier(25,20)(25,30)(35,30)
\qbezier(45,20)(45,30)(35,30)
\qbezier(25,10)(25,0)(35,0)
\qbezier(45,10)(45,0)(35,0)
\put(15,8){\line(1,0){15}} \put(40,18){\line(1,0){15}}
}



}


\put(330,0){

\put(-60,83){\mbox{$\Tr_q  \hat\Pi^n   \ :=$}}
\put(20,0){\vector(0,1){170}}
\put(50,170){\vector(0,-1){170}}

\qbezier(20,170)(20,185)(5,185)
\qbezier(50,170)(50,185)(65,185)
\qbezier(-10,170)(-10,185)(5,185)
\qbezier(80,170)(80,185)(65,185)
\qbezier(20,0)(20,-20)(5,-20)
\qbezier(50,0)(50,-20)(65,-20)
\qbezier(-10,0)(-10,-20)(5,-20)
\qbezier(80,0)(80,-20)(65,-20)

\put(80,0){\vector(0,1){170}}
\put(-10,170){\vector(0,-1){170}}

\put(25,10){\vector(0,1){10}}
\put(45,20){\vector(0,-1){10}}
\qbezier(25,20)(25,30)(35,30)
\qbezier(45,20)(45,30)(35,30)
\qbezier(25,10)(25,0)(35,0)
\qbezier(45,10)(45,0)(35,0)
\put(15,13){\line(1,0){15}} \put(40,18){\line(1,0){15}}

\put(0,40){
\put(25,10){\vector(0,1){10}}
\put(45,20){\vector(0,-1){10}}
\qbezier(25,20)(25,30)(35,30)
\qbezier(45,20)(45,30)(35,30)
\qbezier(25,10)(25,0)(35,0)
\qbezier(45,10)(45,0)(35,0)
\put(15,13){\line(1,0){15}} \put(40,18){\line(1,0){15}}
}

\put(0,80){
\put(25,10){\vector(0,1){10}}
\put(45,20){\vector(0,-1){10}}
\qbezier(25,20)(25,30)(35,30)
\qbezier(45,20)(45,30)(35,30)
\qbezier(25,10)(25,0)(35,0)
\qbezier(45,10)(45,0)(35,0)
\put(15,13){\line(1,0){15}} \put(40,18){\line(1,0){15}}
}

\put(30,120){\mbox{$\ldots$}}

\put(0,140){
\put(25,10){\vector(0,1){10}}
\put(45,20){\vector(0,-1){10}}
\qbezier(25,20)(25,30)(35,30)
\qbezier(45,20)(45,30)(35,30)
\qbezier(25,10)(25,0)(35,0)
\qbezier(45,10)(45,0)(35,0)
\put(15,13){\line(1,0){15}} \put(40,18){\line(1,0){15}}
}


\put(15,13){\line(1,0){15}} \put(40,18){\line(1,0){15}}

\put(20,3){\vector(0,1){2}}  \put(50,140){\vector(0,-1){2}}
\put(80,3){\vector(0,1){2}}  \put(-10,140){\vector(0,-1){2}}
}

\end{picture}
\caption{\footnotesize  Vertical chain of projectors.
Shown are its closures, which are relevant for twist knots.
$n$ is the number of $\hat\Pi$, which is the same as the number of small cycles in the picture.
The lock vertices would stand in between the cycles, thus, there would be $(n-1)$ of them in $\Pi_n$
and $(n+1)$ ones in $\Tr_q \hat{\Pi}^n$. These contractions of projectors contribute to the $[2]$-colored HOMFLY polynomial as~\eqref{pi-coef}.
} \label{fig:vertchain2}
\end{figure}
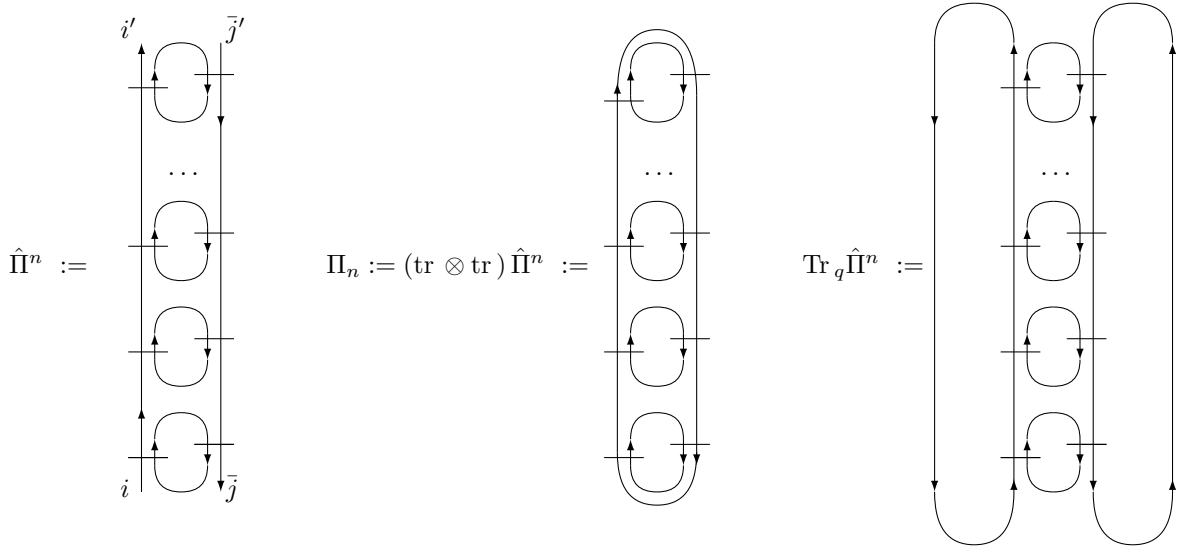

\begin{figure}[h!]
\centering
\includegraphics[width=6cm]{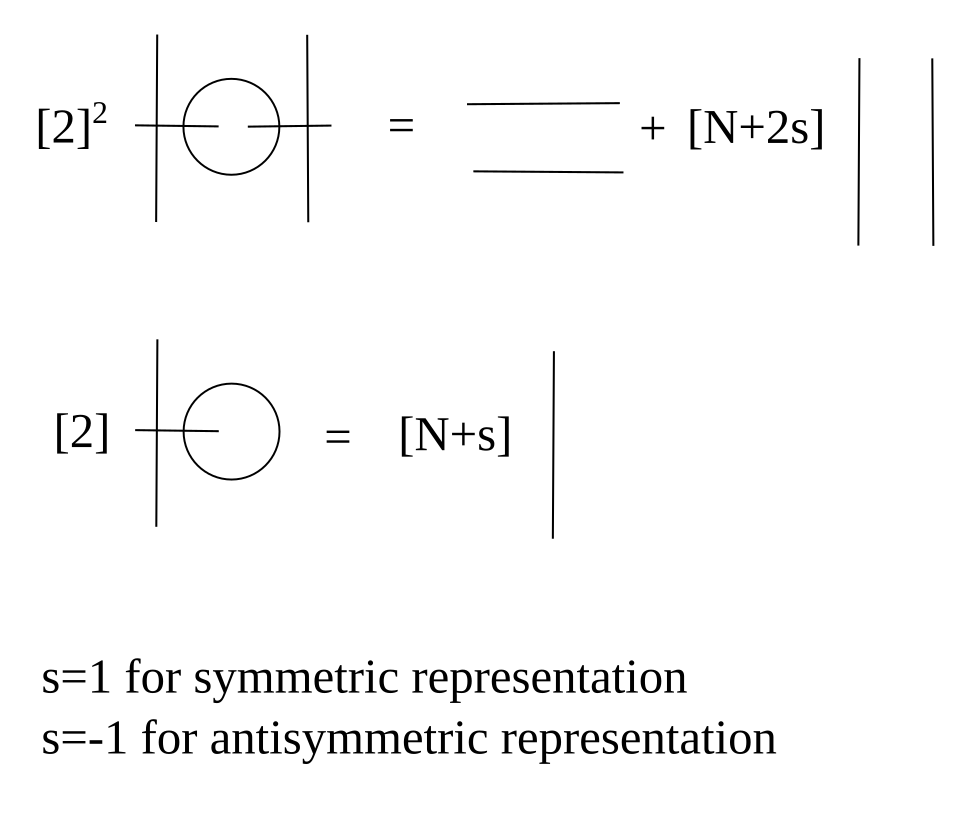}
\caption{\footnotesize Erasure of single circles -- as implied by projector properties. At the l.h.s. we have contractions of projectors, while at the r.h.s. there are only single lines carrying the fundamental representation.}\label{fig:proj-prop}
\end{figure}

In~\cite{bipsym}, we examined that, for example, for twist knots such contractions can be classified: only combinations of projectors from Fig.\,\ref{fig:vertchain2} contribute to the answer for the $[2]$-colored HOMFLY polynomial. However, already for the torus knot $5_1$ (see~\eqref{ex-[2]}) these contractions of projectors are not enough. A clue to the classification of the contractions comes from the reduction rules in Fig.\,\ref{fig:proj-prop}. After the erasing of all circles, we are left with combinations of monomials constructed from just three quantities $\rho =  \frac{[N+2]}{[2]^2}$, $\sigma=\frac{1}{[2]^2}$, $D_{[1]}=D=\frac{\{A\}}{\{q\}}$ forming a {\it multiplicative basis} in the space of contractions of the projectors. Thus, a possible hypothesis is that in the representation $[2]$ the following decomposition is valid
for the reduced HOMFLY polynomial for {\it chiral knots}:
\be
\label{[2]-PD}
\begin{aligned}
      H_{[2]}^{\cal K} &= (A^2 q^2)^{-2n_\bullet}\left( 1 + \sum\limits_{n=1}^l \sum\limits_{k=0}^n \phi_2^k \psi_2^{n-k} \sum\limits_{m=0}^{l} \sum\limits_{j=0}^m \sum\limits_{i\,=\,\max(m-j-1,0)}^{l+m-j} M_{n,k,m,j,i} \rho^j \sigma^{m-j} D_{[1]}^i\right)\,.
\end{aligned}
\ee
Here $l$ is the number of bipartite vertices, and $M_{n,k,m,j,i}$ are non-negative integers. The examples of this decomposition are in~\eqref{ex-[2]}, they were calculated via the discussed planar expansion technique. One can easily get sure that the planar expansion in the representation $[2]$ is much more complicated than in the representation $[1]$. Thus, there again comes an idea of deriving the PD in the representation $[2]$~\eqref{[2]-PD} without using a bipartite diagram (and the meaning of this PD for non-bipartite knots). However, one can check that the result is ambiguous already for 2-locks knots and the ambiguity drastically grows with the growth of the number of bipartite vertices $l$. 

Thus, we need some {\it additional constraints} or other smart ideas on the PD for the representation $[2]$ already for chiral knots. One of such restrictions is in order. As one can easily understand from \cite{bipsym} (comparing Figs.\,\ref{fig:pladeco} and \ref{fig:proje2lock}),
the $\psi_2$-independent pieces of the HOMFLY BEs in the representation $[2]$ can be obtained from BE expressions for $H_\Box$ of the same knots
by substitutions $\phi\longrightarrow \phi_2$ and $D_{[1]}\longrightarrow D_{[2]}$.
These terms are independent of auxiliary parameters like $\Tr_q  \hat\Pi^n$ or $\rho$ and $\sigma$.
Moreover, in $\psi_2$-linear terms these parameters appear just as factors $\Pi_2$.

\subsection{Simple examples}\label{simplexa}

Exhaustive description of PD for twist knots is given in Section 3.6 of \cite{bipsym}, but we prefer to
rewrite it a little differently getting rid of denominators to obtain the expression for the {\it reduced} HOMFLY polynomial.
For this, we need a whole bunch of projectors, see Fig.\,\ref{fig:vertchain2} --
instead of just a single dimension $D_{[2]}:=\frac{\{Aq\}\{A\}}{\{q^2\}\{q\}}$:
in the notation of \cite{bipsym}

\begin{equation}\label{pi-coef}
\begin{aligned}
\pi_n := \frac{\Tr\!_q \hat\Pi^n}{D_{[2]}} &= \frac{1}{[2]^{2n-1}}\left([N+1][N+2]^{n-1}
+ \sum_{i=0}^{n-2} [2]^{n-1-i}[N+1]^{n-2-i}[N+2]^i\right)\,, \\
\frac{\Pi_n}{D_{[2]}}&=\left(\frac{[N+1]}{[2]}\right)^{n-1}=\pi_1^{n-1}\,,
\end{aligned}
\end{equation}
where $\{x\} :=x-x^{-1}$ and $[N+s]:=\frac{\{Aq^s\}}{\{q\}}$.
However, these $\pi_n$ are not sufficient, we need also
its simpler constituents $\rho =  \frac{[N+2]}{[2]^2}$, $\sigma=\frac{1}{[2]^2}$ and $D_{[1]}=D=\frac{\{A\}}{\{q\}}$. The projector-induced variables $\pi_n$ are expressed through the new ones $\rho$, $\sigma$, $D_{[1]}$ as follows:
 \begin{equation}\label{rho-sigma-D}
\begin{aligned}
    D_{[2]} &= \sigma  D_{[1]}^2+\rho D_{[1]} =(\rho+\sigma D_{[1]})D_{[1]} = \pi_1D_{[1]}\,, \\
    \pi_1 & 
    =  \rho+\sigma D_{[1]}\,,   \\
    \pi_2 & 
    = \rho^2+\rho \sigma D_{[1]} +\sigma   = \rho\pi_1 +\sigma\,,  \\
    \pi_3 & 
    = \rho^3+\rho^2\sigma D_{[1]} +\sigma ^2D_{[1]}+2\sigma \rho = \rho\pi_2+\sigma \pi_1\,, \\
    \pi_4 & 
    = \rho^4+\rho^3\sigma D_{[1]} +3\sigma ^2\rho D_{[1]}+3\sigma \rho^2+ \sigma ^3D_{[1]}^2
    = \rho\pi_3+\sigma \pi_1^2\,, \\
    \pi_5 &= \rho \pi_4 + \sigma \pi_1^3\,, \\
    \ldots \\
    \pi_n &= \rho \pi_{n-1} + \sigma \pi_1^{n-2}\,, \\
    \ldots &
\end{aligned}
 \end{equation}
Like in \cite{bipsym}, for representation $[2]$ we will use a slightly special notation for the coefficients
of bipartite expansion:
\begin{equation}\label{planar-coefs-[2]}
   \phi_2 = A^2q^3\{q^2\}\{q\},\qquad \psi_2 = Aq^2[2]\{q^2\}\,. 
\end{equation}
In these terms, a few explicit examples of BE of the reduced HOMFLY polynomials for the representation $[2]$:
\begin{equation}\label{ex-[2]}
\begin{aligned}
    H_{[2]}^{\bigcirc} &=\frac{1}{(A^2 q^2)^2}\left( 1 + D_{[2]} \phi_2 +\pi_1 \psi_2 \right)=1\,, \\
    H^{3_1}_{[2]} &=  \frac{1}{(A^2q^2)^4}\Big(1+ 2D_{[2]}\phi_2  + 2\pi_1\psi_2 + \phi_2^2+ 2\pi_1\phi_2\psi_2 + \pi_2\psi_2^2 \Big)\,, \\
    H_{[2]}^{5_1} &= \frac{1}{(A^2q^2)^8}\bigg(1 + 4 D_{[2]}  \phi_2   + 4 \pi_1 \psi_2  + 3 (D_{[2]}^2+1)\phi_2^2 + 6(D_{[2]}+1) \pi_1\psi_2 \phi_2
+ 3(\pi_2+\pi_1^2)\psi_2^2 + \\
    &+4D_{[2]} \phi_2^3  + 6   (D_{[2]}+1)\pi_1 \phi_2^2 \psi_2
    + 2\Big((D_{[2]}+1)\pi_2+4\pi_1^2\Big) \phi_2\psi_2^2  + 2(\pi_3+ \pi_2\pi_1)\psi_2^3    + \\
     &+\phi_2^4    + 4 \pi_1\phi_2^3 \psi_2    + 3 (\pi_2+\pi_1^2) \phi_2^2 \psi_2^2
      + 2( \pi_3 + \pi_2 \pi_1)  \phi_2\psi_2^3
     + (\underbrace{ \pi_3  \rho  + \pi_2   \sigma}_{\pi_4-\pi_3\pi_1+\pi_2^2}) \psi_2^4 \bigg)\,, \\
     H_{[2]}^{5_2}  &= \frac{1}{(A^2q^2)^6}\Big(1+3\phi_2 D_{[2]}+3\psi_2 \pi_1
 + \phi_2^2(2 +D_{[2]}^2)+2 \phi_2\psi_2 \pi_1 (2+D_{[2]})+\psi_2^2(2 \pi_2 + \pi_1^2)+ \\
 &+\phi_2^3 D_{[2]} + \phi_2^2 \psi_2\pi_1(2 + D_{[2]})+\phi_2\psi_2^2(\pi_2 + 2\pi_1^2)+\psi_2^3 \pi_3\Big)\,, \\
 H^{7_2}_{[2]} &=  \frac{1}{(A^2q^2)^8}\Big(1 + 4D_2\phi_2  + 4\pi_1 \psi_2
+ 3(D_2^2 + 1)\phi_2^2 + 6(D_2+1)\pi_1\phi_2\psi_2  + 3(\pi_2+\pi_1^2)\psi_2^2 + \\
&+ (D_2^2+3)D_2 \phi_2^3  + 3(D_2^2 + D_2 + 2)\pi_1\phi_2^2\psi_2 + 3(\pi_2+\pi_1^2 D_{[2]}+2\pi_1^2)\phi_2\psi_2^2
+ \big(\pi_1^3+3\pi_3\big)\psi_2^3  + \\
&+ D_2^2\phi_2^4 + D_2(D_2+3)\pi_1\phi_2^3\psi_2  + 3(D_2+1)\pi_1^2 \phi_2^2\psi_2^2
+  \big(\pi_3+3\pi_1^3\big)\phi_2\psi_2^3 + \pi_4\psi_2^4\Big)\,.
\end{aligned}
\end{equation}
These examples illustrate the statements of Section~\ref{sec:planar-dec-[2]}. Namely, the twist knots and the unknot of 1 bipartite vertex have BEs in the representation $[2]$ including only powers of $\pi_n$ and $D_{[2]}$. But in general, these combinations of projectors are not enough: one can see the appearance of the last factor $(\pi_3  \rho  + \pi_2   \sigma)$ in the HOMFLY polynomial for the knot $5_1$. It is not expressible through $\pi_n$ and $D_{[2]}$ with positive coefficients\footnote{Note that in this concrete case, one can express this factor in terms of $\pi_n$ but loose positivity instead.} and needs for variables $\rho$, $\sigma$. However, due to~\eqref{rho-sigma-D}, all the present examples can be rewritten in the $\rho$, $\sigma$, $D_{[1]}$ basis.

\subsection{A need for selection rules}

As we see, unlike the case of fundamental representation~\eqref{chirexpanswer}, it seems that there is no substitution of variables in the $[2]$-colored HOMFLY polynomials for chiral knots that makes it expanded in $\psi_2$, $\phi_2$, $\rho$, $\sigma$, $D_{[1]}$ (or other suitable multiplicative basis) as dictated by~\eqref{[2]-PD}. However, one can try to substitute $(A,q)$-monomials in the HOMFLY polynomial in the representation $[2]$ to monomials in $\psi_2$, $\phi_2$, $\rho$, $\sigma$, $D_{[1]}$. A subtlety is that the variables $\rho$, $\sigma$ and $D_{[1]}$ have denominators, so that one should multiply the HOMFLY polynomial by an appropriate factor to avoid dealing with denominators. We also use the discussed in Section~\ref{sec:planar-dec-[2]} fact that $\psi_2$-independent monomials in the BE for the representation $[2]$ can be obtained for the BE in the fundamental representation by the changes $\phi\rightarrow\phi_2$, $D\rightarrow D_{[2]}$ and that $\psi_2$-linear monomials must be proportional to $\pi_1$.

Let us provide the example of the trefoil. And seek for its PD in the representation $[2]$ for the number of bipartite vertices $l=2$ in~\eqref{[2]-PD}.

\medskip

    \noindent {\bf 1.} To get rid of denominators and negative powers in $A$, $q$, we multiply the HOMFLY polynomial by $A^{10} q^{10}\{q^2\}^3$ and also multiply it by the inverse of the framing factor ${\rm Fr}_{[2]}^{-1}=(A^2 q^2)^4$:
    \begin{equation}
    \begin{aligned}
        A^{10} q^{10} \{q^2\}^3 (A^2 q^2)^4 H_{[2]}^{3_1}&=A^{14} q^{28}-3 A^{14} q^{24}+A^{14} q^{22}+\dots
    \end{aligned}
    \end{equation}

    \medskip
    
    \noindent {\bf 2.} The highest degree monomial is $A^{14} q^{28}$, the same one comes only from $\phi_2^2$. The HOMFLY polynomial for {\it knots} also always has the unity. We also know from the BE for the fundamental representation~\eqref{trefoil}, that there must be the addend $2\phi_2 D_{[2]}$ in the BE for the representation $[2]$. We subtract all these monomials and get:
    \begin{equation}
        A^{10} q^{10} \{q^2\}^3 \left((A^2 q^2)^4 H_{[2]}^{3_1}-(1+2\phi_2 D_{[2]}+\phi_2^2)\right)=2 A^{14} q^{26}+A^{14} q^{24}-9 A^{14} q^{22}+\dots
    \end{equation}

    \medskip

    \noindent {\bf 3.} The only allowed combination with the highest degree monomial equal $A^{14} q^{26}$ corresponds to $\phi_2 \psi_2 \pi_1$. It must be subtracted with the highly possible multiple $2$, so there is no more multiples of $\phi_2 \psi_2$ to be subtracted. The resulting expression is:
    \begin{equation}
        A^{10} q^{10} \{q^2\}^3 \left((A^2 q^2)^4 H_{[2]}^{3_1}-(1+2\phi_2 D_{[2]}+\phi_2^2+2\phi_2 \psi_2 \pi_1)\right)=A^{14} q^{24}+A^{14} q^{22}+\dots
    \end{equation}

    \medskip

    \noindent {\bf 4.} There are already $3$ suitable monomials to be subtracted: $D_{[1]}^2 \psi_2^2$, $D_{[1]}\rho \psi_2^2$, $\rho^2 \psi_2^2$. We combine them with undetermined coefficients:
    \begin{equation}
    \begin{aligned}
        A^{10} q^{10} \{q^2\}^3 \left((A^2 q^2)^4 H_{[2]}^{3_1}-(1+2\phi_2 D_{[2]}+\phi_2^2+2\phi_2 \psi_2 \pi_1+a_1 D_{[1]}^2 \psi _2^2+a_2 D_{[1]} \rho \psi _2^2+\left(-a_1-a_2+1\right) \rho^2 \psi _2^2)\right)=\\
        =\left(-4 a_1-2 a_2+1\right) A^{14} q^{22}+\dots
    \end{aligned}
    \end{equation}

\medskip 

\noindent {\bf 5.} At this step, there is $2$ allowed combinations: $D_{[1]}^2 \sigma \psi_2^2$ and $D_{[1]}\sigma \rho \psi_2^2$.

\begin{equation}
\begin{aligned}
    A^{10} q^{10} \{q^2\}^3 \big((A^2 q^2)^4 H_{[2]}^{3_1}-(1+2\phi_2 D_{[2]}+\phi_2^2+2\phi_2 \psi_2 \pi_1+a_1 D_{[1]}^2 \psi _2^2+a_2 D_{[1]} \rho \psi _2^2+\left(-a_1-a_2+1\right) \rho^2 \psi _2^2 + \\ 
    +a_3 D_{[1]}^2 \sigma \psi_2^2+\left(-4 a_1-2 a_2-a_3+1\right) D_{[1]}\sigma \rho \psi_2^2)\big)=\left(-6 a_1-a_2-2 a_3\right) A^{14} q^{20}+\dots
\end{aligned}
\end{equation}

\medskip 

\noindent {\bf 6.} The only $\left(-6 a_1-a_2-2 a_3\right) D_{[1]}^2 \sigma^2 \psi _2^2$ can be subtracted:
\begin{equation}
\begin{aligned}
    A^{10} q^{10} \{q^2\}^3 \big((A^2 q^2)^4 H_{[2]}^{3_1}-(1+2\phi_2 D_{[2]}+\phi_2^2+2\phi_2 \psi_2 \pi_1+a_1 D_{[1]}^2 \psi _2^2+a_2 D_{[1]} \rho \psi _2^2+\left(-a_1-a_2+1\right) \rho^2 \psi _2^2 + \\ 
    +a_3 D_{[1]}^2 \sigma \psi_2^2+\left(-4 a_1-2 a_2-a_3+1\right) D_{[1]}\sigma \rho \psi_2^2+\left(-6 a_1-a_2-2 a_3\right) D_{[1]}^2 \sigma^2 \psi _2^2)\big)=-4 a_1 A^{14} q^{18}+\dots
\end{aligned}
\end{equation}

\medskip

\noindent {\bf 7.} Now we subtract $D_{[1]}^2 \pi_1 \psi_2$ and $D_{[1]} \rho \pi_1 \psi _2$ with appropriate coefficients:
\begin{equation}
\begin{aligned}
    A^{10} q^{10} \{q^2\}^3 \big((A^2 q^2)^4 H_{[2]}^{3_1}-(1+2\phi_2 D_{[2]}+\phi_2^2+2\phi_2 \psi_2 \pi_1+a_1 D_{[1]}^2 \psi _2^2+a_2 D_{[1]} \rho \psi _2^2+\left(-a_1-a_2+1\right) \rho^2 \psi _2^2 + \\ 
    +a_3 D_{[1]}^2 \sigma \psi_2^2 + \left(-4 a_1-2 a_2-a_3+1\right) D_{[1]}\sigma \rho \psi_2^2+ \left(-6 a_1-a_2-2 a_3\right) D_{[1]}^2 \sigma^2 \psi _2^2+\\
    +a_4 D_{[1]}^2 \pi_1 \psi_2+(-4 a_1-a_3-a_4)D_{[1]} \rho \pi_1 \psi _2)\big)=\left(3 a_1+a_3-2 a_4\right) A^{14} q^{16}+\dots
\end{aligned}
\end{equation}

\medskip

\noindent {\bf 8.} The only one monomial fits: $D_{[1]}^2 \pi_1 \sigma \psi _2$.
{\small \begin{equation}
\begin{aligned}
    A^{10} q^{10} \{q^2\}^3 \big((A^2 q^2)^4 H_{[2]}^{3_1}-(1+2\phi_2 D_{[2]}+\phi_2^2+2\phi_2 \psi_2 \pi_1+a_1 D_{[1]}^2 \psi _2^2+a_2 D_{[1]} \rho \psi _2^2+\left(-a_1-a_2+1\right) \rho^2 \psi _2^2 + \\ 
    +a_3 D_{[1]}^2 \sigma \psi_2^2 + \left(-4 a_1-2 a_2-a_3+1\right) D_{[1]}\sigma \rho \psi_2^2+ \left(-6 a_1-a_2-2 a_3\right) D_{[1]}^2 \sigma^2 \psi _2^2+\\
    +a_4 D_{[1]}^2 \pi_1 \psi_2+(-4 a_1-a_3-a_4)D_{[1]} \rho \pi_1 \psi _2+(3 a_1+a_3-2 a_4)D_{[1]}^2 \pi_1 \sigma \psi _2)\big)=\\
    =\left(5 a_1+a_3-a_4\right) A^{14} q^{14}+\left(2 a_1-a_4\right) A^{14} q^{12}+\left(-9 a_1-2 a_3+a_4\right) A^{14} q^{10}+\left(a_4-3 a_1\right) A^{14} q^8+\left(4 a_1+a_3\right) A^{14} q^6+a_1 A^{14} q^4+\\
    +\left(2 a_1+a_2+1\right) A^{12} q^{24}+\dots
\end{aligned}
\end{equation}} 

\medskip

\noindent {\bf 9.} The monomials in the pre-last line cannot be cancelled via our previous method. Thus, we must set $a_1=a_3=a_4\equiv 0$. But then, we get both monomials with coefficients $-a_2$ and $a_2$, and the BE can be positive iff $a_2\equiv 0$ too. The monomial in the last line can be vanished by $\sigma \psi_2^2$:
\begin{equation}
\begin{aligned}
    A^{10} q^{10} \{q^2\}^3 \big((A^2 q^2)^4 H_{[2]}^{3_1}-(1+2\phi_2 D_{[2]}+\phi_2^2+2\phi_2 \psi_2 \pi_1+ \rho^2 \psi _2^2 + D_{[1]}\sigma \rho \psi_2^2+\sigma \psi_2^2)\big)=2 A^{12} q^{20}+\dots
\end{aligned}
\end{equation}

\medskip

\noindent {\bf 10.} The subtraction of $2 \pi_1 \psi_2$ gives us zero at the r.h.s. Thus, the resulting BE is
\begin{equation}
    H_{[2]}^{3_1}=(A^2 q^2)^{-4}(1+2\phi_2 D_{[2]}+\phi_2^2+2\phi_2 \psi_2 \pi_1+ \rho^2 \psi _2^2 + D_{[1]}\sigma \rho \psi_2^2+\sigma \psi_2^2+2 \pi_1 \psi_2)
\end{equation}
what gives exactly the BE for the trefoil from~\eqref{ex-[2]} when substituting $\pi_2=\rho^2+\rho \sigma D_{[1]} +\sigma$. We have summarized all the appeared $(A,q)$-monomials and the corresponding combinations that might appear in the PD for the trefoil in Table~\ref{tab:mon-BE-[2]}.

\medskip

To deal with more complicated examples,
we need to continue the list of $(A,q)$-monomials and the corresponding combinations of $\psi_2$, $\phi_2$, $\rho$, $\sigma$, $D_{[1]}$ including this monomial as the highest degree one (as in Table~\ref{tab:mon-BE-[2]}) to higher powers of $A^2$.
However, then we run into problems.
The number of monomials with the same asymptotics grows fast and this does not allow to expand arbitrary polynomial
so simply.
One could impose further restrictions on the monomials -- say, leaving just one for each asymptotics.
However, then we do not obtain a positive polynomial even in chiral case.
A restriction (selection rule) needs to be {\it clever}, not arbitrary -- and it still remains to be found.
At this moment, we cannot exclude that the {\it chiral} case for higher representations
will appear to be no less ambiguous than the {\it non-chiral} one in the fundamental case:
at the end, there can be different ways to realize projector factors and no canonical choice 
applicable simultaneously to all knots, even to chiral ones.

\begin{table}[h!]
    \centering
    \begin{tabular}{|c|c|}
        \hline \raisebox{-2pt}{$(A,q)$-monomials} & \raisebox{-2pt}{combinations in the PD for $3_1$}  \\ [1.1ex] \hline \hline
        \raisebox{-2pt}{$A^{14} q^{28}$} & \raisebox{-2pt}{$\phi^2$} \\ [1.1ex] \hline
        \raisebox{-2pt}{$A^{14} q^{26}$} & \raisebox{-2pt}{$\pi_1 \phi_2 \psi_2$} \\ [1.1ex] \hline 
        \raisebox{-2pt}{$A^{14} q^{24}$} & \raisebox{-2pt}{$D_{[1]}^2 \psi_2^2$, $D_{[1]}\rho \psi_2^2$, $\rho^2 \psi_2^2$} \\ [1.1ex] \hline
        \raisebox{-2pt}{$A^{14} q^{22}$} & \raisebox{-2pt}{$D_{[1]}^2 \sigma \psi_2^2$, $D_{[1]}\sigma \rho \psi_2^2$} \\ [1.1ex] \hline
        \raisebox{-2pt}{$A^{14} q^{20}$} & \raisebox{-2pt}{$D_{[1]}^2 \sigma^2 \psi _2^2$, $\phi_2 D_{[2]}$} \\ [1.1ex] \hline 
        \raisebox{-2pt}{$A^{14} q^{18}$} & \raisebox{-2pt}{$D_{[1]}^2 \pi_1 \psi_2$, $D_{[1]} \rho \pi_1 \psi _2$} \\ [1.1ex] \hline 
        \raisebox{-2pt}{$A^{14} q^{16}$} & \raisebox{-2pt}{$D_{[1]}^2 \pi_1 \sigma \psi _2$} \\ [1.1ex] \hline 
        \raisebox{-2pt}{$A^{14} q^{14}$, $A^{14} q^{12}$, $A^{14} q^{10}$, $A^{14} q^{8}$, $A^{14} q^{6}$, $A^{14} q^{4}$} & \raisebox{-2pt}{--} \\ [1.1ex] \hline 
        \raisebox{-2pt}{$A^{12} q^{24}$} & \raisebox{-2pt}{$\sigma \psi_2^2$} \\ [1.1ex] \hline 
        \raisebox{-2pt}{$A^{12} q^{20}$} & \raisebox{-2pt}{$\pi_1 \psi_2$} \\ [1.1ex] \hline
    \end{tabular}
    \caption{\footnotesize Correspondence between the $(A,q)$-monomials coming from $A^{10} q^{10} \{q^2\}^3 (A^2 q^2)^4 H_{[2]}^{3_1}$ and the combinations of bipartite variables having such monomials of the highest degree ones.}
    \label{tab:mon-BE-[2]}
\end{table}

\setcounter{equation}{0}
\section{Summary
\label{summary}}

A bipartite knot/link diagram is entirely made from antiparallel lock tangles.
As explained in \cite{bipfund}, the corresponding bipartite expansion (BE) of the reduced HOMFLY polynomial is
\be
H_\Box = A^{-2(n_\bullet- n_\circ)}\sum_k \sum_{i=0}^{n_\bullet} \sum_{j=0}^{n_\circ} {\cal N}_{ijk} D^k \phi^i\bphi^{\,j}
\label{sumBE}
\ee
where $n_\bullet$ and $n_\circ$ are the numbers of tangles with two different orientations, see Fig.\,\ref{fig:pladeco}, 
which we call positive and negative, and ${\cal N}_{ijk}$ are non-negative integers.
If $n_\circ=0$, we call a diagram and the corresponding knot chiral.
Only some knots possess bipartite diagrams, and only some of them are chiral.
Parameters in BE (\ref{sumBE}) are not independent: $G=\phi + \bphi + \phi\bphi D = 0$,
thus, such polynomials should be considered modulo $G$, but after such factorization
the coefficients can become different from the unities and even be negative.

In this paper, we have considered the question if arbitrary symmetric (under the change $q\leftrightarrow -q^{-1}$) polynomial --
in particular, a fundamental HOMFLY polynomial of a given knot -- can be represented in the form
(\ref{sumBE}) with no reference to a bipartite knot diagram.
Our primary answer is the formula/algorithm (\ref{chirexpanswer}), it provides a {\it chiral}
($\bphi$-independent) expansion like (\ref{sumBE}), but for generic polynomials the coefficients
are not obligatory positive.
Then we have the set of possibilities:
\begin{itemize}
\item{} For either $H_\Box(A,q)$ or $H_\Box(A^{-1},q)$ the coefficients are non-negative.
Then we say that the HOMFLY polynomial possesses chiral {\it positive decomposition} (PD).
When a knot is chiral bipartite, PD coincides with BE.
Somewhat unexpectedly, there are also non-bipartite knots, which possess chiral PD.

\item{} If both polynomials have some negative coefficients, this can be cured by addition of
multiples of $G$.
Moreover, this can be done in infinitely many different ways, because all the coefficients in $G$
are positive.
In this way we associate infinitely many non-chiral PD with a given knot and its reduced fundamental HOMFLY.

\item{} If the knot possesses a bipartite realization/diagram, then the corresponding BE
is reproduced by one of such PD.
Different (Reidemeister-equivalent) bipartite diagrams can have different BE, thus some different
PD from the infinite set can appear in the role of BE.

\item{} Still, if the knot is not bipartite, its HOMFLY polynomial still possesses infinitely many non-chiral PD.
One can wonder, if knowing PD, deduced from (\ref{chirexpanswer}), one can decide if it is also a BE,
i.e., that there is a bipartite diagram behind it.
We have discussed two necessary conditions, coming from $D=1$ restriction and from the {\it precursor} Jones polynomial.
Sufficient condition is unknown.

\item{} We have considered one interesting origin of non-bipartite PD, when the fundamental HOMFLY appears
to be a sum of several HOMFLY polynomials for different bipartite knots. Another possibility for a non-bipartite knot is to have a bipartite clone with the same HOMFLY polynomial possessing this PD.

\item{} Beyond fundamental representation BE becomes more involved: there are more planar diagrams,
more parameters like $\phi$, and coefficients are positive polynomials not only in $D$ but of more diverse
structures, coming from conversions of projectors \cite{bipsym}.
As the simplest try, we have attempted to find an analogue of algorithm (\ref{chirexpanswer}) for the representation $[2]$
(already in this case, the colored HOMFLY polynomials are no longer symmetric).
We have provided some evidence, that it can exist, but the algorithm is not yet conclusive.
Moreover, it is not fully clear if such PD and BE are unique even in the chiral case,
or the ambiguity problem is now extended from the non-chiral situation to the chiral one as well.

\end{itemize}

Thus, a seemingly simple story of bipartite knots acquires a new significance --
the planar decomposition, which it implies for arbitrary $N$, gets extended from a special
class of knots (bipartite ones) to all, but in a still not-quite-controllable way.

\section{Conclusion
\label{conc}}

In this paper, we have extended the study of {\it positive decomposition} (PD) of the fundamental HOMFLY polynomials,
lifting it to a universal characteristic of all knots, not only bipartite.
While it is a direct generalization of the Kauffman planar decomposition for bipartite knots,
for other ones it is a new structure requiring a clear definition and understanding.
We have provided a way to get this decomposition, which turns unique for {\it chiral} knots
and exhibits an interesting ambiguity for non-chiral ones.
It is an open question if this ambiguity corresponds to that in the choice of bipartite diagrams.
We have described some restrictions of this kind in Section \ref{obsta}, but it is not yet clear if they are exhaustive.
A {\it canonical} definition of PD in the non-chiral case is still not found.

Next, we have looked for a criterium to distinguish knots which possess bipartite realizations
looking only at their fundamental HOMFLY polynomials.
It looks like consideration of the {\it precursor Jones polynomials} from \cite{bipfund},
also outlined in Section \ref{obsta}, is quite efficient for this purpose.
In variance with consideration of the {\it Alexander ideals}, it uses nothing but the fundamental HOMFLY. 
However, this method requires the knowledge of the Jones polynomials for all links which are ``smaller'' than dictated by a PD under consideration.

The existence of non-chiral PD for  non-bipartite knots is, perhaps, not too striking because of
being ambiguous.
It is much more puzzling that the well-defined {\it chiral} PD can also exist
beyond biparticity.
An amusing possibility has been considered in Section \ref{sec:sum} -- that the HOMFLY polynomial of a non-bipartite knot can be a sum of
the HOMFLY polynomials for bipartite diagrams.
This is an essentially new structure for the theory of knot polynomials,
and can have other interesting applications.

There are plenty of other questions, raised by the discovery of PD -- direct extension of the Kauffman calculus
from $N=2$ to arbitrary $N$.
One of them is the possibility to use it for a generalization of the Khovanov homologies \cite{Kho,Kho2,Kho3}, probably, different
from the Khovanov--Rozansky approach \cite{Kho2,Kho3}.
We hope to return to this and other questions in future publications.

\section*{Acknowledgements}

Our work is partly
funded within the state assignment of NRC Kurchatov institute. 
It is partly supported by the grants of the Foundation for the Advancement of Theoretical Physics and Mathematics “BASIS” (A.A. and A.M.). The work of E.L. is partly supported by the Ministry of Science and Higher Education of the Russian Federation (agreement no. 075–15–2022–287).


\printbibliography

\end{document}